\documentclass[12pt,preprint,trackchanges]{aastex62}

\shorttitle{Discovery of a Third Planet in the Kepler-47 System}
\shortauthors{Orosz et al.}

\begin{document}

\title{Discovery of a Third Transiting Planet in the Kepler-47 Circumbinary 
System}

\correspondingauthor{Jerome A.\ Orosz}
\email{jorosz@sdsu.edu}

\author{Jerome A. Orosz}
\affil{Astronomy Department, San Diego State University,
 5500 Campanile Drive, San Diego, CA 92182-1221, USA}

\author{William F.\ Welsh}
\affil{Astronomy Department, San Diego State University,
 5500 Campanile Drive, San Diego, CA 92182-1221, USA}

\author{Nader Haghighipour} 
\affil{Institute for Astronomy, University of Hawaii-Manoa,
2680 Woodlawn Dr., Honolulu, HI 96822, USA}

\author{Billy Quarles}
\affil{Homer L. Dodge Department of Physics \& Astronomy, 
University of Oklahoma, 440 W. Brooks Street, Norman, OK 73019, USA}

\author{Donald R.\ Short} 
\affil{Astronomy Department, San Diego State University,
 5500 Campanile Drive, San Diego, CA 92182-1221, USA}

\author{Sean M. Mills}
\affil{The Department of Astronomy and Astrophysics, 
The University of Chicago, 5640 S. Ellis Ave., Chicago, IL 60637, USA}
\affil{Department of Astronomy, California Institute
of Technology, 1200 E. California Blvd., Pasadena, CA 91125, USA}

\author{Suman Satyal} 
\affil{The Department of Physics, University of Texas at 
Arlington, 502 Yates Street, Arlington, TX 76019, USA}

\author{Guillermo Torres}
\affil{Harvard-Smithsonian Center for Astrophysics, 60 Garden 
Street, Cambridge, MA 02138, USA}

\author{Eric Agol} 
\affil{Department of Astronomy, University of Washington,
Box 351580, 3910 15th Ave NE, Seattle, WA 98195, USA}

\author{Daniel C.\ Fabrycky}
\affil{The Department of Astronomy and Astrophysics, 
The University of Chicago, 5640 S. Ellis Ave., Chicago, IL 60637, USA}

\author{Daniel Jontof-Hutter} 
\affil{Department of Astronomy \& Astrophysics, The Pennsylvania 
State University, 525 Davey Lab, University Park, PA 16082, USA}
\affil{Department
of Physics, University of the Pacific, 3601 Pacific Avenue, 
Stockton, CA 95211, USA}

\author{Gur Windmiller}
\affil{Astronomy Department, San Diego State University, 
 5500 Campanile Drive, San Diego, CA 92182-1221, USA}

\author{Tobias W. A. M\"uller}
\affil{Institute for Astronomy and Astrophysics, 
University of T\"ubingen,  Auf der Morgenstelle 10, D-72076 Tuebingen, Germany}

\author{Tobias C. Hinse} 
\affil{Department of Astronomy and Space Science,
Chungnam National University, Daejeon 34134,
Republic of Korea}
\affil{Korea Astronomy \& Space Science Institute,
776 Daedukdae-ro, Yuseong-gu, 305-348 Daejeon, Republic of Korea}

\author{William D.\ Cochran}
\affil{The University of Texas at Austin,
Department of Astronomy, 2515 Speedway, Stop C1400, 
Austin, TX 78712-1205, USA}
\affil{McDonald Observatory, The University of Texas at Austin, 
Austin, TX 78712-0259, USA}

\author{Michael Endl} 
\affil{McDonald Observatory, The University of Texas at Austin, 
Austin, TX 78712-0259, USA}

\author{Eric B.\ Ford}
\affil{Department of Astronomy \& Astrophysics, The Pennsylvania 
State University, 525 Davey Lab, University Park, PA 16082, USA}

\author{Tsevi Mazeh}
\affil{School of Physics and Astronomy, Tel Aviv University,
    Tel Aviv 69978, Israel}

\author{Jack J.\ Lissauer}
\affil{Space Science and Astrobiology Division,
MS 245-3,  NASA Ames Research Center, Moffett Field, CA 94035, USA}

\begin{abstract}
Of the nine confirmed transiting circumbinary planet systems, only
Kepler-47 is known to contain more than one planet.  Kepler-47 b (the
``inner planet'') has an orbital period of 49.5 days and a radius of
about $3\,R_{\oplus}$.  Kepler-47 c (the ``outer planet'') has an
orbital period of 303.2 days and a radius of about $4.7\,R_{\oplus}$.
Here we report the discovery of a third planet, Kepler-47 d (the
``middle planet''), which has an orbital period of 187.4 days and a
radius of about $7\,R_{\oplus}$.  The presence of the middle planet
allows us to place much better constraints on the masses of all three
planets, where the $1\sigma$ ranges are less than $26\,M_{\oplus}$,
between $7-43\,M_{\oplus}$, and between $2-5\,M_{\oplus}$ for the
inner, middle, and outer planets, respectively. The middle and outer
planets have low bulk densities, with $\rho_{\rm middle} <
0.68$ g cm$^{-3}$ and $\rho_{\rm outer} < 0.26$ g cm$^{-3}$ at the
$1\sigma$ level.  The two outer planets are ``tightly packed,''
assuming the nominal masses, meaning no other planet could
stably orbit between them. All of the orbits have low eccentricities
and are nearly coplanar, disfavoring violent scattering scenarios and
suggesting gentle migration in the protoplanetary disk.
\end{abstract}


\section{Introduction}

Binary stars are common: for example, roughly half of all Sun-like
stars are found in pairs \citep{Raghavan2010}.  Circumbinary planets
(hereafter, CBPs), i.e., planets that orbit an entire binary star
system, can reveal their presence in a variety of different ways.
Known as P-type systems \citep{Dvorak_1982}, these planets, if
sufficiently large, may affect radial velocities of the stars or the
timing of stellar eclipses \citep{Schwarz_2011}.  Several claims of
the detection of CBPs using eclipse timing variations have been made
\citep[for example][]{Qian_2012a, Qian_2012b,Qian_2013,Lee_2014},
but, the validity of these detections is currently under question
\citep[e.g.,][]{Marsh_2017}.  A CBP may also transit one or
both stars in the binary.  When the system is viewed edge-on (i.e., as
in an eclipsing binary), transits in the light curve provide the most
secure detection of the planet.  To date, ten transiting CBPs have
been discovered with this method
\citep{Doyle_2011,Orosz_2012a,Orosz_2012b,Kostov_2013,
  Kostov_2014,Kostov_2015,Welsh_2012,Welsh_2015,Schwamb_2013} 
in the set of $\sim$2900
eclipsing binary stars\footnote{http://keplerebs.villanova.edu}
\citep{Prsa_2011,Slawson_2011,Kirk_2016} observed with the {\em
  Kepler} telescope \citep{Borucki_2010a, Koch_2010}.  Transits are
particularly valuable because they allow the planet's radius to be
measured, and the large variations in transit durations and the
deviations from periodicity unambiguously demonstrate that the body is
in a circumbinary orbit \citep{Orosz_2012b}.

Kepler-47, the topic of this paper, is the only known multi-planet
circumbinary system, and the multi-planet nature of this system allows
us to study an ensemble of star--planet and planet--planet dynamics. Its
binary stars have an orbital period of $\sim7.5$ days, the shortest
known in any CBP system.  The innermost planet (Kepler-47~b) has an
orbital period of $\sim49.5$ days, and is the smallest of the known
CBPs ($\sim3\, R_{\oplus}$).  The Uranus-size outer planet
(Kepler-47~c) has a $\sim303$ day orbit, placing it well within the
habitable zone \citep{Haghighipour_2013}, i.e.,
the region surrounding a
star where water could exist in a liquid state on a terrestrial planet
\citep{Kopparapu_2013}.  
The existence of the Kepler-47 system, particularly with the newly
discovered middle planet
described herein, shows that multi-planetary systems can form
and survive around close binary stars. This despite theory
suggesting that
strong perturbations from the binary on the protoplanetary
disk most likely inhibit in situ formation in regions close to the
binary
\citep{Meschiari_2012,Paardekooper_2012,
Marzari_2013,Dunhill_2013,Pierens_2013,
Rafikov_2013, Lines_2014,Lines_2015,Lines_2016}.


In the system discovery paper \citep{Orosz_2012b}, eighteen transits
of the inner planet and three of the outer planet were detected. An
additional 0.2\% deep ``orphan'' transit event was noted, a feature
that could not be attributed to the two planets known at that time.
Further {\em Kepler} observations revealed two more transits not
attributable to the two known planets,
suggesting the existence of a
third planetary body and allowing for a 
robust determination of its orbital period.  This, in turn,
allowed us to identify three much weaker transits in the earlier data.
In total, six transits were observed that can be attributed to an
additional planet, Kepler-47~d, orbiting between planets b and c.  In
this work, we provide an updated analysis of the five-body dynamics.
The paper is organized as follows.  We discuss the available
observational material in \S\ref{obsmat}.  The transits of the new
planet and transit times for all of the planets are discussed in
\S\ref{newtransits}.  Measurements of times when the primary
and secondary stars eclipse one another are given in
\S\ref{eclipsetimesection}.  Our five-body photodynamical model is
described in \S\ref{photodynamical}.  The mass estimates for the
planets, the long-term dynamical stability of the system, the
evolutionary status of the two stars, and the locations of the
middle and outer planets to the habitable zone of the binary are
discussed in \S\ref{discussion}.  We summarize our results in
\S\ref{summary}.

\section{Observational Data}\label{obsmat}

\subsection{Optical Light Curves}

Kepler-47 (KIC 10020423, KOI 3154, 2MASS J19411149+4655136)
was observed in long-cadence mode
($\approx 30$ minute samples) from {\em
  Kepler} Quarter 1 (BJD 2,454,964.51 or 2009 May 13) through the end
of the nominal spacecraft operation in the second month of Quarter 17
(BJD 2,456,424.00 or 2013 May 11).  There were a total of 65,428
long-cadence observations during this span (including flagged data),
for a duty cycle of 91.6\%.  Starting in Quarter 14 (BJD 2,456,107.13
or 2012 June 28), Kepler-47 was on the list of short-cadence targets
($\approx 1$ minute samples), where it remained until the cessation of
normal spacecraft operations.  There were a total of 409,046
short-cadence observations, including flagged data.  We downloaded the
FITS tables from the Mikulski Archive for Space Telescopes
(MAST). These files were processed using the SOC release 9.0.3, and
they include a $\approx 1$ minute correction to the time stamps that
was not available at the time of original work reported in
\citet{Orosz_2012b}.

The {\em Kepler} light curves of Kepler-47 required detrending, 
owing to
instrumental trends and modulations due to star spots.  We
arrived at our final detrended light curves using a two-step process.
First, an interactive technique described in \citet{Orosz_2012b} and
\citet{Bass_2012} was used.  In this step, the light curve was broken
up into segments using data gaps and sudden jumps in the flux as end
points. For each segment, data in the eclipses and transits were
masked out, and cubic splines were fit to the remaining data.  The
segments were normalized by the spline fits, and pieced together to
produce a detrended light curve.  While effective, this technique is
not entirely reproducible; it is also difficult to use with the
short-cadence data, owing to limitations in our implementation of the
method.  Thus, an automated and reproducible detrending algorithm was
devised as follows: The light curve that was 
detrended manually was
modeled as described in \S\ref{photodynamical}.  Once a good fit was
found, the model was used to  precisely determine the times of the
eclipses and transits and the durations of these events.  The
times and durations were then
used to determine which data points occurred during an
eclipse or a transit event, which allowed us to assign those points
zero weight during the normalization.

In our original work, only long-cadence {\em Kepler} data were
available to us.  As noted above, Kepler-47 was on the list of 
short-cadence targets for {\em Kepler} Quarters Q14 through Q17.  In our
view, using short-cadence data (when available) instead of 
long-cadence data is preferable because the ingress and egress phases of
eclipses and transits are better-resolved.  Thus, a ``raw'' light
curve using long-cadence data from Quarter 1 through Quarter 13 and
short-cadence data from Quarter 14 through Quarter 17 was made, and
data with quality flags with values greater than 16 were eliminated
(in our experience, obviously poor cadences almost always have
data quality flags greater than 16).  The overall duty cycle of
the retained data was 91.6\%.  Segments centered at the mid-eclipse
and transit times with widths of three times the duration of the event
in question were extracted from the raw light curve.  In a few cases,
these segments included both an eclipse event and a transit event; 
also
in a few cases, the segments were truncated owing to gaps in the data.
A fifth-order Legendre polynomial was fit to each segment,
where data that occurred during an eclipse or a transit event were
given zero weight.  The normalized segments were assembled to give the
final normalized and trimmed light curve.  We also produced a
normalized light curve consisting entirely of long-cadence
observations for the purposes of plotting the relatively weak
transits.

\subsection{Radial Velocities}

\begin{figure}[t]
\begin{center}
\includegraphics[width=0.65\textwidth,angle=-90]{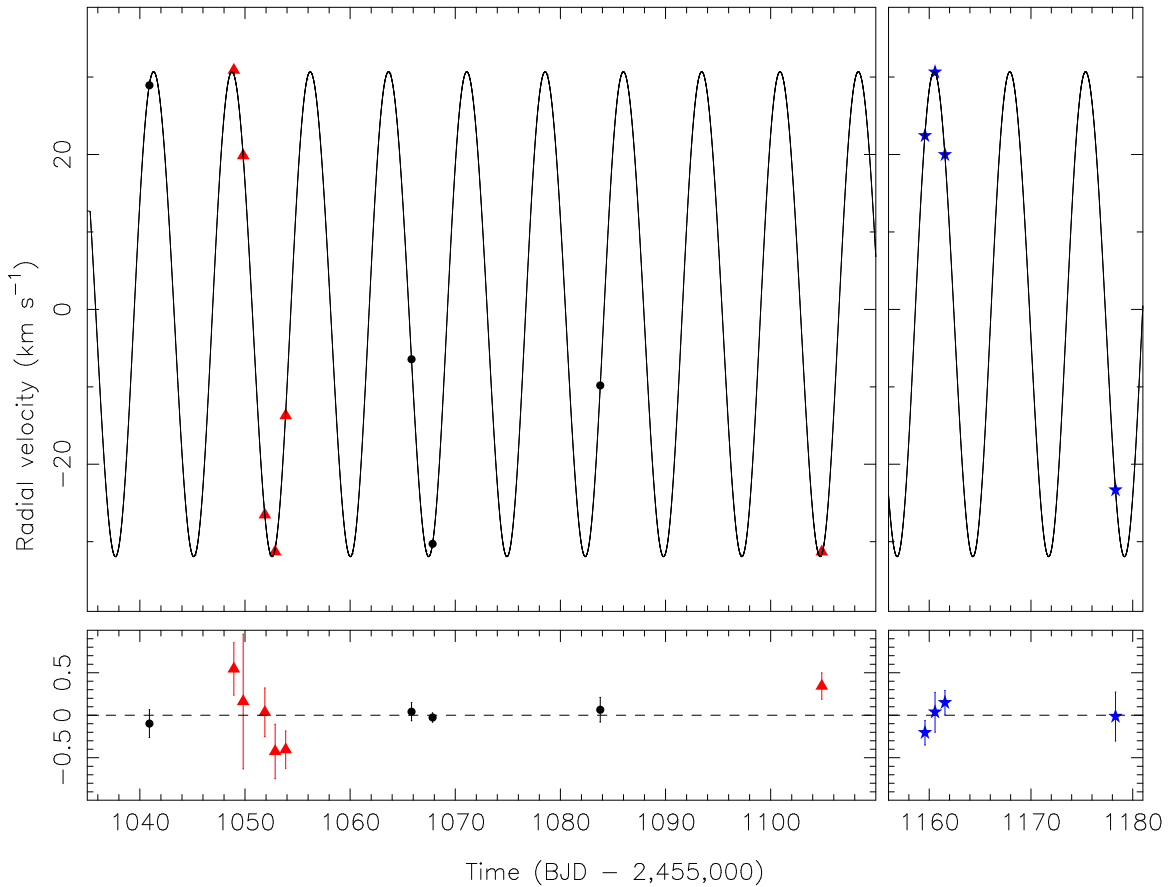}
\caption{Top: The radial velocity measurements (in km
s$^{-1}$) of the Kepler-47 primary are shown along with the
best-fitting model. The horizontal axis gives the time in
days, where the abbreviation BJD used here and in subsquent
figures means ``barycentric Julian date.''  The black
filled circles denote measurements from the Hobby--Eberly Telescope,
the red filled triangles denote measurements from the
McDonald 2.7m, and the blue filled stars denote measurements
taken from Kostov et al.\ (2013).  For clarity, the model
curve in the $\approx 50$ day gap between the radial velocities
given in \cite{Orosz_2012b} and those given in \cite{Kostov_2013}
is not shown.  Bottom: The residuals of the model fit,
where the symbols have the same meaning as in the top
panels.\label{showRV} }
\end{center}
\end{figure}

Owing to the relative faintness of the secondary star (it is
about 0.5\% as bright as the primary in the optical),
Kepler-47 is a single-lined binary, so only radial velocity
measurements for the primary star are available. We used 11
radial velocity measurements of the primary in the original work
\citep{Orosz_2012b}.  Four of these came from the HRS spectrograph on
the Hobby--Eberly Telescope (HET), six of these came from the Tull
Coud\'e spectrograph on the Harlan J. Smith 2.7m telescope (HJST) at
McDonald Observatory, and one measurement came from the HIRES
spectrograph on the Keck I telescope.  Since that time,
\citet{Kostov_2013} have
provided an additional four measurements with the
SOPHIE spectrograph on the 1.93m telescope at Haute-Provence
Observatory.  These measurements are summarized in Table \ref{tabRV}
and displayed in Figure \ref{showRV}. 
Small differences were found in the velocity zero-points between the different
observatories, so these were fitted for and removed.  To arrive at our
final radial velocity curve, we subtracted systemic offsets of 4.5680
km s$^{-1}$ for the HJST radial velocities, 4.5999 km s$^{-1}$ for the
HET radial velocities, and 4.2685 km s$^{-1}$ for the
\citet{Kostov_2013} velocities.  Given that we have only one
measurement from Keck, we cannot determine the systematic offset and
we do not use that measurement in our analysis.  The
uncertainties in the individual measurements in each set were
scaled to give $\chi^2=N$, where $N=14$ is the number of radial
velocity measurements.  The offset velocities with the scaled
uncertainties are also given in Table \ref{tabRV}.

\section{New Transits and Transit Times}\label{newtransits}

\begin{figure}
\begin{center}
\includegraphics[width=0.6\textwidth,angle=-90]{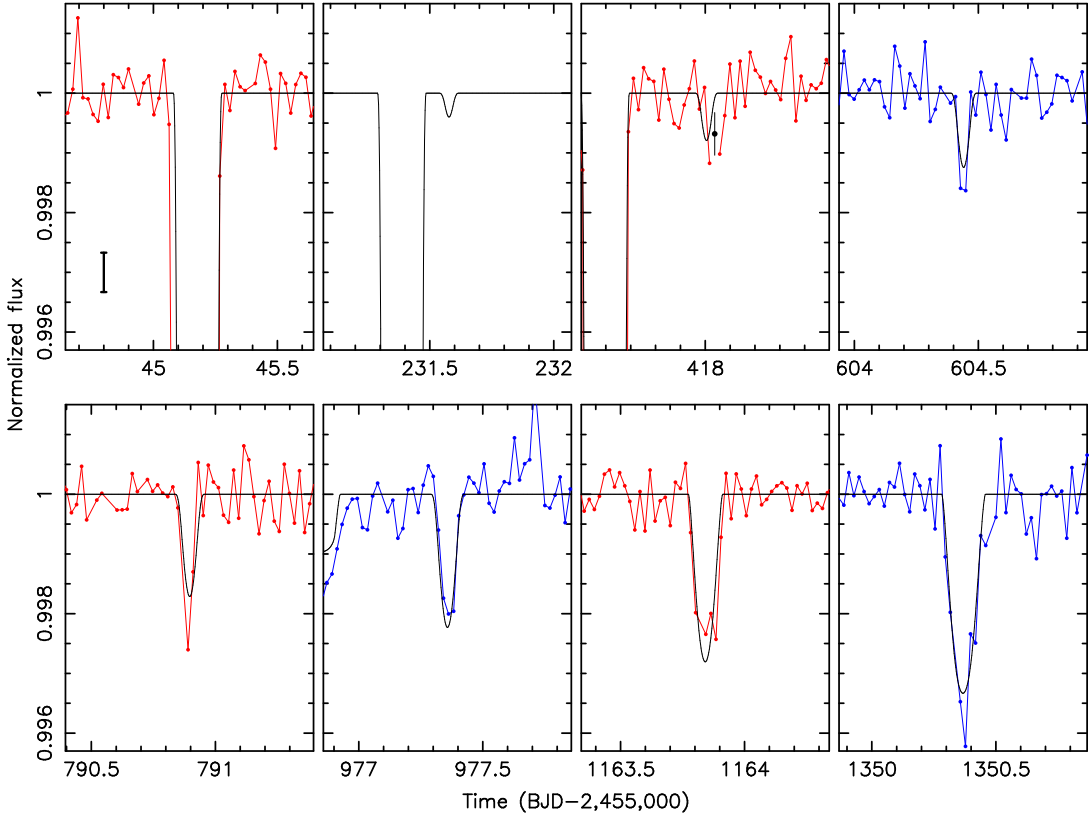}
\caption{The detrended and normalized {\em Kepler} light curves are
shown for the six observed transits of the middle planet
(across the primary star).  The color of the points denotes
the observing season:  red points are
used for Q2, Q6, Q10, and Q14; 
blue points are used
for Q4, Q8, Q12, and Q16.  For clarity, the error bars on
the points are not shown (a representative error bar is shown
in the upper left panel).  A total of 
eight primary transits
occurred during the span of the {\em Kepler} observations, but one
transit with a very small amplitude occurred during a primary
eclipse (first panel) and another transit was missed during a gap in the
data (second panel).  The models for the missed events are shown for
completeness.  The ``orphan'' transit from the discovery paper is
the one at day 977.4 (a transit of the inner planet across
the primary occurred at near day 976.9).  The transit near day
418.0 had one point affected by a cosmic ray (data quality flag
8192), so that point was not used in the analysis.  The black point
with the error bar shows where the cosmic-ray-corrected flux would
have been had it not been excluded.  The last two transits were
observed in short-cadence mode (one-minute sampling), but 
we show the long cadence data here for consistency. 
\label{showmiddle} }
\end{center}
\end{figure}

Figure \ref{showmiddle} 
shows the transits of the middle planet across the primary star. 
During the nominal {\em Kepler} mission, this planet
transited the primary eight times. However, we were able to 
observe only six of these transits. The first transit  
near day 45.15 was blended with an eclipse of the primary, and 
the second transit near day 231.58 was lost due to a gap 
in the data.
As discussed
further below, the primary transits of the middle planet were
getting deeper with time, owing to nodal precession of its
orbit.  The three weak primary transits near days 418.0,
604.5, and 790.9 were not identified until a preliminary orbital
solution based on the last three observed primary transits was
available.  According to our best-fitting model discussed
below, the middle planet did not transit the secondary star
during the nominal {\em Kepler} mission.  Even if transits of the
secondary star did occur, they would be nearly impossible to see in
the data, owing to the faintness of the secondary star.

\begin{figure} 
\begin{center}
\includegraphics[angle = -90,width=0.85\textwidth]{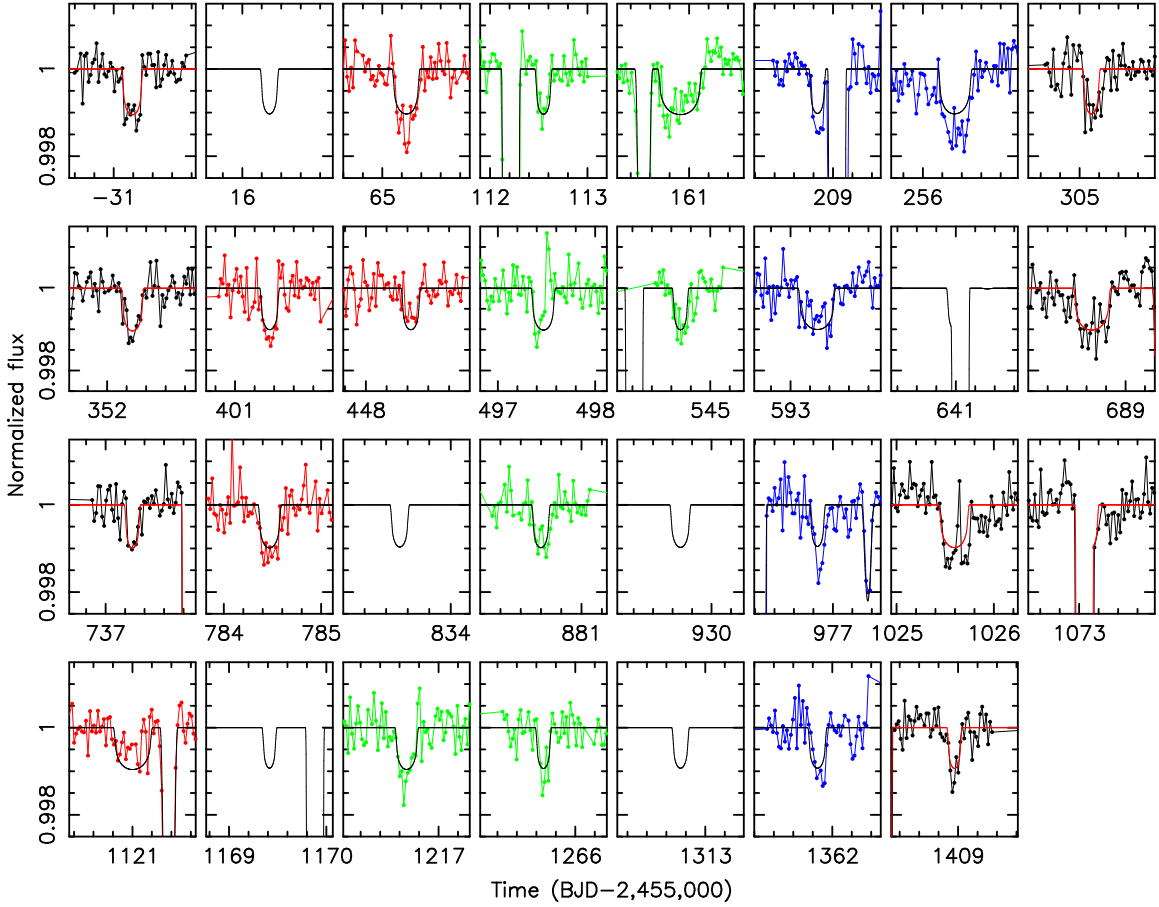}
\caption{Normalized light curves displaying the 25 observed transits
of the inner planet across the primary and the best-fitting
model are shown. The color scheme is similar to that in Figure
\ref{showmiddle}:  black points for Q1, Q5, Q9, Q13, and Q17;
red for Q2, Q6, Q10, and Q14; green for Q3, Q7, Q11, and Q15; and
blue for Q4, Q8, Q12, and Q16.  A total of 31 primary
transits occurred during the span of the {\em Kepler} observations,
but six were missed during gaps in the data.  The models for
the missed events are shown for completeness.  The last five
transits were observed in short-cadence mode (one-minute sampling),
but for consistency we show the long-cadence data
here for consistency.
\label{showinner}}
\end{center}
\end{figure} 

\begin{figure} 
\begin{center}
\includegraphics[angle = -90,width=0.85\textwidth]{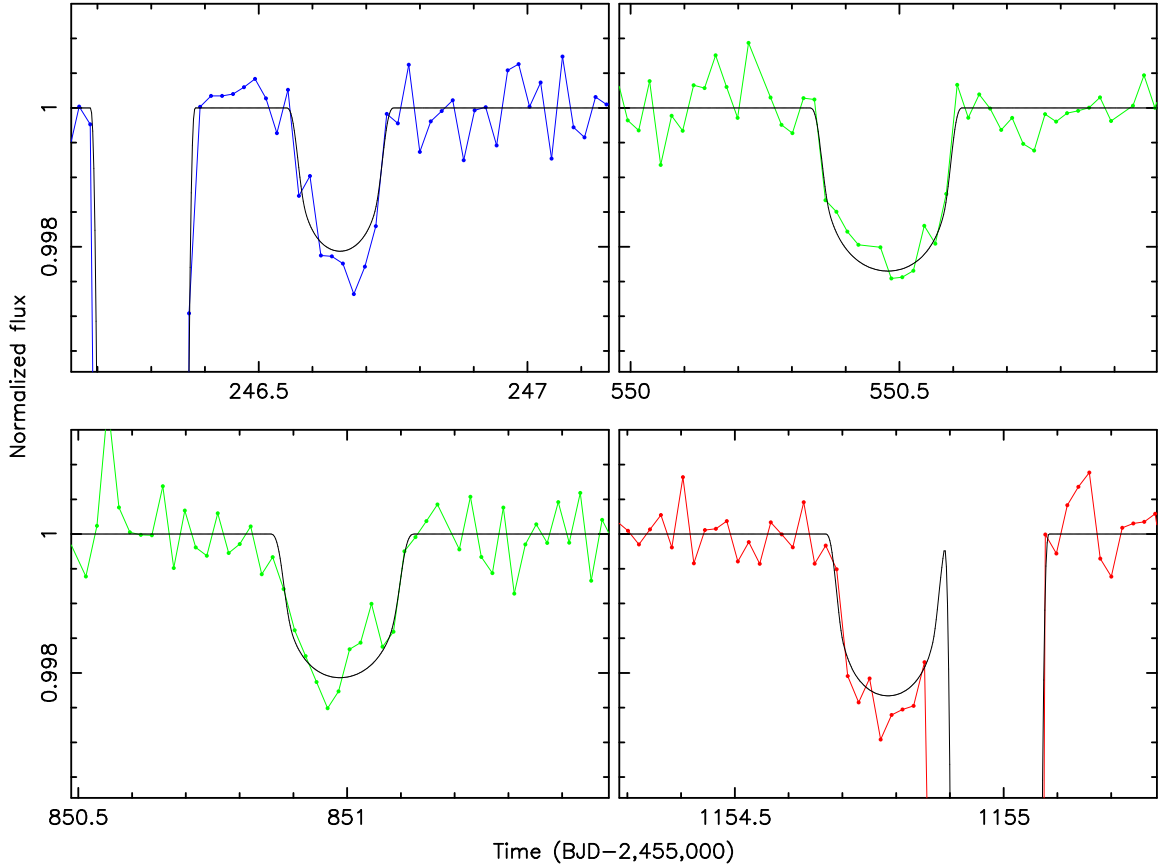}
\caption{Normalized light curves displaying the four 
observed transits of
the outer planet across the primary 
on top of the best-fitting
model. The color scheme is the same as in Figure
\ref{showinner}.  The last transit was observed in short-cadence
mode (one-minute sampling), but  we show the
long cadence data here for consistency.
\label{showouter}}
\end{center}
\end{figure} 

There were a total of 25 transits of the inner planet Kepler-47 b
across the primary observed in the complete {\em Kepler} data
set, seven of which are shown here for the first time (see Figure
\ref{showinner}).  Six primary transits of the inner planet
were missed due to gaps in the data.  A total of four transits of the
outer planet Kepler-47c across the primary are available in
the complete {\em Kepler} data set (see Figure \ref{showouter}), and
fortunately, none were missed.  According to our best-fitting
model discussed below, the inner planet had seven transits across
the secondary star, and the outer planet had one transit across the
secondary star during the nominal {\em Kepler} mission.  All of
these secondary transits had impact parameters larger than 0.75
and
are not seen above the noise level in the data, owing to the extreme
faintness of the secondary star.  Hereafter, when we use the term
``transit,'' it is understood to be a transit of a planet across the
primary star.

We measured the transit times for each of the planets using the ELC
code \citep{Orosz_2000}.  Separate data files were made from the
normalized light curve for each event, where between 0.5 and 1.0 days
of out-of-eclipse data were kept on either side of the transit event,
when possible.  In cases where the transit events occurred near
eclipses, the eclipses were trimmed from the data.

To  model the profile of a planet transit across a single star
(assuming a circular orbit),
one needs to specify the time of mid-transit, the orbital
period, the inclination (or impact parameter), the radius of the star,
the ratio of the stellar and planetary radii, and the
limb darkening
parameters.  Given these parameters, the
\citet{Mandel_2002} algorithm can be used to produce a model light
curve.  Obviously, in the case of Kepler-47, the primary is part of a
binary system and not a single star.  Nevertheless, the shapes of the
transits can still be matched with a suitable change in the orbital
period of the simple model.  Because the transit
shapes are well-matched, our model of a single
planet transiting a single-star
should produce reliable times of mid-transit.

\begin{figure}
\begin{center}
\includegraphics[width=0.65\textwidth,angle=-90]{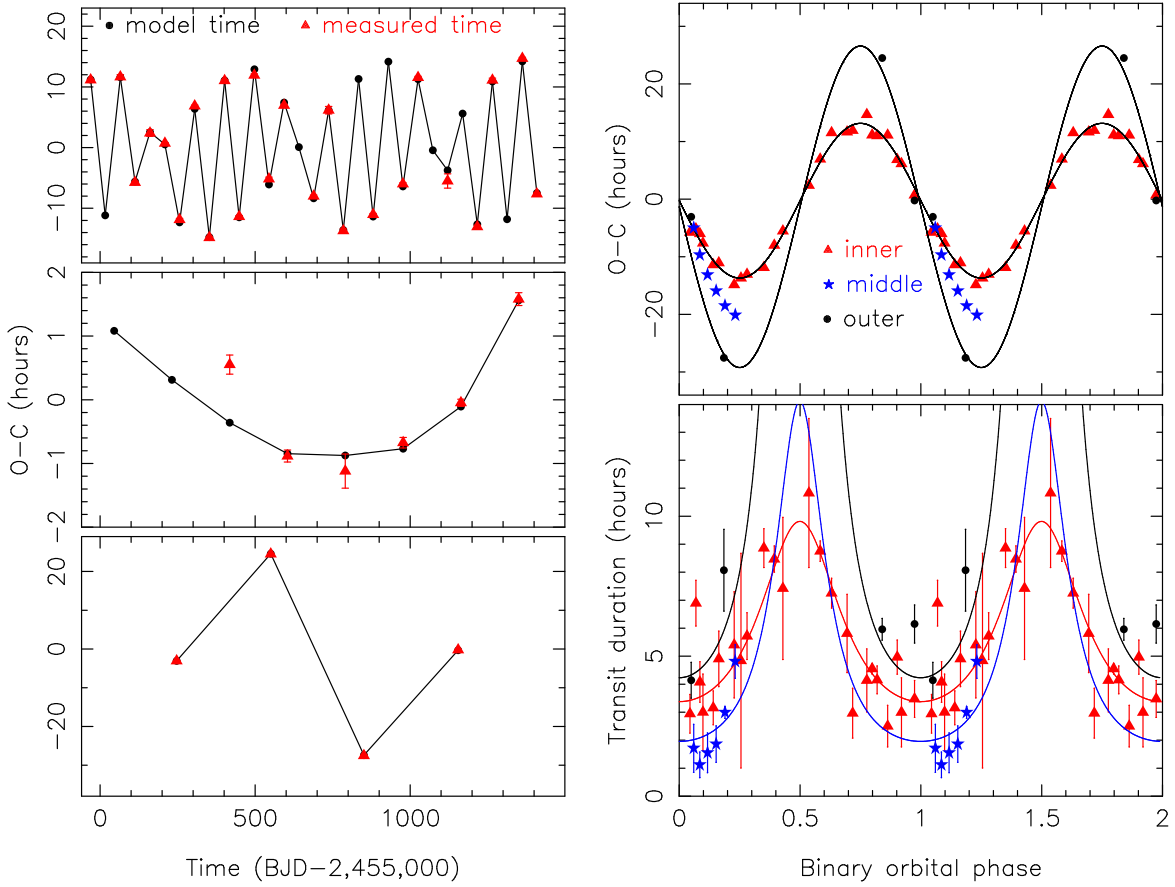}
\caption{Left: The $O-C$ times of transit
are shown for the inner, middle, and outer planets (top to
bottom). The ``computed'' times are the expected times based on a
linear ephemeris.  The red triangles show the measured deviations,
and the filled black circles show the photodynamical model
predictions.  Top Right: The $O-C$ values are now plotted vs.\ the
binary orbital phase (two cycles are shown for clarity).  The black
curves are sine functions fitted to the empirical 
$O-C$ points,
which closely match the predicted variations of the inner
and outer planets assuming circular, edge-on orbits.  Because the
transits of the middle planet span a limited range in binary phase,
due to the proximity of the planet's orbit to the 25:1 mean
motion resonance with the binary, we do not use the $O-C$
values shown in the middle left panel, and we do not
show the
fitted sine curve.  Instead, we used a model covering 3500 days
to determine a linear ephemeris that was then used to compute the
$O-C$ values that are actually shown.  Bottom right: Transit durations
are plotted against binary phase, using the same colors and
symbols used for the upper right panel.  The solid curves are
expected durations using the analytic expressions given in
\cite{Kostov_2014}, which assume circular orbits for the planets
and constant impact parameters (we used $b=0.45$, 0.63, and 0.92
for the inner, middle, and outer planets, respectively).  
\label{plotmultiplanetOC}}
\end{center}
\end{figure}

The model fits to each individual transit were optimized using both a
genetic algorithm and a Differential Evolution Monte Carlo Markov
Chain (DE-MCMC) routine \citep{terBraak_2006}.  Once a best-fitting
model was found, the uncertainties on the observations were scaled to
give a reduced $\chi^2$ of 1 and the model was re-optimized.  The
lower ($\sigma_{\rm low}$) upper ($\sigma_{\rm high}$)
$1\sigma$ uncertainties were taken to be the interval in the
time where $\chi^2=\chi^2_{\rm min}+1$  on the
low and high
sides of the best-fitting time, respectively.  Table
\ref{measuredtransits} gives the measured times and durations 
along with
their uncertainties, where the adopted uncertainty on each time was
the larger of $\sigma_{{\rm low}}$ or $\sigma_{{\rm high}}$.  For
completeness, we also give the corresponding times and durations found
from the best-fitting photodynamical model described below.

There is a point that was affected by a cosmic ray (data quality 
flag
8192) close to the middle of the transit of the middle planet, 
near day
418.0 (see Figure \ref{showmiddle}).  That point was not used in the
analysis.  When that transit is plotted with all of the points
connected by lines and without the bad point, it appears the transit
is $\approx 30-45$ minutes late relative to the model.  The formal
uncertainty on the measured transit time with the bad point excluded
is about none minutes.  There is a corrected flux value for the time in
question, and the black point with the error bar in Figure
\ref{showmiddle} shows where that corrected flux would have been had
that observation been in the detrended light curve. 
The transit looks less convincing 
 when that point
is shown,
because its flux value is a
bit higher than the flux values on either side.  We did some Monte
Carlo simulations where the bad point was put back into the light
curve, but with a flux value drawn from a Gaussian distribution
centered on the corrected value and with a standard deviation equal to
the reported uncertainty.  The measured transit time was shifted
earlier by about three minutes from where it was without the
questionable point included.

As discussed in \citet{Orosz_2012b}, the planet transit times are not
expected to follow a simple linear ephemeris--owing mostly to
the motion of the primary star around the system barycenter (because
the location of the primary star changes as it orbits, transits can
occur early or late relative to a stationary star), and to a
lesser extent, perturbations from the binary \citep{Agol_2005}.  We
fitted each set of transit times to linear ephemerides, and formed
``Observed'' minus ``Computed'' (``$O-C$'') 
diagrams that are shown in Figure
\ref{plotmultiplanetOC}.  The middle planet has 
$O-C$ variations of a
few hours, which is strong evidence that the transits are due to a
circumbinary body and are not from a background blend (a common
false-positive scenario for planets around single stars).  
The inner
planet has maximum $O-C$ variations of about 25 hr, and the outer
planet has maximum $O-C$ variations of nearly 50 hr.  Moreover, the
$O-C$ variations for all three planets vary cyclically on the {\em
  binary star's} orbit.  We define phase 0.0 as the time of the
primary eclipse.  Given this definition, the 
$O-C$ variation is zero at
phases 0.0 and 0.5 because the projected displacement of the star from
the center of mass is at a minimum.  Near phases 0.25 and 0.75, the
star is at the projected ends of its orbit, and the timing variations
are largest.  If the binary system and the planet both have circular
orbits, then the $O-C$ variation with orbital phase will be sinusoidal;
see Figure \ref{plotmultiplanetOC}.  Note that the
transits of the middle planet span a limited range in binary phase
due to the roughly 25:1 ratio
of the  planet's orbital period to the
binary orbital period,
so only a small portion
of the expected sinusoidal variation is seen.  This limited
range in binary phase has no adverse effect on the mass determination
from the photodynamical model discussed below.

Finally, the planet transit durations depend on the relative projected
velocities of the primary star and the planet.  If a transit occurs
near primary eclipse, the bodies are moving in opposite directions,
resulting in a narrow (short-duration) transit.  For transits near
secondary eclipse, the bodies are moving in the same direction,
yielding a longer duration transit; 
see Figure \ref{plotmultiplanetOC}.
Note that, if the planet transits near the stellar limb rather than
the
center, then the duration can be short at any phase--this is the case for
the middle planet.  \cite{Kostov_2014} give an analytic
expression to compute the transit durations as a function of binary
phase, assuming a circular orbit for the planet and a constant
impact parameter.  Figure \ref{plotmultiplanetOC} shows the model
duration curves for each planet, where we used $b=0.45$, 0.63, and
0.92 for the inner, middle, and outer planets, respectively.  The
model curves do a reasonable job of fitting the observed durations,
although the scatter is somewhat large given that individual
transits can have an impact parameter that is quite different than
the mean; see Table \ref{measuredtransits}.

\section{Stellar Eclipse Times and 
Corrections}\label{eclipsetimesection}

We measured the times of the primary and secondary eclipses using the
technique outlined in \citet{Welsh_2012} and \citet{Orosz_2012b}, and
the results are given in Table \ref{eclipsetimes}.  Note 
that the cycle
numbers given for the secondary eclipses are not exactly half integers
because the orbit is eccentric.

\begin{figure} 
\begin{center}
\includegraphics[scale=0.58]{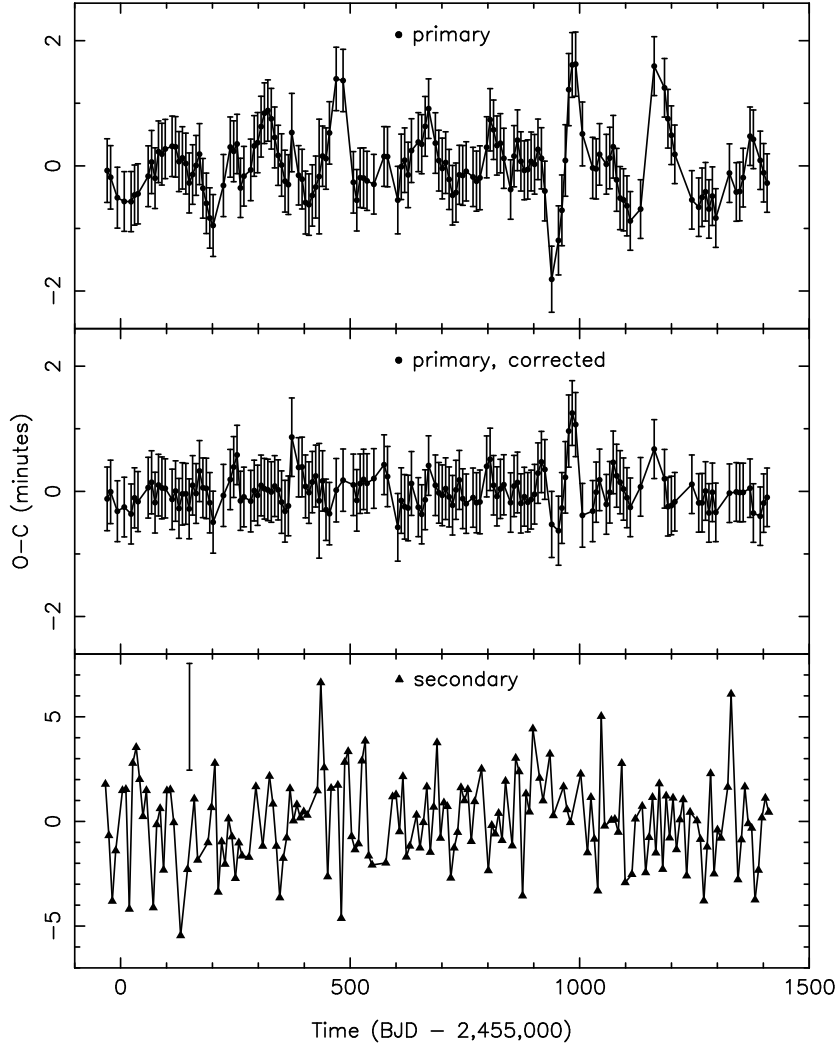}
\caption{Top: The $O-C$ residuals from the linear ephemeris fit for the
primary eclipses.  There are coherent deviations of up to two
minutes seen, with a quasiperiod of about 178 days as discussed in
\citet{Orosz_2012b}.  Middle: The $O-C$ residuals from the linear
ephemeris fit for the primary eclipse times that have been corrected
for the effects of star spots.  The coherent deviations seen in the
top panel have been mostly removed.  Bottom: The $O-C$ residuals from
the linear ephemeris fit for the secondary eclipse times.  Note the
change in the vertical scale. For clarity, the error bars for
the individual points have been omitted.  Instead, the average
uncertainty of 2.54 minutes is shown by the isolated error bar.
\label{plotOC}}
\end{center}
\end{figure} 

\begin{figure} 
\begin{center}
\includegraphics[angle = -90,scale=0.5]{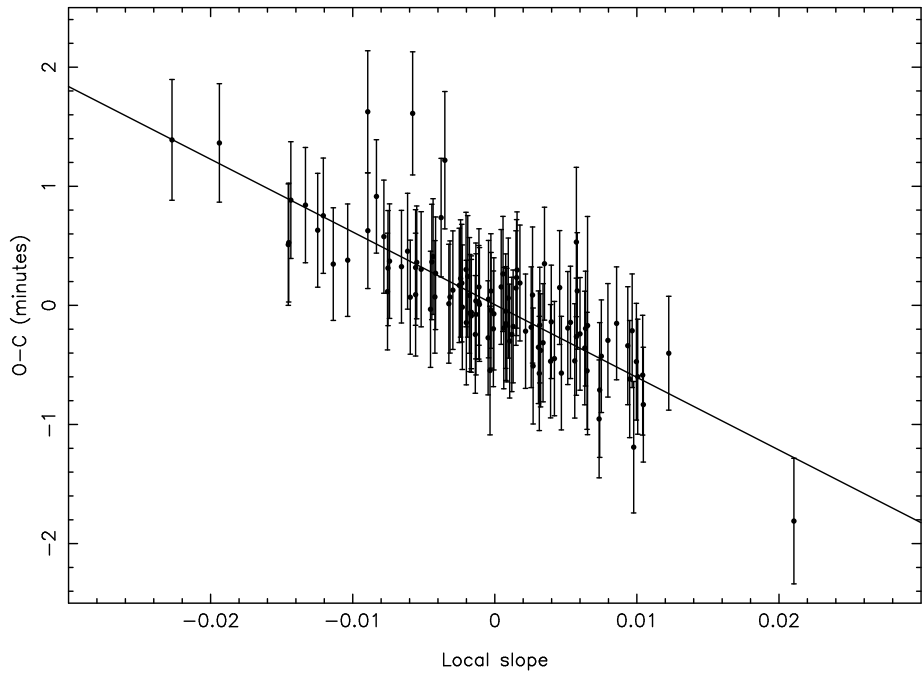}
\caption{The correlation of the residual $O-C$ time and the local light
curve slope near primary eclipse. The correlation between the
primary eclipse $O-C$ times and the local slope in the SAP light curve
is shown.  The best-fitting line has a coefficient of correlation 
$r=-0.84$.
\label{slopevsoc}}
\end{center}
\end{figure}

The times of the primary eclipses were fitted to a linear ephemeris.
The $O-C$ residual times were computed and are shown in Figure
\ref{plotOC}.  As was the case in our earlier work
\citep{Orosz_2012b}, there are coherent modulations of up to about two
minutes, with a quasiperiod near 178 days.  These timing anomalies are
caused by star spots on the primary that are partially or fully
covered during the primary eclipse.  When this occurs, the eclipse
profile is not symmetric in time.  As discussed in \citet{Orosz_2012b}
and in \citet{Welsh_2015}, the shift in the measured eclipse time will
depend on where the spot is seen on the face of the primary star and
when the secondary passes over it.  Also, the slope of the light curve
near primary eclipse depends on where the spot is on the primary as it
rotates into and out of view.  Therefore, 
a correlation is expected 
between the local light curve slope and the $O-C$ residual
\citep{Holczer_2015,Mazeh_2015}, which is shown in Figure
\ref{slopevsoc}.  A linear function was fitted to the data shown in
Figure \ref{slopevsoc} and used to statistically correct the times of
the primary eclipses. A new linear ephemeris was fitted to the
corrected eclipse times, and the $O-C$ residuals are shown in the middle
panel of Figure \ref{plotOC}.  Apart from a feature near day 1000, the
scatter in the residuals has been greatly reduced compared to the the
residuals shown in the top panel of the figure.  No spot-induced
variations were seen in the times of the secondary eclipse, so no
correction to those times was applied.

The best-fitting ephemerides for the corrected primary eclipse times and
the secondary eclipse times are: 

\vspace{1em}
\begin{tabular}{rcr@{\,$\pm$\,}lll}
$P_A$ & = & $7.44837568$ & $0.00000029$ d    & Kepler-47 primary & (1) \\
$P_B$ & = & $7.44837596$ & $0.00000193$ d   & Kepler-47 secondary & \\
$T_0(A)$ &=& BJD $2,454,963.246137$ &    $0.000032$ & Kepler-47 primary & \\
$T_0(B)$ &=& BJD $2,454,959.427964$  &  $0.000228$ & Kepler-47 secondary & \\
\end{tabular}

\vspace{1em}
\noindent The difference between the primary and secondary periods is
$0.02\pm 0.17$ s.  The fitted period from the secondary eclipses
is formally longer than the fitted period from the primary eclipses,
although the difference is not significant.  
This lack of a significant
difference between the two periods means the effect of the planets on
the eclipse timings is unmeasurable, and the masses of the planets are
constrained by planet-planet interactions, not planet--binary
interactions (see \S\ref{constraints}).

\section{Photodynamical Modeling of the Light and Velocity 
Curves}\label{photodynamical}

The complete Kepler-47 light and velocity curves were modeled using
the ELC code. Given five bodies with some initial
positions and velocities, the Newtonian equations of motion can
be
integrated, thereby giving the positions and velocities of each body
as a function of time.  Then, given the position of each body on the
plane of the sky at a specific time (corrected for light travel time)
and information about their radiative properties, the light curve can
be computed.  In the discussion below, we give details of the
photodynamical model and its application to the Kepler-47 data.

\subsection{ODE Integrator}

The numerical integrator that
ELC uses is a symplectic, 12th-order Gaussian
Runge-Kutta (GRK) routine based on methods and codes devised by
\citet{Hairer_2006}.  For situations where only the mutual
gravitational forces between the five bodies are considered, the
ordinary differential equations (ODEs) are second-order and can be
written such that only the positions of the bodies appear explicitly.
GRK is a collocation method in which a system of equations based on
the Gaussian quadrature nodes must be solved at each time step.  In
particular, for the 12th-order method, there are six nodes that
determine the coefficients of the collocation polynomial.  These
coefficients are solved for iteratively.  With a suitable initial
guess, the convergence of this iteration is quadratic--and hence, very
fast.  If needed, the equations of motion can be modified to account
for extra effects such as precession due to General Relativity (GR)
and/or precession due to tidal bulges on the stars
\citep{Eggleton_1998,Mardling_2002}.  In the case of the GR
correction, the velocities of the bodies appear explicitly in the
equations of motion, and the first-order ODEs must be solved using a
different iteration scheme.  This iteration scheme is only to
first order, so the convergence is somewhat slower than the
second-order scheme.

Many of the symplectic integrators used in the field of solar system
dynamics \citep[e.g.,][]{Wisdom_1991} rely on a factorization (or
splitting) of the Hamiltonian in a Lie algebra sense.  These
integrators are very fast, but are best applied to problems where
there is one dominant central mass.  One advantage of GRK integrators
is that they do not approach the issue as a Lie algebra factorization
and hence do not have the restriction of just one large mass.  The
trade-off is that the step size is determined by the shortest period,
which results in longer run-times (we typically use a time step $h$
that is about 400 times smaller than the smallest period in the
system).  However, we note that, even with the relatively low speed
compared to other integrators, it takes much less than one second to
solve the coupled equations of motion for the five bodies in Kepler-47
over a 1500 day time span using the GRK method.

The GRK scheme does have another advantage, in that it is
possible to find positions and velocities at intermediate times to
high accuracy with relatively few computations.  In our
implementation, the solutions at the six internal nodal points used
for the collocation for each time step are saved.  Using the
collocation points with divided differences, it is easy to determine
the positions and velocities of each body at intermediate times to the
same order of accuracy (namely, 12th) as at the integration times (e.g.,
$t_{\rm start}$, $t_{\rm start}+h$, $t_{\rm start}+2h$, etc.).  The
number of operations needed to achieve the 12th-order accuracy is
comparable to what one would need for a spline interpolation scheme,
which would give the intermediate values with a much lower-order
accuracy.

The integrator works in Cartesian coordinates relative to the system's
barycenter.  The unit of mass is the solar mass, the unit of distance
is the astronomical unit, 
the unit of time is the day, and the unit of velocity is 
astronomical units
per day.  The adopted value of the Gaussian gravitational constant is
$k=0.01720209895$ \citep{Clemence_1965}.  Newton's gravitational
constant, expressed in $({\rm AU})^2{M_{\odot}}^{-1}({\rm day})^{-2}$,
is then $G=k^2$.  For convenience, instantaneous Keplerian parameters
valid at some reference epoch for each body are used to specify the
initial conditions (our adopted reference time is $T_{\rm ref}
=$ BJD 2,454,965.000).  These parameters are the orbital period
$P$, the time of barycentric transit (which is the same as the
time of inferior conjunction) $T_{\rm conj}$, the inclination $i$,
the eccentricity $e$, the argument of periastron $\omega$, and the
nodal angle $\Omega$.  The coordinate system is Jacobian, so that the
orbital parameters of the secondary star are given relative to the
primary, the orbital parameters of the first planet are given relative
to the binary's center of mass, and so on.  Because of this, the
specified conjunction time for the binary occurs very near a primary
eclipse, whereas nothing observable has to happen at the conjunction
times of the planets (the actual transits can occur either earlier or
later than the barycentric conjunction times).  The Keplerian
parameters are converted to Cartesian coordinates using the algorithms
given in \citet{Murray_1999}.

After the ODE is solved, the plane-of-sky positions of each body
are corrected for light travel time following the method outlined in
\citet{Carter_2011}.  The times of the primary eclipses and
secondary eclipses, the times of transits of each planet across the
primary  secondary
are 
then computed by finding the times when the projected separation
between two given bodies in the plane of the sky is minimized.  For
computational convenience, we minimize the square of the
plane-of-sky distance vector, which is just the dot product of
that distance vector with itself.  That dot product is minimized when
its derivative is zero.  That derivative is two 
times the dot product of
the distance vector with its velocity vector (which itself is readily
found using the appropriate components of the velocities that are
output from the ODE solver), and is zero when the distance vector
itself is zero or when the distance vector is perpendicular to its
velocity vector.  Several iterations of the secant method are used to
find the zeros of the derivative.  By comparing the results found
using the first-order ODE solver with the results found using the
second-order ODE solver, we estimate the times of conjunction 
to be
accurate to $\approx 1\mu$s or better.  An eclipse or a transit will
occur at a conjunction if the projected minimum separation on the sky
is less than the sum of the radii of the two bodies.

For the specific case of Kepler-47, the corrections due to GR result
in an apsidal advance of about 0.00017 degrees per orbital period.
The rate of apsidal advance due to tidal bulges on each star is
between 10 and 50 times smaller, depending on the values of the tidal
Love numbers used (we considered $0.0 \le k_2 \le 0.01$ for the
primary and $0.0 \le k_2\le 0.2$ for the secondary).  In what follows,
we have included the GR correction in the ODEs and neglected
the corrections
for the apsidal advance due to tidal bulges on the stars.

\subsection{Light Curve Synthesis}

For situations where the bodies are spherical with linear or quadratic
limb darkening laws, ELC can use the algorithm of \citet{Mandel_2002}
or the algorithm of \citet{Gimenez_2006a} to compute the light curves
during eclipses or transits.  In the present work, we have used the
\citet{Mandel_2002} routine with a quadratic limb darkening law,
because it is much faster than the \citet{Gimenez_2006a}
routine.

\subsection{Model Setup for Kepler-47}

Our model as applied to Kepler-47 has 42 free parameters in total.
Each orbit needs six Keplerian parameters for a total of 24.  However,
the nodal angle of the binary is fixed at zero, so there are really
only 23 free parameters to describe the orbits.  For each orbit, we
used
the combinations $e\cos\omega$ and $e\sin\omega$ rather than $e$
and $\omega$ separately, because the former pair of variables
usually are less correlated with each other compared to the latter
pair of variables.  Because there are five bodies,
five parameters describing the masses of the bodies and five
parameters describing the sizes of the bodies
are needed.  For the binary, we
used the primary mass $M_1$ and the binary mass ratio $Q_{2,1}\equiv
M_2/M_1$.  For the two stars, we 
used the fractional radii $R_1/a$ and
$R_2/a$ to parameterize their sizes. To parameterize the
planetary radii, we 
used the ratio $R_1/R_p$, which is the ratio of
the primary star's radius to the particular planet radius
in question.  There are two radiating bodies in the system, so
a total of six parameters are required
to specify the radiative
properties: the two temperatures and two limb darkening coefficients
for each star.  In this case, the temperature of the primary was fixed
at 5636 K \citep{Orosz_2012b}, so only five free parameters were used.
For convenience, the temperature ratio $T_2/T_1$ was used instead
of directly using $T_2$.  ELC has a table of model atmosphere specific
intensities (for solar metallicity) derived from the NextGen models
\citep{Allard_1997,Hauschildt_1999}, and these 
were used to compute the
baseline fluxes of the primary and secondary star (in the {\em Kepler}
bandpass) given their temperatures and gravities.  The standard
quadratic limb darkening law given by
\begin{equation}
I(\mu)/I_0 = 1 - u_1(1-\mu)-u_2(1-\mu)^2 \nonumber
\end{equation}
was used (where $\mu=\cos\theta$ is the projected distance from the
center of the stellar disk), but with the ``triangular'' sampling
technique of \citet{Kipping_2013} with coefficients given by
$q_1=(u_1+u_2)^2$ and $q_2=0.5u_1(u_1+u_2)^{-1}$.  Finally, to account
for light from other sources in {\em Kepler's} aperture, four seasonal
contamination parameters were
used.

\subsection{Model Optimization}\label{modeloptimization}

As was shown in \S\ref{eclipsetimesection}, many of the primary
eclipse profiles show clear evidence of spot crossing events. Because
our light curve model does not include star spots, fitting the primary
eclipses where there are clear spot crossing events might lead to
biased results.  To avoid this
possible bias in the fitting, we fit the
statistically corrected primary eclipse times along with only a small
subset of the light curve.  We chose ten primary eclipse profiles
(seven in long cadence and three in short cadence) that do not have
spot crossing events to fit, as well as ten secondary eclipses that
are close in time to the ten selected primary eclipses.  We also fit
the 25 transits of the inner planet, the six transits of the middle
planet, and the four transits of the outer planet.  Finally, we also
included the radial velocities of the primary star in the fit.

\begin{figure} 
\begin{center}
\includegraphics[angle = -90,scale=0.6]{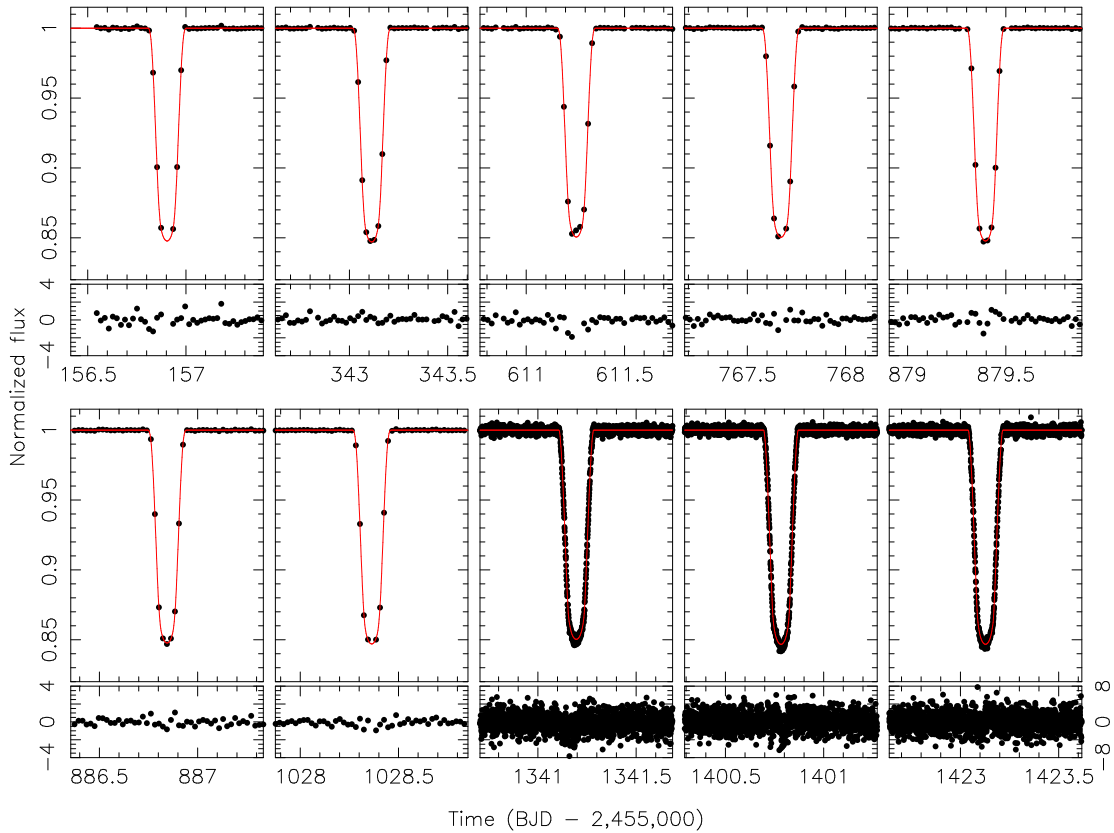}
\caption{Model fits to the ``clean'' primary eclipse
profiles. The fits to the 10 primary eclipse profiles that have
no strong spot crossing events are shown.  The vertical scale
on the panels showing the residuals are parts per thousand.
Note the scale change on the three panels showing the residuals
of the fits to the short-cadence data. 
\label{showcleanprimary}}
\end{center}
\end{figure} 

\begin{figure} 
\begin{center}
\includegraphics[angle = -90,scale=0.6]{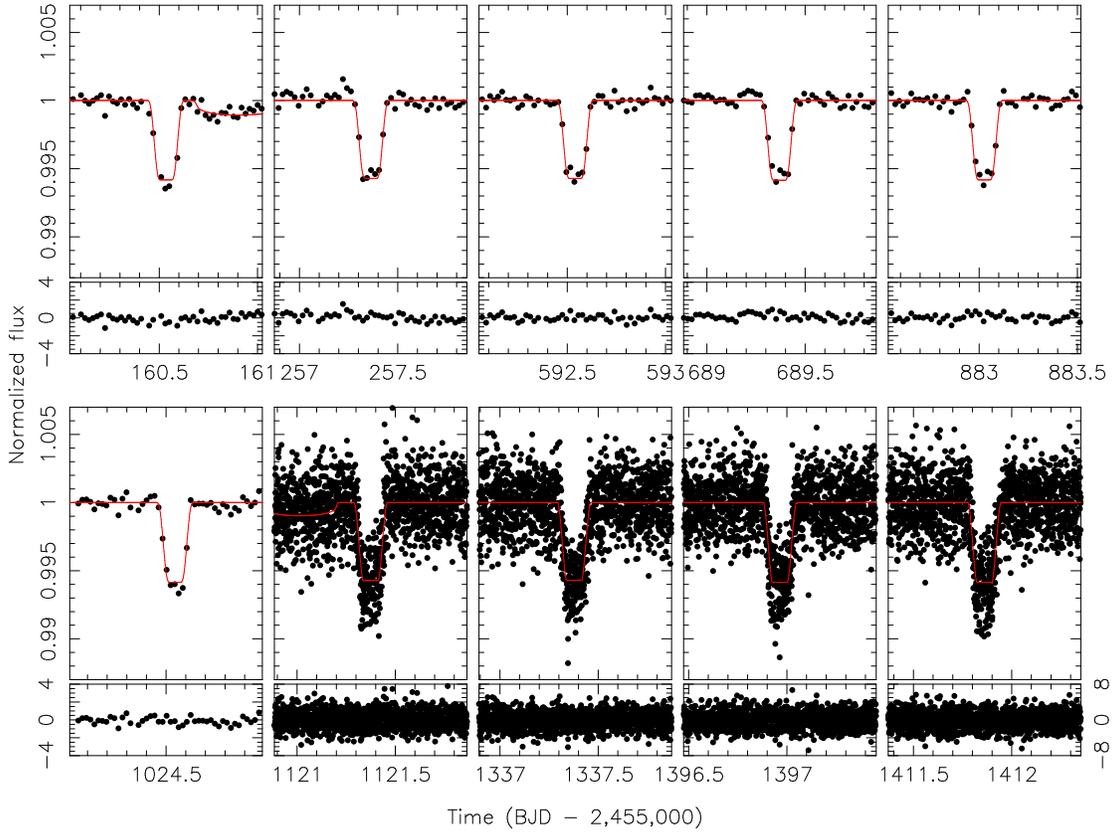}
\caption{Model fits to the ``clean'' secondary eclipse profiles.  The
fits to the 10 secondary eclipse profiles included in the fitting
are shown. The vertical scale on the panels showing the residuals
are parts per thousand.  Note the scale change on the four
panels showing the residuals of the fits to the short cadence data.
Transits of the inner planet can be seen near day 160.7 and day
1121.0.
\label{showcleansecondary}}
\end{center}
\end{figure} 

We used four algorithms to explore parameter space: an adaptation of a
simple grid-search algorithm \citep{Bevington_1969}, a genetic
algorithm \citep{Charbonneau_1995}; a simple search based on Monte
Carlo Markov Chain (MCMC) algorithm \citep{Tegmark_2004}, and a
Differential Evolution MCMC (DE-MCMC) algorithm \citep{terBraak_2006,
Nelson_2014}.  All four of these algorithms are bounded, meaning the
fitting parameters never wander outside predefined ranges.  In
addition, the grid-search code, the genetic code, and the DE-MCMC code
can be run in parallel on multiple CPU cores.  We 
assumed the
likelihood function follows a $\chi^2$ distribution, where $\chi^2$ is
based on the sum of squares of differences between observed quantities
(a photometric measurement, a radial velocity measurement, or an
eclipse or transit time) and predicted quantities normalized by the
measurement uncertainty for each quantity.  Each of these algorithms
needs to have fixed ranges for each free parameter.  Some parameters,
like some of the orbital parameters for the binary, are constrained
reasonably
well.  We also had some good starting solutions for the
parameters of the inner planet from our previous work
\citep{Orosz_2012b}, whereas some of the orbital parameters 
(for example, the
eccentricity parameters)
for the
outer planet were constrained much less well.  
Apart from the rough value of the orbital
period and the fact that the inclination of the orbit must be
near $90^{\circ}$ in order for transits to occur, the orbital
parameters for the middle planet were initially not known.

Because the parameter space is vast, we looked for optimal solutions in
three main stages: (i) updating the model from the previous work by
adding the middle planet to the model and finding some initial fits;
(ii) exploring parameter space to find something close to the ``best''
model, get some rough idea of the uncertainties in the
fitted and derived parameters, and also look for 
possible
correlations between various
parameters; and (iii) finding realistic
uncertainties on the fitted and derived parameters.  The three stages
are described below.

{\em Stage i}: 
The inner planet has seven additional transits and the outer planet has
one additional transit since the previous work, so the orbital
parameters for those two planets were revised first using the DE-MCMC
code.  Next, the middle planet was added to the model, and initial
orbital parameters were found ``by hand.''  Educated guesses for the
time of barycentric conjunction, the inclination, the nodal angle,
etc. were used to initialize the simple MCMC code, and the model fits
were inspected by eye after 
running a few iterations of that optimizer.
It was not unduly difficult to find a model that produced
transits of the middle planet at roughly the correct times.

{\em Stage ii}:
After a few decent models were found, longer runs using the genetic
algorithm and the DE-MCMC were done.  Both of these codes require an
initial ``population'' of models (we typically used population sizes
between 100 and 200), and we have various ways of generating the
initial population: (a) totally random values between the specified
lower and upper parameter bounds for each free parameter; 
(b) one or
more ``elite'' models (e.g.\ ones that are already a good match to the
data) along with randomly generated models; and 
(c) one or more
``elite'' models along with ``mutated'' copies of the elite models
where a few randomly chosen parameters are ``tweaked'' by adding or
subtracting small offsets.  In practice, options 
(b) and (c) work
the best for cases where one has a few dozen or more free parameters.

For this intermediate stage of fitting, we adopted ranges for the free
parameters given in Table \ref{priors}.  We ran the genetic algorithm
or the DE-MCMC code for a few pilot runs of several thousand
generations to confirm that the prior ranges included support for the
entire range with nontrivial likelihood.  After each pilot run, plots
of parameter values versus the generation number and plots of the
$\chi^2$ versus each parameter were inspected to find instances where
the allowed range of a parameter was either too large or too small.
We verified that the likelihood falls to extremely small values by the
time model parameters reach any of these boundaries (with the
possible
exception of hard physical boundaries).
  
After this intermediate state of optimization, several million
models have been computed.  The genetic algorithm and the DE-MCMC
code explore parameter space in very different ways; and after several
long runs of each, we can be reasonably sure the optimal region
in parameter space has been found.  

{\em Stage iii}:
We used a brute-force ``stepper'' to generate initial ``seed'' models
for the final run of the DE-MCMC code.  The stepper works 
as follows. Take the best overall model and choose a key fitting
parameter, such as the primary mass.  Offset that parameter by a small
amount (say, 0.5\% of the parameter value) and hold it fixed while
optimizing the other parameters using the simple parallel grid-search
code, where the parameter values from the optimal model are used as
the initial guess.  After that optimization is done, offset the same
key parameter again and repeat the optimization, using the previous
solution as the initial guess.  After the stepper is done, you have a
set of reasonably optimal models with a range of different values of
the chosen key parameter.  We ran steppers on the mass parameters
(primary mass, binary mass ratio, and the planet masses), the radius
parameters (primary and secondary fractional radii and planet radius
ratios), and the planet orbital periods.

For the last run of the DE-MCMC code, we used a population size of
1600 models per generation with 900 ``seed'' models found from the
steppers.  The other 700 models were mutated copies of randomly
selected seed models with five parameters in each model randomly
tweaked.  The code was ran for 17,000 generations, giving about 27
million models in total.  Judging from plots of the parameters
vs.\ the generation number, convergence was achieved after about 4000
generations.  However, to be conservative, the first 5000 generations
were discarded, leaving about 19 million models.  Figure \ref{showRV}
shows the best-fitting model radial velocity curve.  The best fits to
the transits of the middle planet, inner planet, and the outer planet
are shown in Figures \ref{showmiddle}, \ref{showinner}, and
\ref{showouter}, respectively.  Figure \ref{showcleanprimary} shows
the best-fitting model with the ``clean'' primary eclipses and Figure
\ref{showcleansecondary} shows the best-fitting model with the
corresponding secondary eclipses.  The best-fitting input parameters
are given in Table \ref{tab:tab2}, and Table \ref{tab:tab3} summarizes
several derived parameters of interest.  In order to allow
others to reproduce our best-fitting model using other codes, we
give the initial Cartesian barycentric coordinates for the
best-fitting model in Table \ref{cartcoords}, and the initial
orbital elements (which are traditionally used in dynamical studies)
for the best-fitting model in Table \ref{elements}.  These
coordinates and orbital elements are valid for the reference time
BJD 2,454,965.000.

\begin{figure}[t] 
\begin{center}
\includegraphics[angle = -0,width=0.66\textwidth]{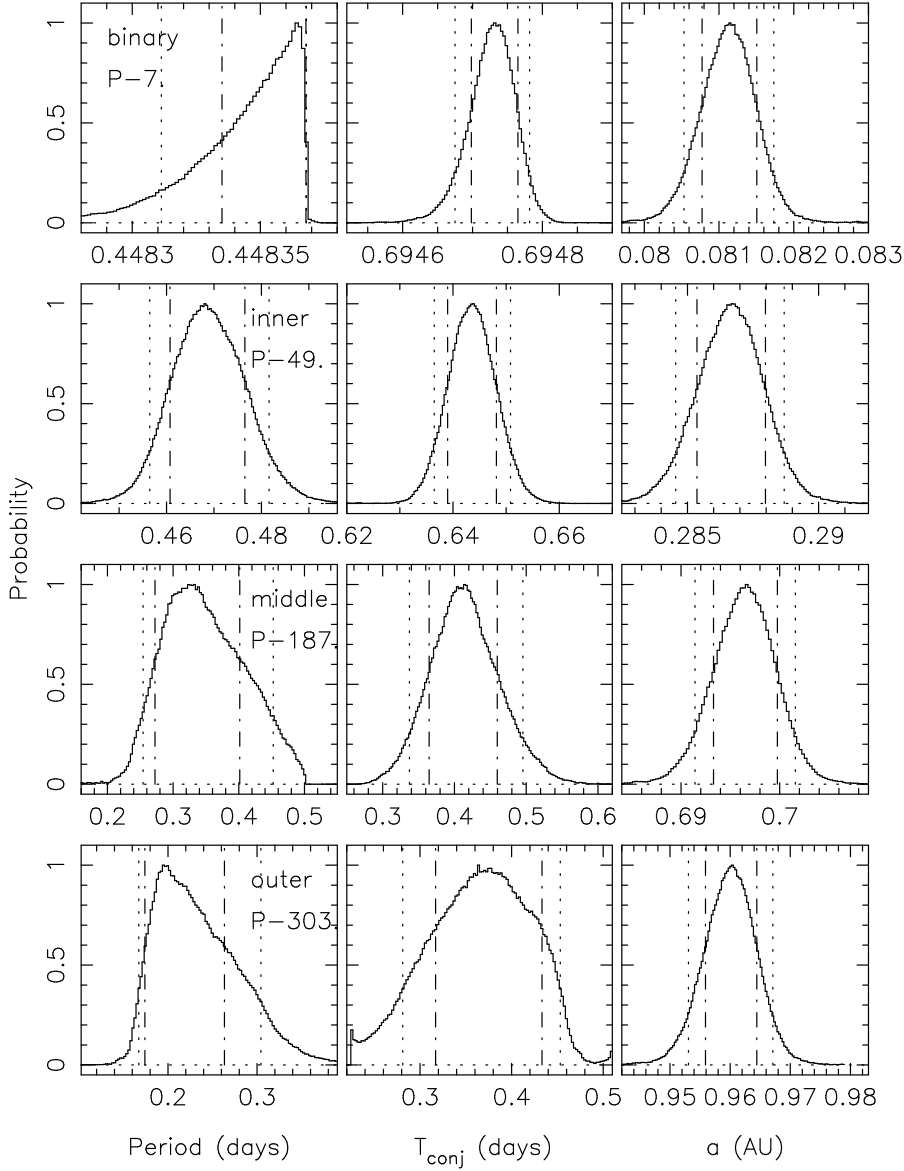}
\caption{The 
posterior distributions from the DE-MCMC run of the orbital periods
(left), times of barycentric conjunction (middle), and semimajor axes
(right).  We show only the decimal parts of the periods and times of
conjunction (see Table \ref{tab:tab2} for parameter values from the
best-fitting model).  The dashed vertical lines denote the lower and
upper boundaries that contain 68.3\% of the area, and the vertical
dotted lines denote the lower and upper boundaries that contain 90\%
of the area.
\label{plothist01}}
\end{center}
\end{figure} 

\begin{figure} 
\begin{center}
\includegraphics[width=0.66\textwidth]{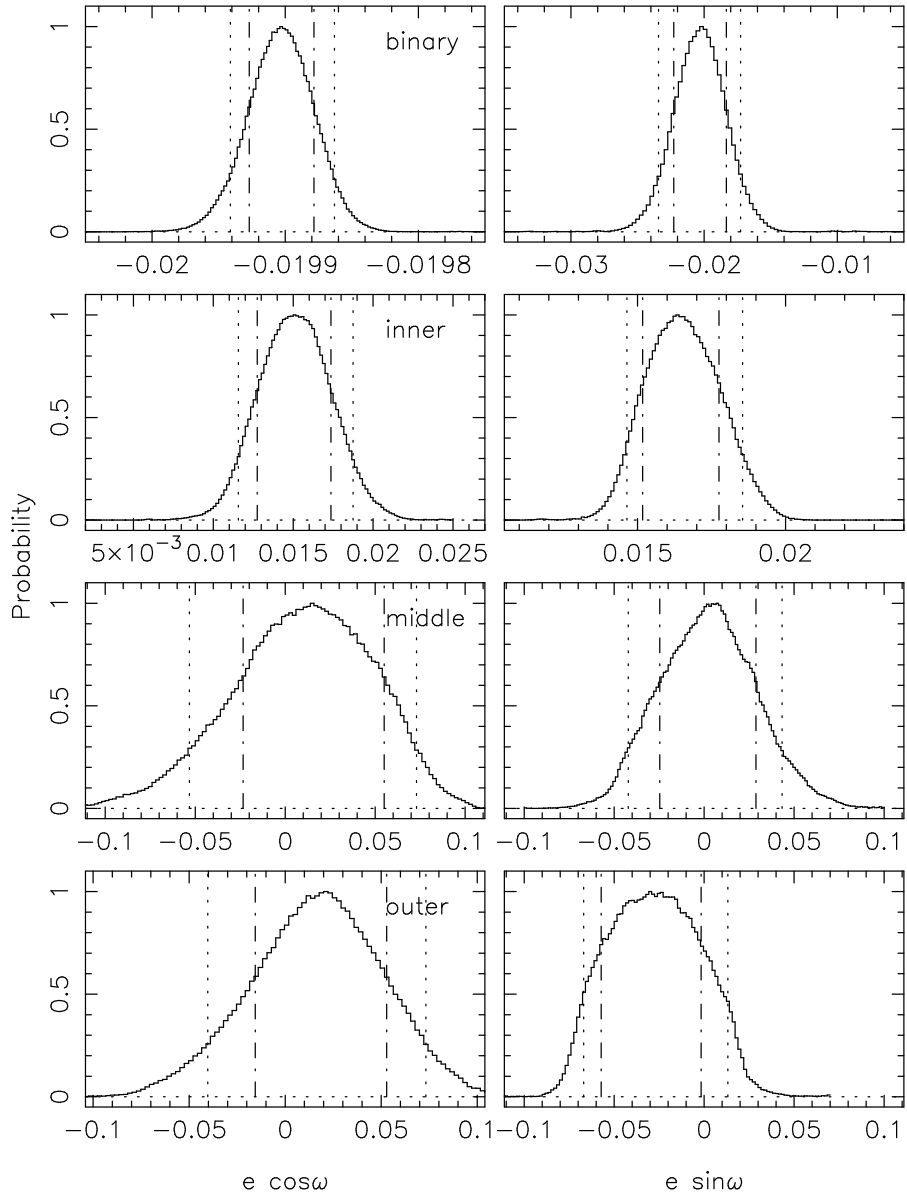}
\caption{Similar to figure \ref{plothist01}, but showing 
the posterior distributions of 
$e\cos\omega$ (left) and $e\sin\omega$  (right).  
\label{plothist02}}
\end{center}
\end{figure}

\begin{figure} 
\begin{center}
\includegraphics[width=0.66\textwidth]{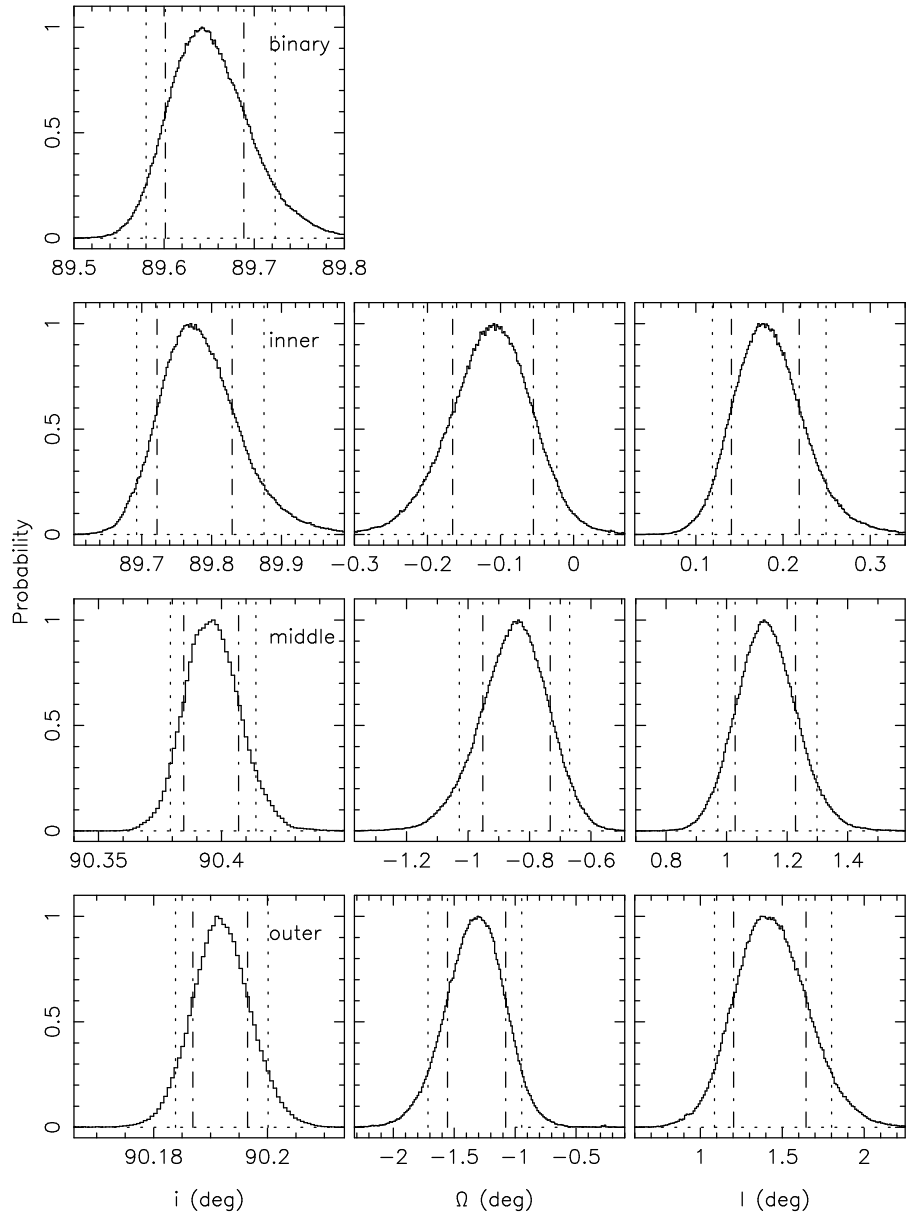}
\caption{Similar to figure \ref{plothist01}, but showing
the posterior distributions of the orbital inclinations (left); 
the nodal angles of 
the orbits--note the nodal angle of the binary
is fixed at 0.0 (middle), and the mutual
inclinations relative to the binary plane (right).
\label{plothist03}}
\end{center}
\end{figure}

\begin{figure} 
\begin{center}
\includegraphics[width=0.66\textwidth]{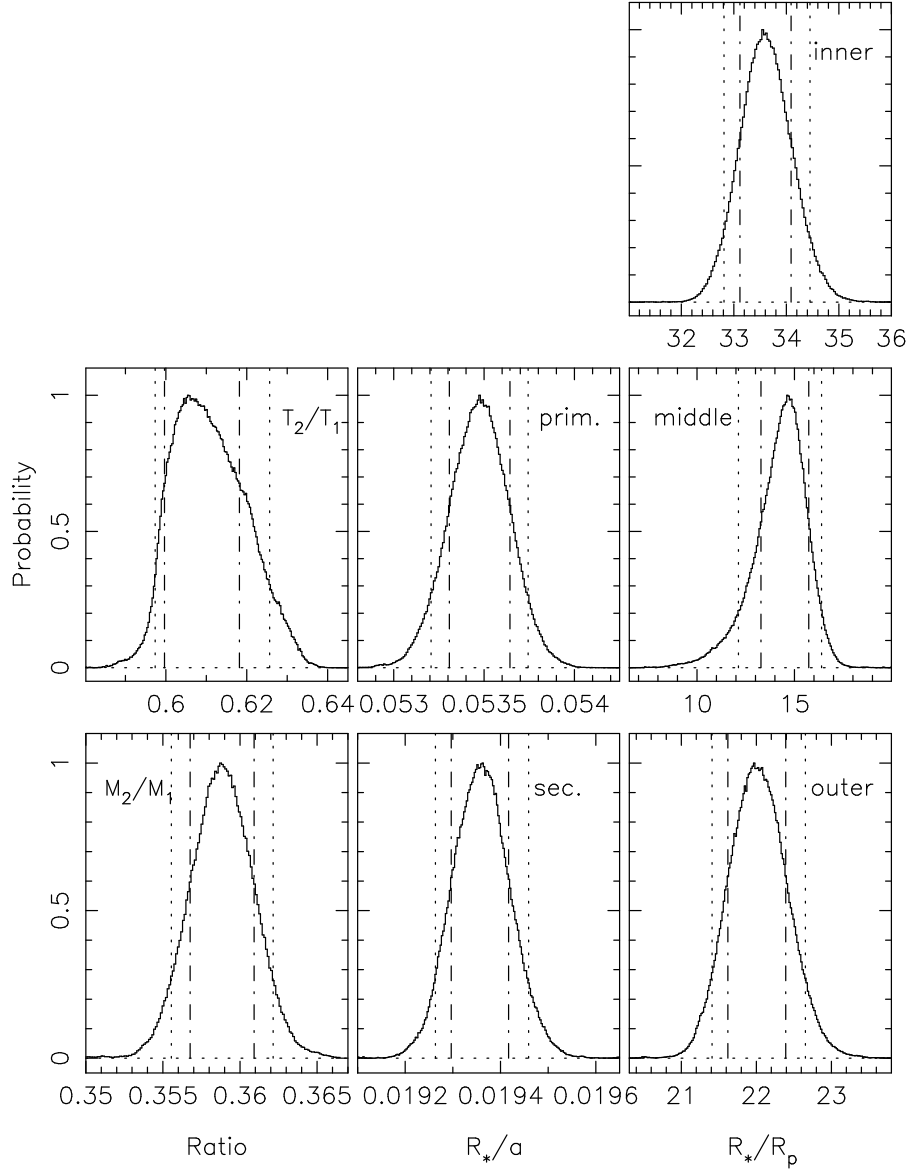}
\caption{Similar 
to figure \ref{plothist01}, but showing the posterior distributions of 
the binary mass and temperature ratios (left), the fractional stellar radii
$R_*/a$  (middle), and the planet radius ratios $R_*/R_p$ (right).  
\label{plothist04}}
\end{center}
\end{figure}

\begin{figure} 
\begin{center}
\includegraphics[width=0.66\textwidth]{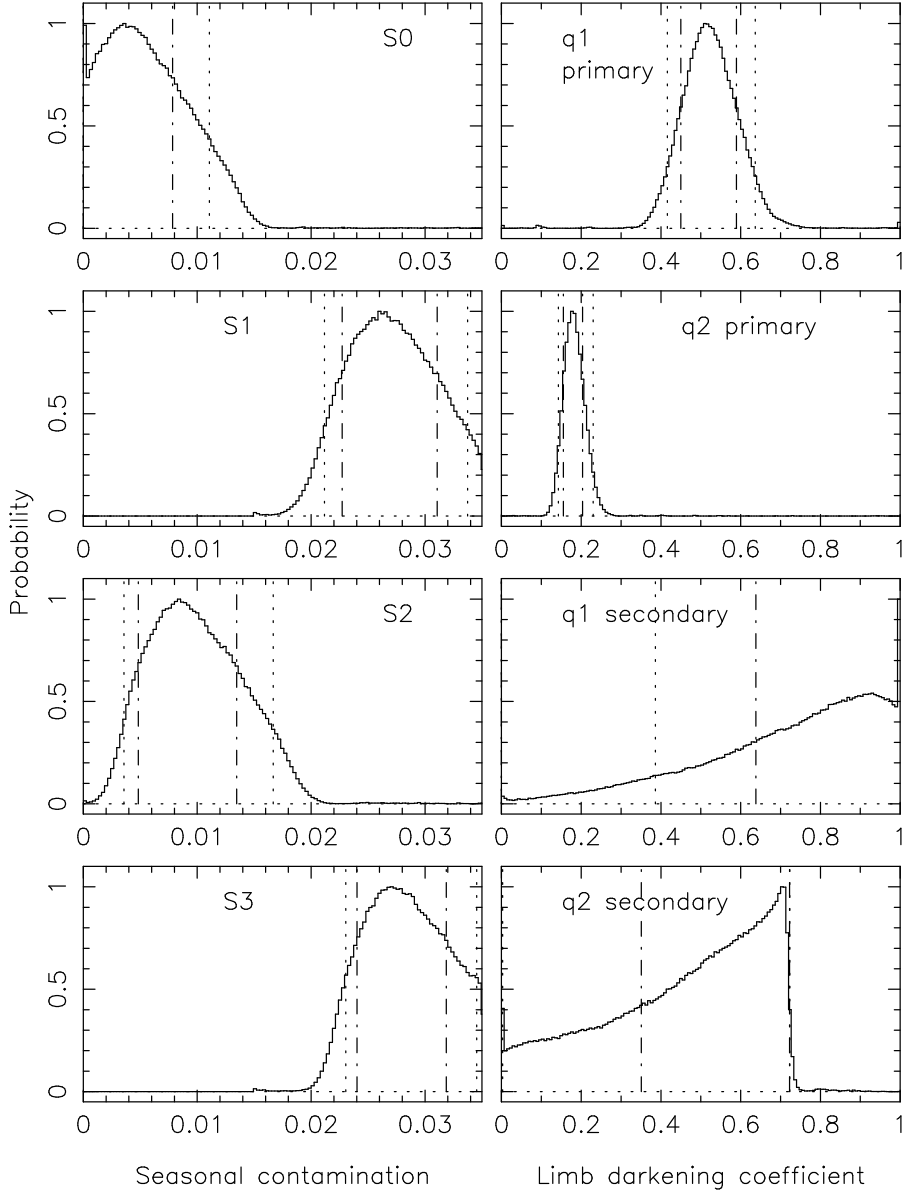}
\caption{Similar to figure \ref{plothist01}, but showing
the posterior distributions of 
the seasonal contamination parameters (left) 
and the  limb darkening parameters (right).  
\label{plothist05}}
\end{center}
\end{figure}

\begin{figure} 
\begin{center}
\includegraphics[width=0.66\textwidth]{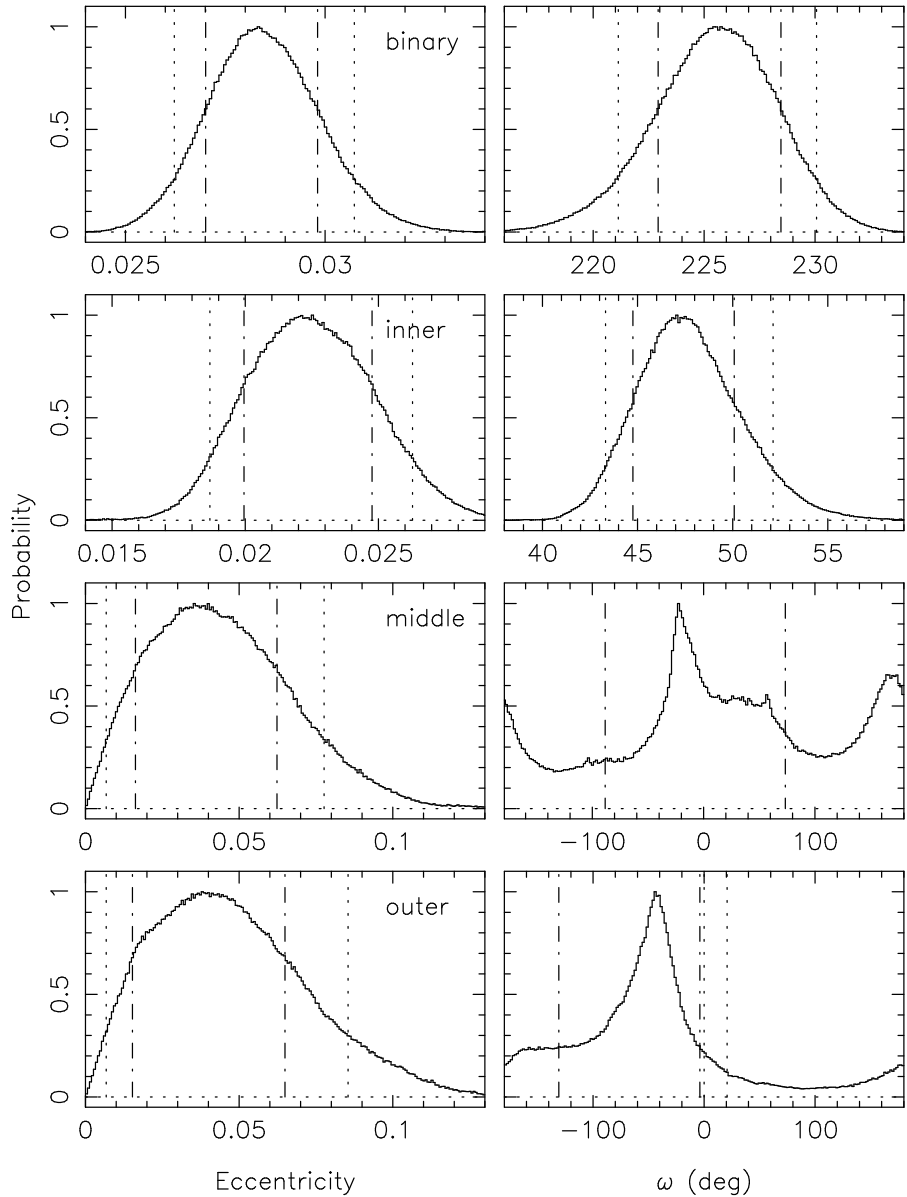}
\caption{Similar to figure \ref{plothist01}, but showing the
posterior distributions of the eccentricities (left) and arguments
of periastron (right).
\label{plothist06}}
\end{center}
\end{figure}

\begin{figure} 
\begin{center}
\includegraphics[width=0.66\textwidth]{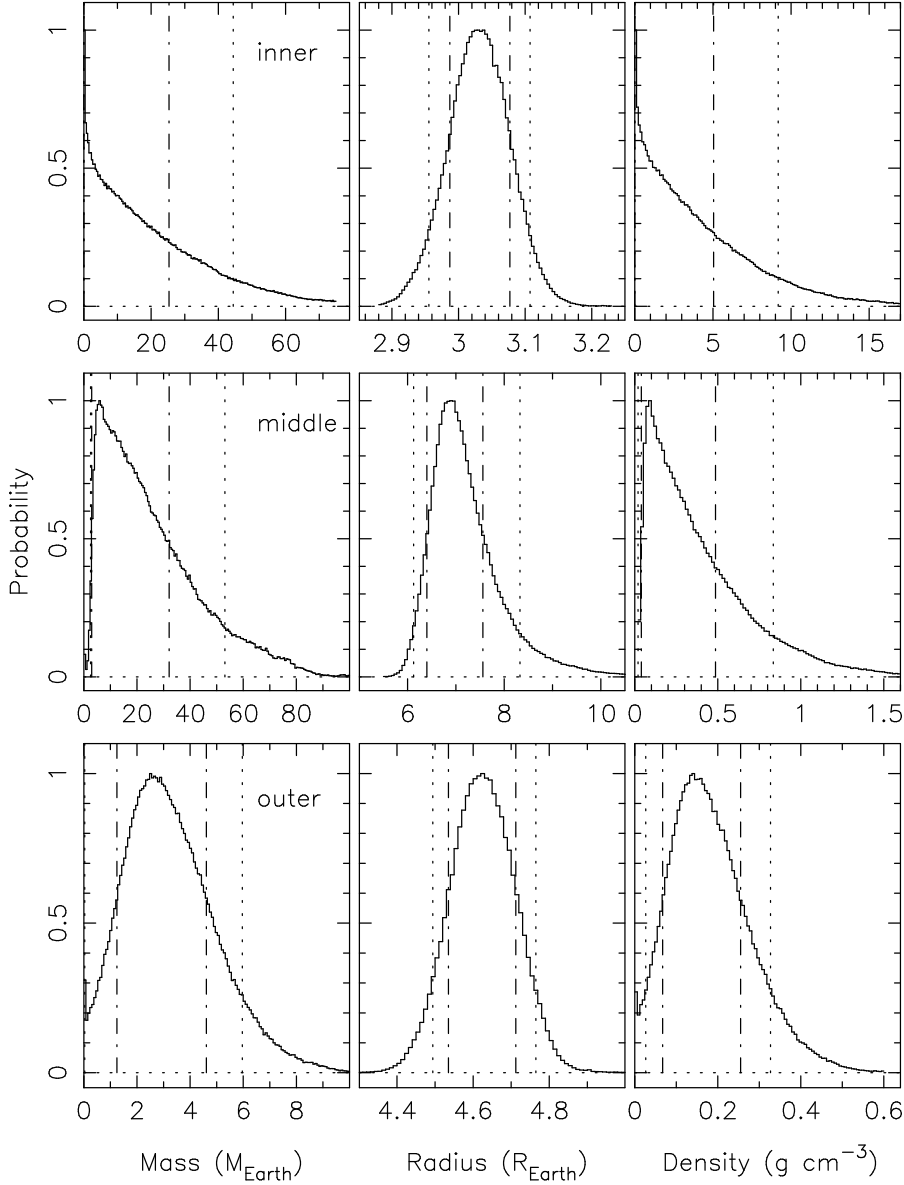}
\caption{Similar to figure \ref{plothist01}, but showing
the posterior distributions of the planet masses (left), 
the planet radii (middle),
and the  planet densities (right).
\label{plothist07}}
\end{center}
\end{figure}

\begin{figure} 
\begin{center}
\includegraphics[width=0.66\textwidth]{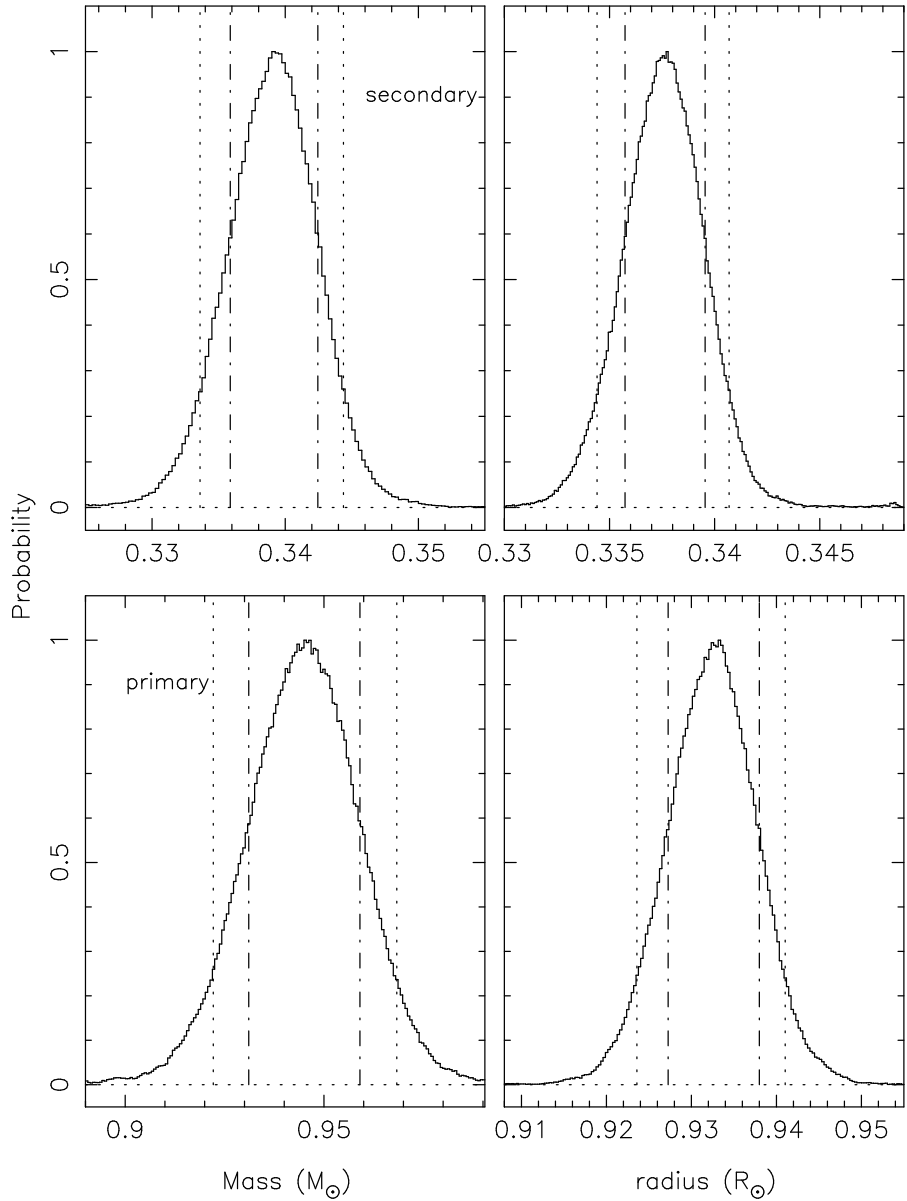}
\caption{Similar to figure \ref{plothist01}, but showing
the posterior distributions of the stellar masses (left) and the
stellar radii (right).
\label{plothist08}}
\end{center}
\end{figure}

\begin{figure} 
\begin{center}
\includegraphics[clip,width=0.95\textwidth]{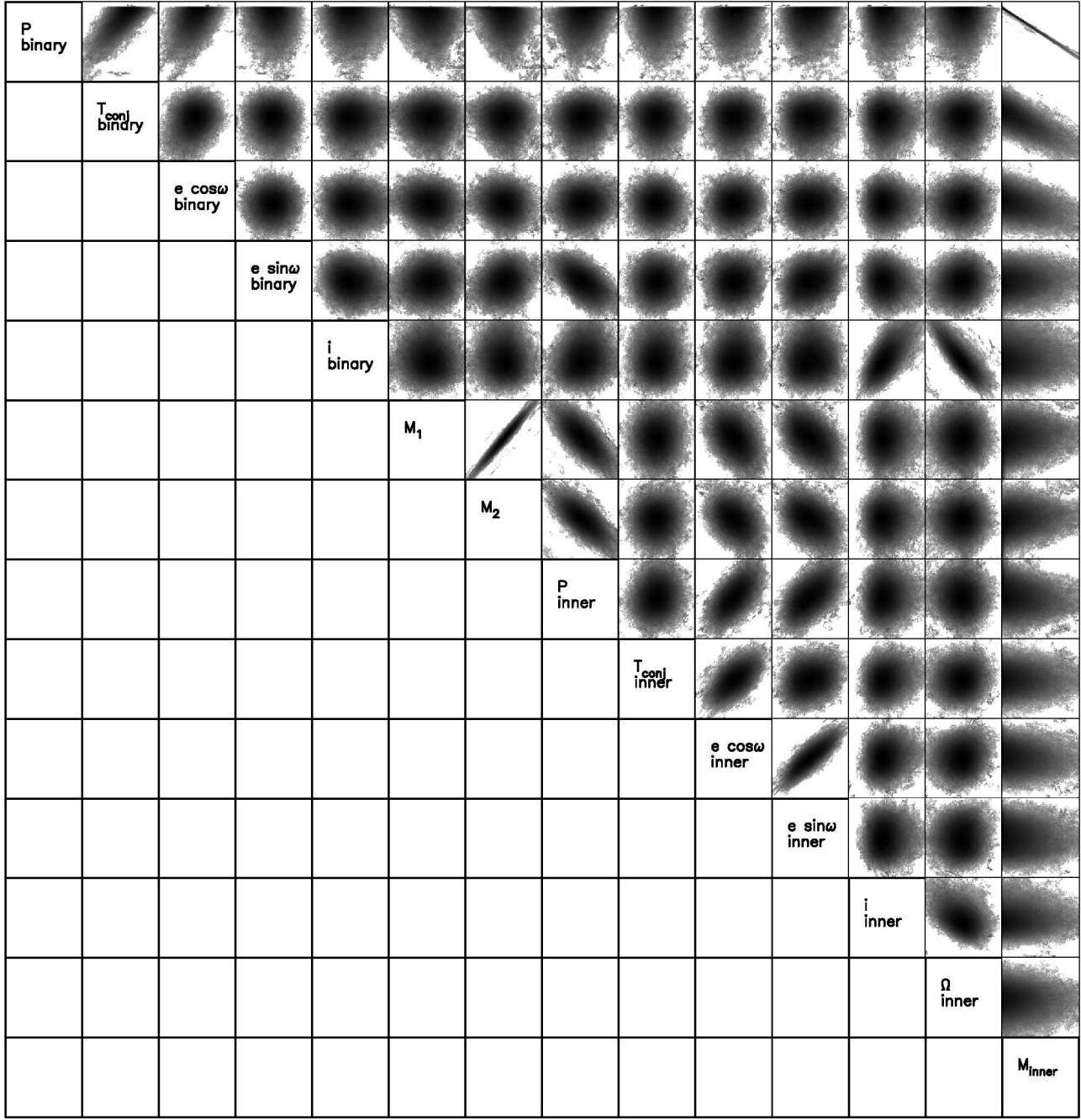}
\caption{The joint posterior distributions for
28 orbital parameters. These parameters and the ranges that are displayed
are shown in Table
\ref{tab_appendix}. The parameter densities are plotted logarithmically.
For clarity, only parameter numbers 1-14 vs.\
parameter numbers 1-14 are shown here.  The following two figures
give the remaining combinations.
\label{quad1}}
\end{center}
\end{figure}

\begin{figure} 
\begin{center}
\includegraphics[clip,width=0.95\textwidth]{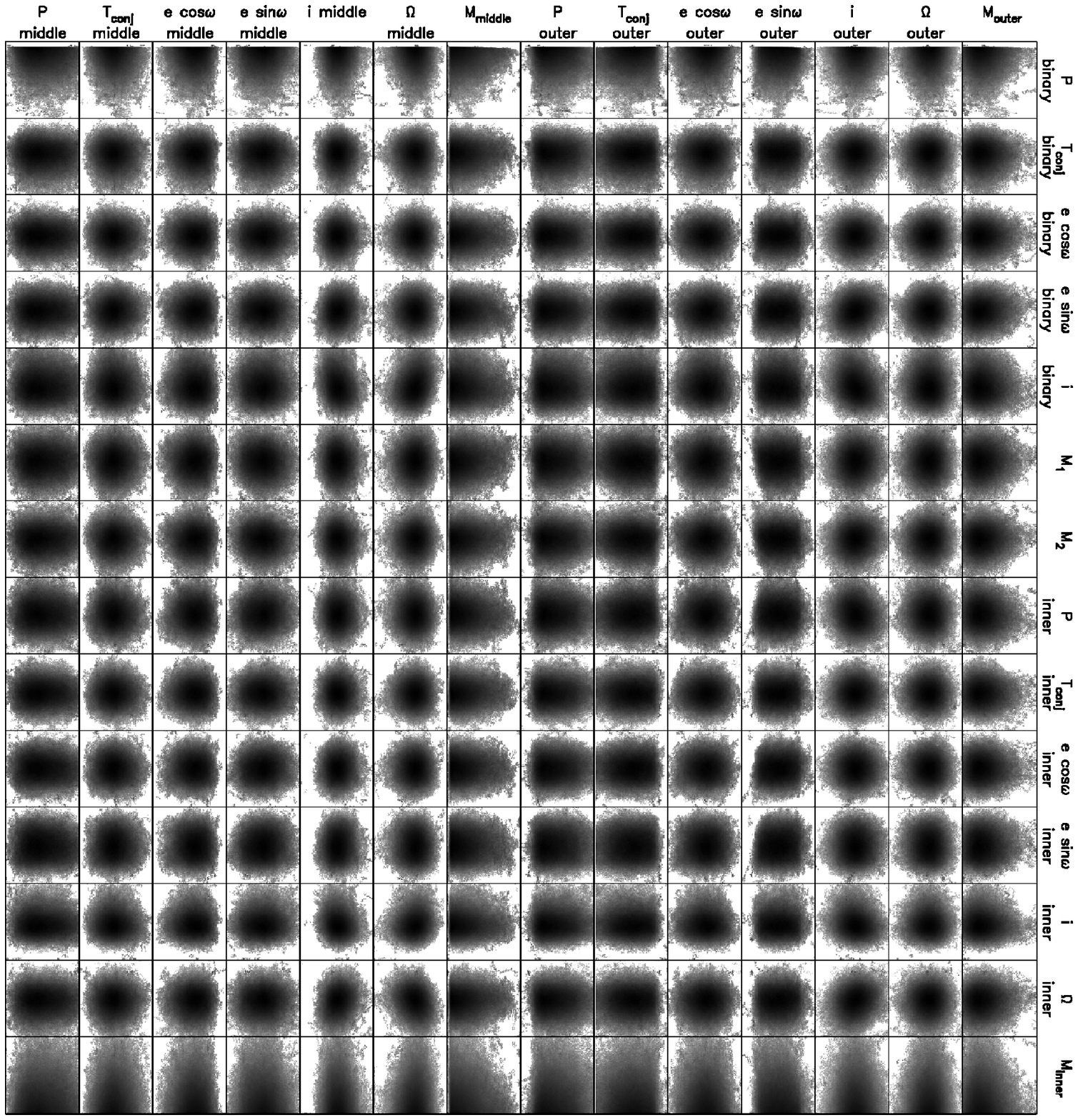}
\caption{Similar to figure \ref{quad1} 
for parameter numbers 15-28 vs.\ parameter numbers 1-14
\label{quad2}}
\end{center}
\end{figure}

\begin{figure} 
\begin{center}
\includegraphics[clip,width=0.95\textwidth]{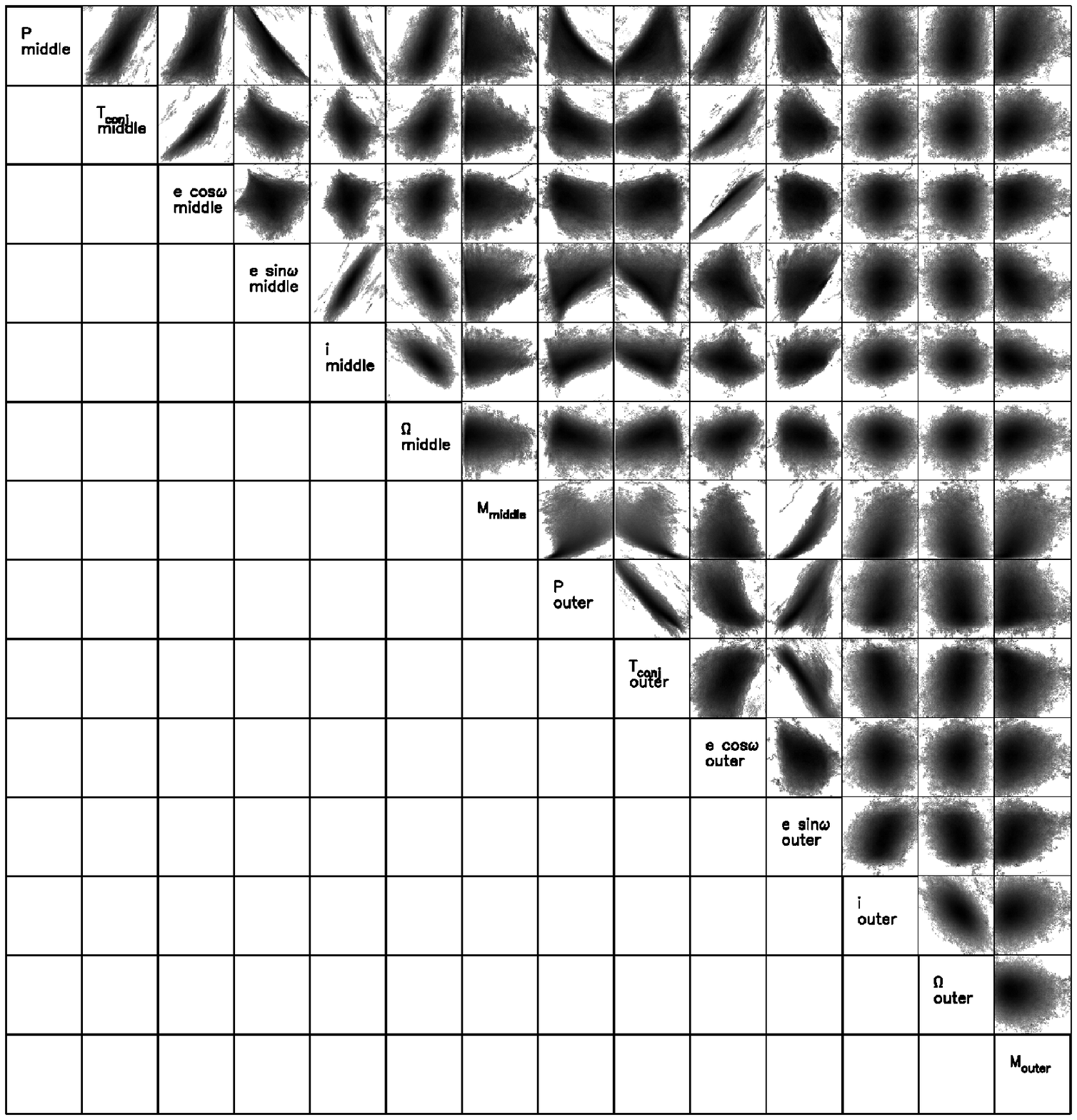}
\caption{Similar to figure \ref{quad1}
for parameter numbers 15-28 vs.\ parameter numbers 15-28.
\label{quad3}}
\end{center}
\end{figure}

For each parameter's posterior distribution, we computed the mode and the
lower and upper boundaries of the region that contains 68.3\% of the
area under the curve.  We
show plots of posterior distributions for several
fitting and derived parameters in Figures 
\ref{plothist01}-\ref{plothist08},  and
plots of the two-parameter joint posterior
distributions for the 28 major orbital parameters in
Figures
\ref{quad1}-\ref{quad3} (Table \ref{tab_appendix}
gives the displayed parameters and their ranges).

When one examines the posterior distribution of a given parameter, the
mode of the distribution might be taken as the ``most likely'' value
of that parameter.   We note, however, that a model constructed
using the modes of each distribution will not necessarily
provide a good fit to the data.  Therefore, for the final adopted
parameters, we take the parameters from the best-fitting model.

\section{Discussion}\label{discussion}

\subsection{Constraints on the Planetary Masses}\label{constraints}

For several CBPs
\citep{Doyle_2011,Welsh_2012,Kostov_2015}, 
the gravitational pull of the 
planet is sufficiently
large  to measurably alter the positions and 
velocities of their host stars.  These perturbations can be 
detected by measuring slight deviations from strict periodicity in 
the times of the stellar eclipses. From these eclipse-timing 
variations, the mass of the planet can be estimated.
For Kepler-47, only 
upper limits on the masses could be made at the time of their discovery:
$<$~2.0 Jupiter masses ($< 635 \ M_{\oplus}$) for the inner planet and
$<$~28 Jupiter masses ($< 8900 \ M_{\oplus}$) for the outer planet 
\citep{Orosz_2012b}.

The presence of the third planet dramatically changes 
this situation.  
The two
outer planets gravitationally perturb one another, resulting in 
dynamical transit-timing variations. Due to their proximity,
the planets perturb each other more than they
affect the stars.
In the discovery paper \citep{Orosz_2012b},
the planets were assumed to be massless 
in the modeling process,
as the
interactions between the planets and the 
binary--and between each other--was 
determined to be small.  In the present model, all five bodies have
mass and the mutual interactions between each body are fully accounted
for.  We find that we can place meaningful constraints on the masses
of the middle and outer planets.
The $1\sigma$ range of the mass of the inner planet is between
zero and $25.8\,M_{\oplus}$,
whereas the mass of the
middle planet is different from zero at the $\approx 5\sigma$ limit
and the mass of the outer planet is different from
zero at the
$\approx 2\sigma$ limit.  

\begin{figure}[t] 
\begin{center}
\includegraphics[angle = -90,scale=0.6]{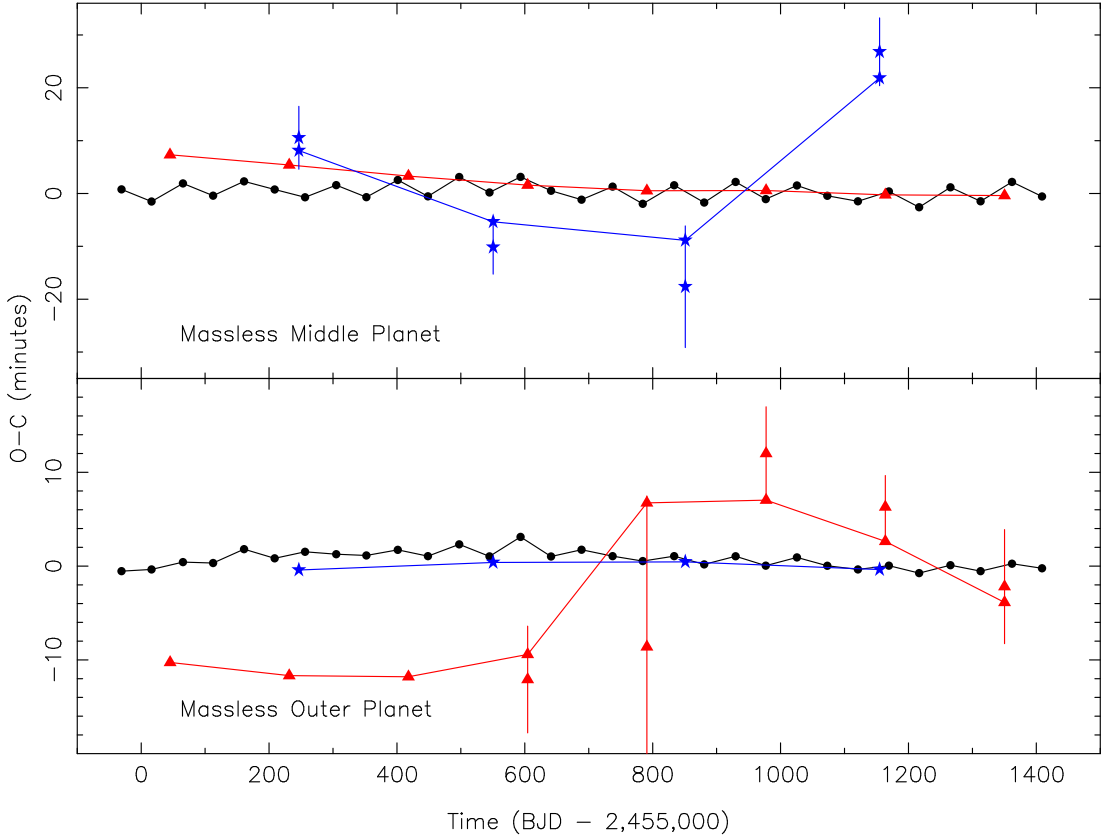}
\caption{Top: 
The $O-C$ diagram showing the change in the times of the planet
transits when the middle planet is forced to have zero mass
(i.e.\ the times from the overall best-fitting  minus the
times from the model with the zero-mass middle planet), using
black
filled circles for the inner planet, 
blue filled stars
for the middle planet, and red filled
triangles for the outer planet.  The filled stars with error bars
are the observed transit times for the outer planet
minus the predicted times from
the model with the zero-mass middle planet.
Bottom: Similar to the top, but where the outer planet is forced
to have zero mass.
The filled triangles with error bars
are the observed transit times for the middle planet
minus the predicted times from
the model with the zero-mass outer  planet.
The $O-C$ value for the first observed transit
(near day 418) is $\approx 41\pm 9$ minutes, but
this value is probably prone to large
systematic errors owing to the missing cadence
near the middle of the transit.
\label{nomassOC}}
\end{center}
\end{figure}

We computed the planet transit times
of the best model with a zero-mass middle planet and the best model with
a zero-mass outer planet, and compared those times to the
transit times from the overall best-fitting model.  The difference in the
respective transit times are shown in Figure \ref{nomassOC}.  In either
case, the transit times of the inner planet do not 
change by more than
$\approx 1$ minute.
On the other hand, a massless middle planet results in changes in the
outer planet's transit times by up to $\approx 20$ minutes.  Likewise, 
a massless outer planet results in changes in the middle planet's transit
times by up to $\approx 10$ minutes.  These changes in the transit times
can be detected at the significance of a few $\sigma$, leading to the
mass constraints that we have
on the middle and outer planets.

\begin{figure}
\center
\includegraphics [height = 3.5 in]{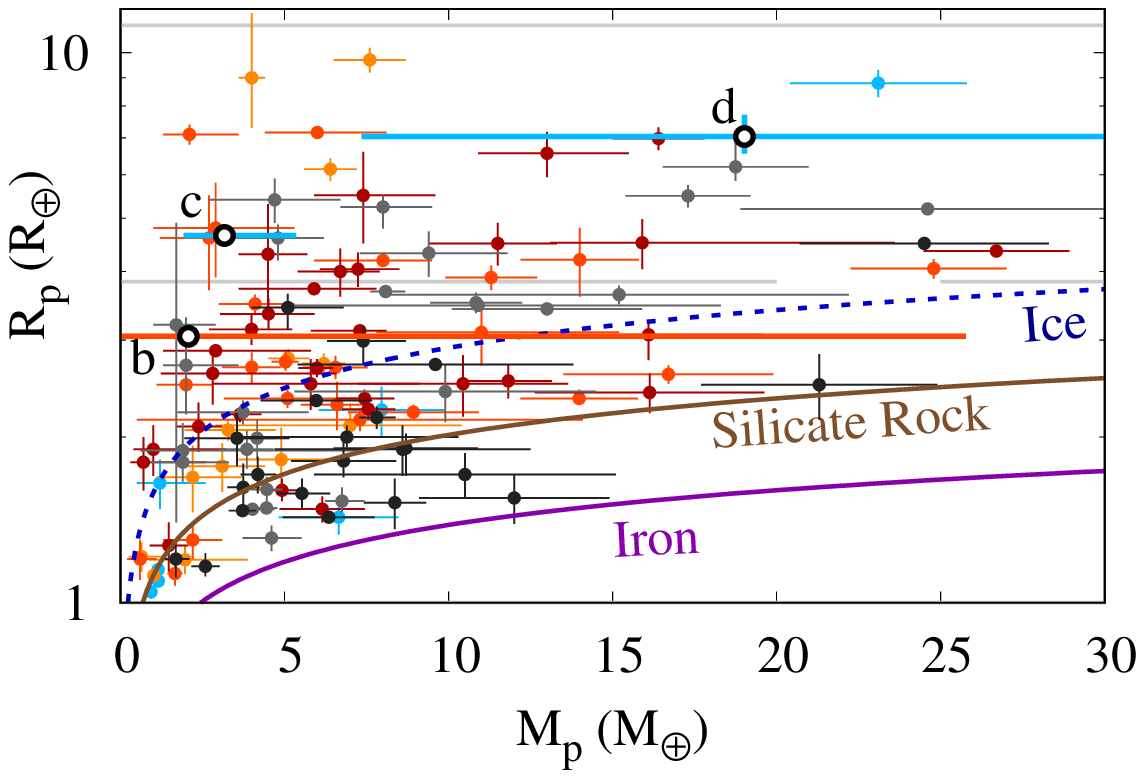}\\
\includegraphics [height = 3.5 in]{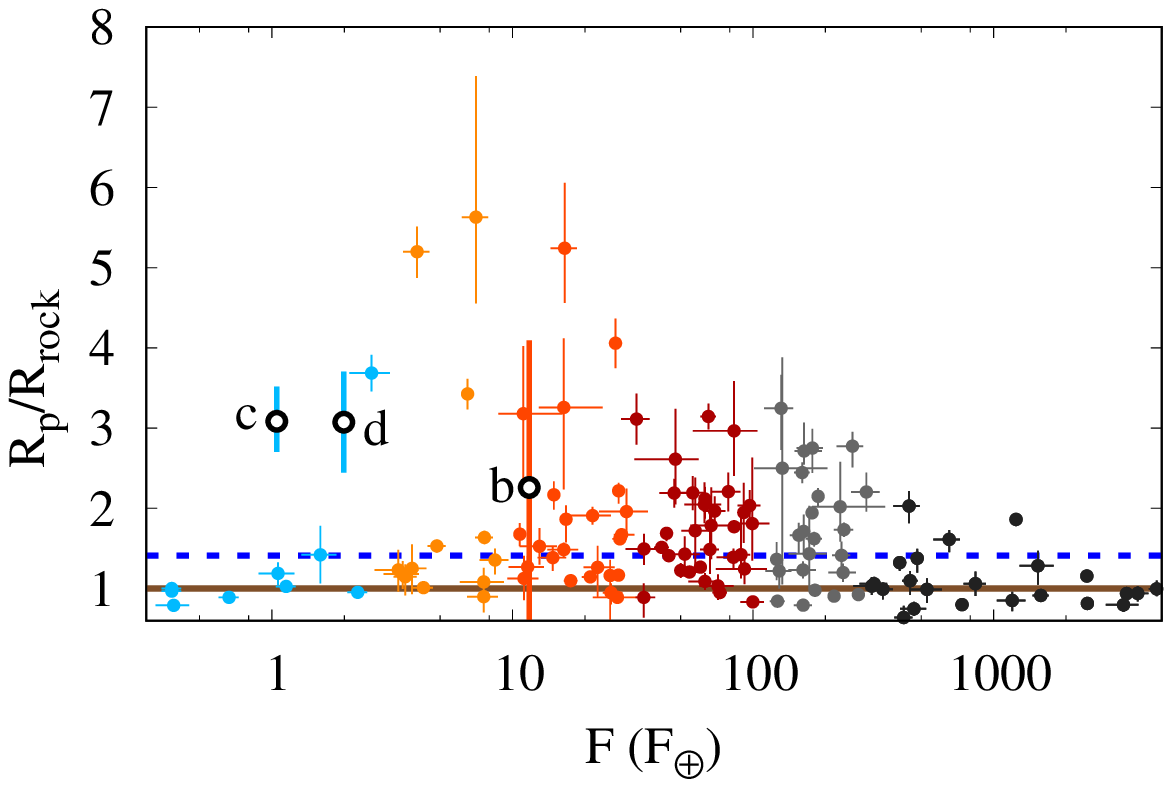}
\caption{Top: Radii and masses for well-measured low-mass 
exoplanets are plotted, 
with the Kepler-47 planets shown with 
open circles along with their $1\sigma$ error bars.
The colors denote the incident flux (S) on each planet relative to the
Sun--Earth insolation, ranging from blue for $S<3$ to black for 
$S>300$ (with orange, red, maroon, and dark gray in-between).  
For comparison, theoretical curves \citep{Lopez_2014}
for planets composed of pure iron, silicate rock, or water are shown.
Bottom: The radii of the planets shown in the upper panel, 
relative to a silicate 
rock world of the same mass, are plotted against the incident flux the 
planet receives from its host star.
Planets larger (less dense) than a pure-rock planet lie above the value
$R_{p}/R_{\rm rock} = 1$. 
Planets larger than a pure water planet lie above the dashed blue line. 
These low-density planets must have deep hydrogen and helium atmospheres. 
Planets that have $R_{p}/R_{\rm rock} < 1$ are 
denser than pure rock and are 
probably a mixture of rock and metal, similar to Earth and Venus.
The middle and outer planets have significantly larger values
of $R_p/R_{\rm rock}$ than other planets with similar
low-incident fluxes.\label{massrad}
}
\end{figure}

Uncertainties in the planet masses have improved by over an order of 
magnitude compared with the discovery paper \citep{Orosz_2012b}.
The mass estimates for the inner  and outer  planets 
are now $< 26 \, M_{\oplus}$ and $\sim 2$-$5 \, M_{\oplus}$, at the
$1\sigma $ level.
The middle planet  has a mass $\sim 7$-$43 \ M_{\oplus}$
($1\sigma$ range).
As usual, the radii are determined much better  than the masses:
3.05, 7.0, and 4.7  $R_{\oplus}$ for the inner, middle, and outer
planets, respectively.
The middle and outer planets have low bulk densities, 
$<0.68$ and $<0.26$ g cm$^{-3}$ at the 
$1\sigma$ level (Saturn's density is 0.69 g cm$^{-3}$).
The densities of all currently known low-mass planets are shown in
Figure \ref{massrad};  models indicate that such low-density planets 
must have substantial hydrogen and helium atmospheres \citep{Lopez_2014,
Jontof_2016}.
There is also an apparent trend that highly irradiated planets 
tend to have high densities while planets with low incident fluxes 
have a range of densities. Kepler-47c and Kepler-47 d fit this 
latter trend.

Several studies have shown that the formation of CBPs at close
distances to the binary is not expected to
proceed efficiently, and that CBPs
are likely
formed at large distances and migrated to their current orbits
\citep{Pierens_2008,Pierens_2013,
Meschiari_2012, Paardekooper_2012, Pelupessy_2013, 
Martin_2013,Rafikov_2013,Kley_2014,Kley_2015,Lines_2014}.
Planets that survive the migration phase are expected to have orbits 
with small eccentricities and small mutual inclinations relative to 
the orbital plane of the binary stars.  
The planets in Kepler-47 have low masses compared with Jupiter or 
Saturn, and the orbits have low eccentricities: 
$e <$ 0.030, 0.024, 0.05, 0.07 (at the +1$\sigma$ level)
for the binary and the inner, middle, and outer planets,
respectively.
All four orbits have mutual inclinations aligned to within 
1.6 degrees of one another (+1$\sigma$ level).
This nearly circular, co-planar, packed configuration is unlikely to 
have arisen as an outcome of strong gravitational scattering
of the planets into their current orbits.
Rather, the observations suggest that this planetary configuration 
is the result of 
relatively gentle migration 
in a circumbinary protoplanetary disk.

\subsection{Stability of the Orbits}\label{stability}

\begin{figure} 
\begin{center}
\includegraphics[width=0.95\textwidth]{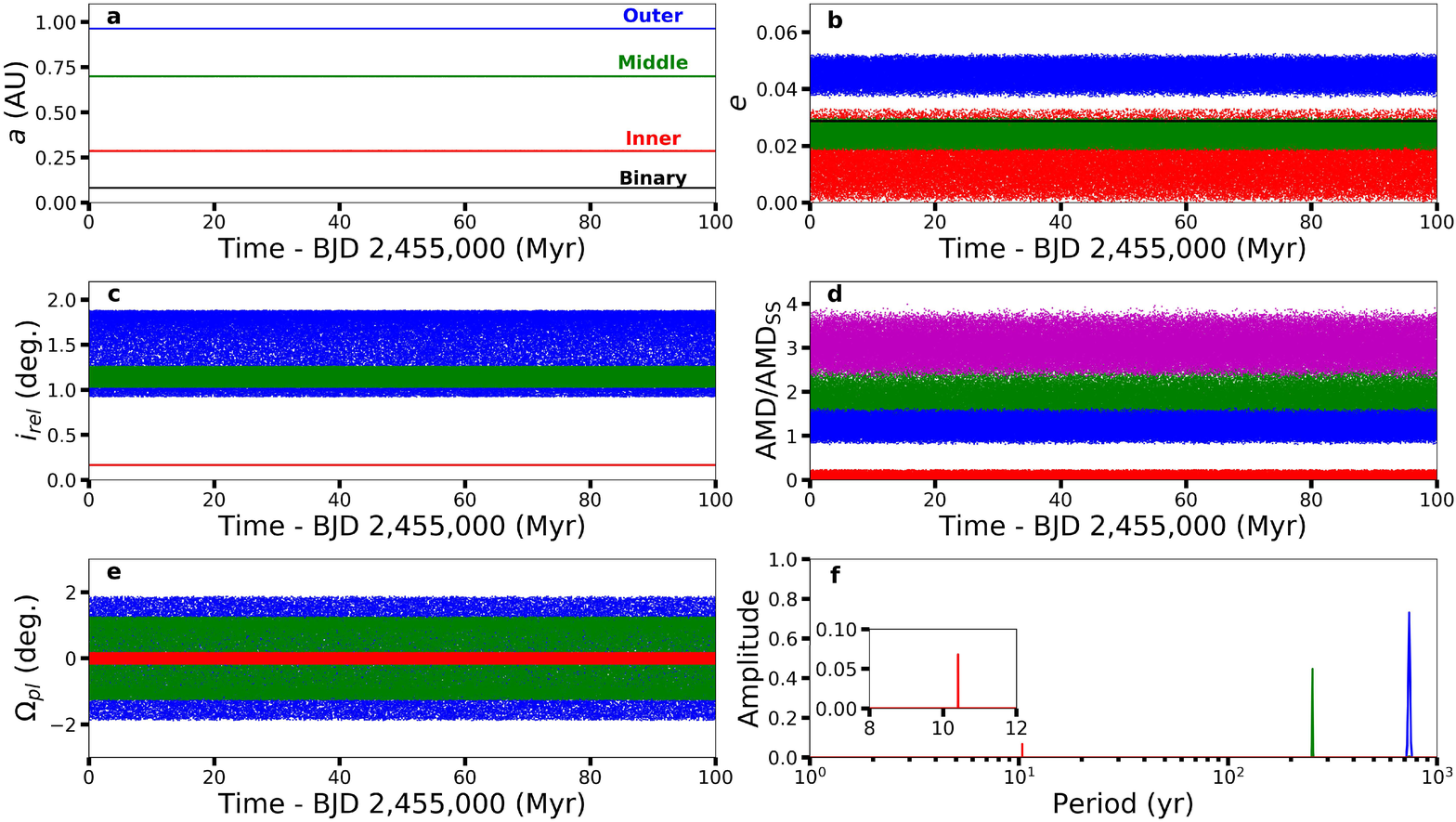}
\caption{Aspects of the long-term 
dynamics of the Kepler-47 system.
(a) The evolution of the semimajor axes for the four orbits
over a 100 Myr simulation.  Values for the binary are plotted
in black, values for the inner planet's orbit are plotted
in red, values for the middle planet's orbit are plotted in
green, and values for the outer orbit are plotted in blue.
(b) The evolution of the orbital eccentricity $e$ of the four orbits.
(c) The evolution of the mutual inclination between the
planetary orbits and the binary orbit.
(d) The evolution of 
the AMD relative to the terrestrial planets
in the solar system. The violet line shows the total AMD.
(e) The evolution of the nodal angle of the planetary orbits.
(f)
A power spectrum of the variations of the mutual inclination, showing
precession time scales of $\sim 10$ yr, 245 yr, and 738 yr for the inner,
middle, and outer planets, respectively.
\label{billy1}}
\end{center}
\end{figure}

\begin{figure} 
\begin{center}
\includegraphics[width=0.90\textwidth]{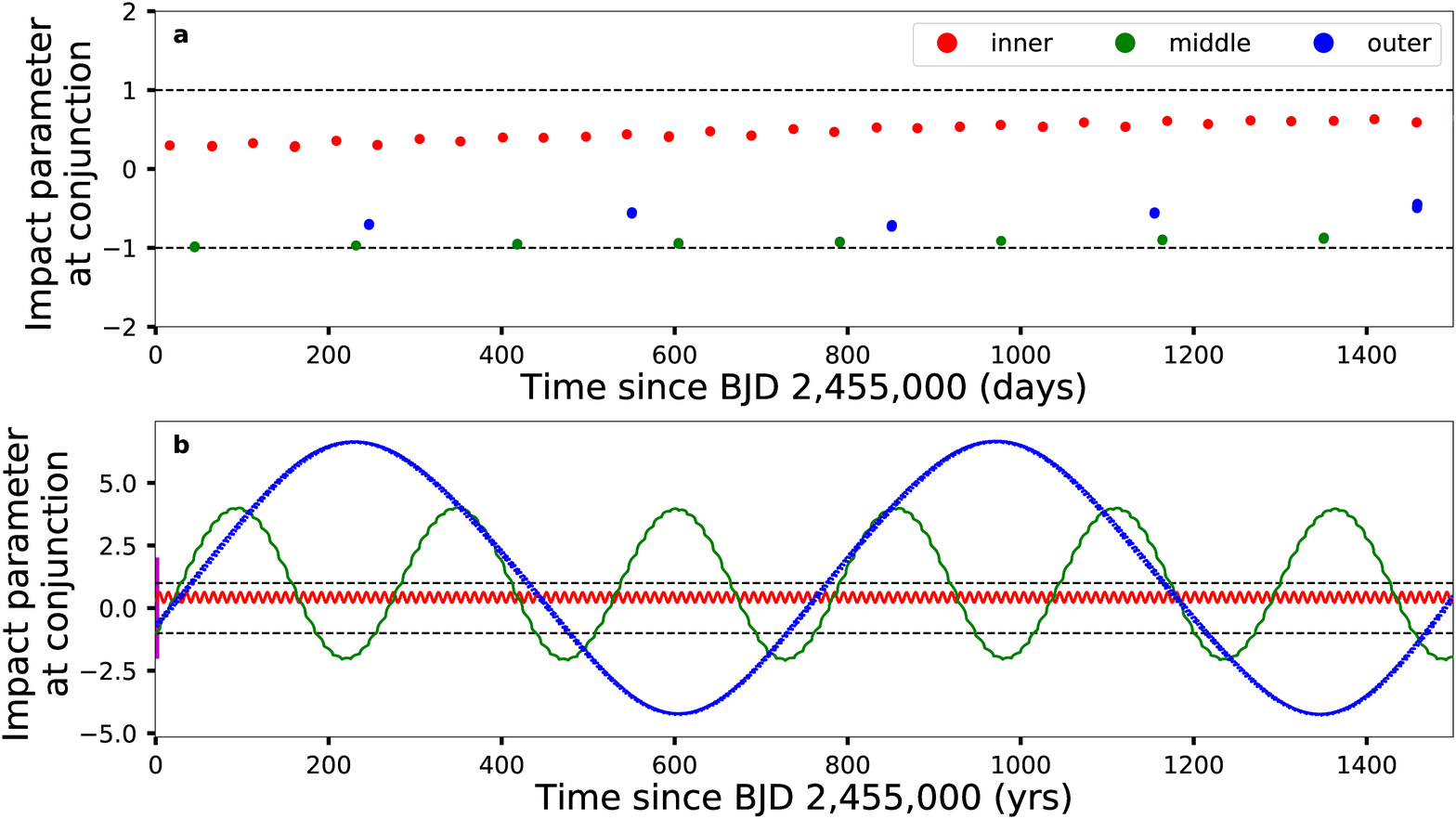}
\caption{Short-term evolution of future conjunctions.  
(a) The evolution of the
impact parameter $b$ at conjunction with the primary star for the
inner (red), middle (green), and outer (blue) planets for the duration
of the {\em Kepler} 
mission in \textit{days}.  (b) The evolution of future
conjunctions ($b\lesssim 1$ for transits) with the primary for each planet
for the next 1500 \textit{years}, demonstrating the precession cycles
of the orbital planes.
\label{billy2}}
\end{center}
\end{figure}

Numerical integrations have shown that the region around the 
binary where the orbit of a planet will become unstable 
extends to $\approx 0.18$ au 
\citep{Hinse_2015}.  Thus 
the semimajor axis of the innermost planet (0.2877 au) 
is much farther beyond the outer edge of the unstable region.
The integrations of the observed system
(using the nominal masses given in Table
\ref{tab:tab3}) also
show evidence of
stability, as the variations in the semimajor axis,
eccentricity, ascending node, and inclination relative to the binary
plane, $i_{\rm rel}$, do not have appreciable secular changes and are
remarkably flat in their maximum and minimum values over the 100 Myr
integration time scale
(see Figure \ref{billy1}).  
The angular momentum deficit (AMD) of the
system is used to estimate the changes of secular perturbations that
could induce instability on a long time scale.  We scale this quantity
relative to the present-day solar system value for the terrestrial
planets.  The Kepler-47 planets are more massive than our terrestrial
system, thereby allowing for higher values in AMD.  Finally, we show a
periodogram for the cyclical variations in inclination within the
system and find the precession time scales of the inner, middle, and
outer planets to be $\sim 10$ yr, 245 yr, and 738 yr, respectively
(see Figure \ref{billy1}).  Figure \ref{billy2}
shows the evolution of the impact parameters (for transits across
the primary star) for the three planets.  The horizontal dashed
lines in the figure denote impact parameters where transits
of the primary star could occur.  As one might expect,
the precessionary motion affects the duration of time for
which a given planet can transit relative to our line of sight
\citep{Schneider_1994}.  

We examined the possibility of additional planets within the
Kepler-47 system. Before exploring the 
possibility of the existence of an
actual planet, we began by integrating the orbits of a large battery
of test particles in between the system's planets.
The numerical scheme for these simulations used a modified
version of the orbital integration package, mercury6, that is
specifically designed to evaluate the orbits of CBPs
efficiently \citep{Chambers_2002}.
We chose to investigate the test particles initially on coplanar
(relative to the binary plane), circular
orbits and varied the initial phase of the test particles between 
$0^{\circ}$ and
$360^{\circ}$, in increments of $2^{\circ}$.  
Additionally we varied the
starting semimajor axis of the test bodies from 0.25 to 1.05 au, such
that we could evaluate whether non-transiting planets between either
the inner-middle pair or the middle-outer pair could be stable for 10 Myr.
Overall, we integrated test particles corresponding to a grid of
$801\times 181$ 
initial conditions.  From this study, we found that virtually
all of the 
test particles between the middle and outer planets became
unstable (experiencing
ejection from the system or a collision with another body),
indicating that no stable (prograde)
planetary orbits exist between these two
planets. In other words, the region between the middle and outer
planets is dynamically full (i.e., these planets are in a packed
orbital configuration).
However, there remained a broad region of parameter space where the
test particles were stable between the inner and middle planets.
Despite this stability, there is 
insufficient evidence to suggest that a massive body  exists,
because
of the many dynamical interactions that would influence the
observations of the transiting planets and thereby alert us to its
presence.  There are many regions where the test particles'
eccentricities were significantly increased due to mean motion resonances
with the middle planet and we expect these locations to be eroded over
time in a similar manner to how the 
Kirkwood gaps in the solar system
are.

Based on
analytical studies, we used the estimate of \citet{Chambers_1996},
assuming that the binary can be approximated as a single star with a
large oblateness, and found
that the separation in mutual Hill radii of
the inner-middle pair and middle-outer pair to be $32.90 \pm 17.44$
$R_{H,m}$ and $12.34 \pm 4.43$ $R_{H,m}$, 
respectively. These values are well
above the critical value of $2\sqrt{3}\approx 3.46$ 
for single-star systems 
\citep{Gladman_1993}
or the critical range of values (5--7) suggested by \citet{Kratter_2014}
for circumbinary systems. 
The mutual Hill radii of the middle-outer pair are quite close,
even when considering the uncertainties, and this proximity does not
allow for any additional planets to orbit between 
0.7 and  0.96 au.  Based on
our numerical tests, 
assuming nominal mass estimates, the middle and outer
planets are ``packed.'' This conclusion is supported by
other numerical studies
\citep{Smith_2009,Kratter_2014,Quarles_2018a}.

\begin{figure} 
\begin{center}
\includegraphics[angle = -90,clip,width=0.95\textwidth]{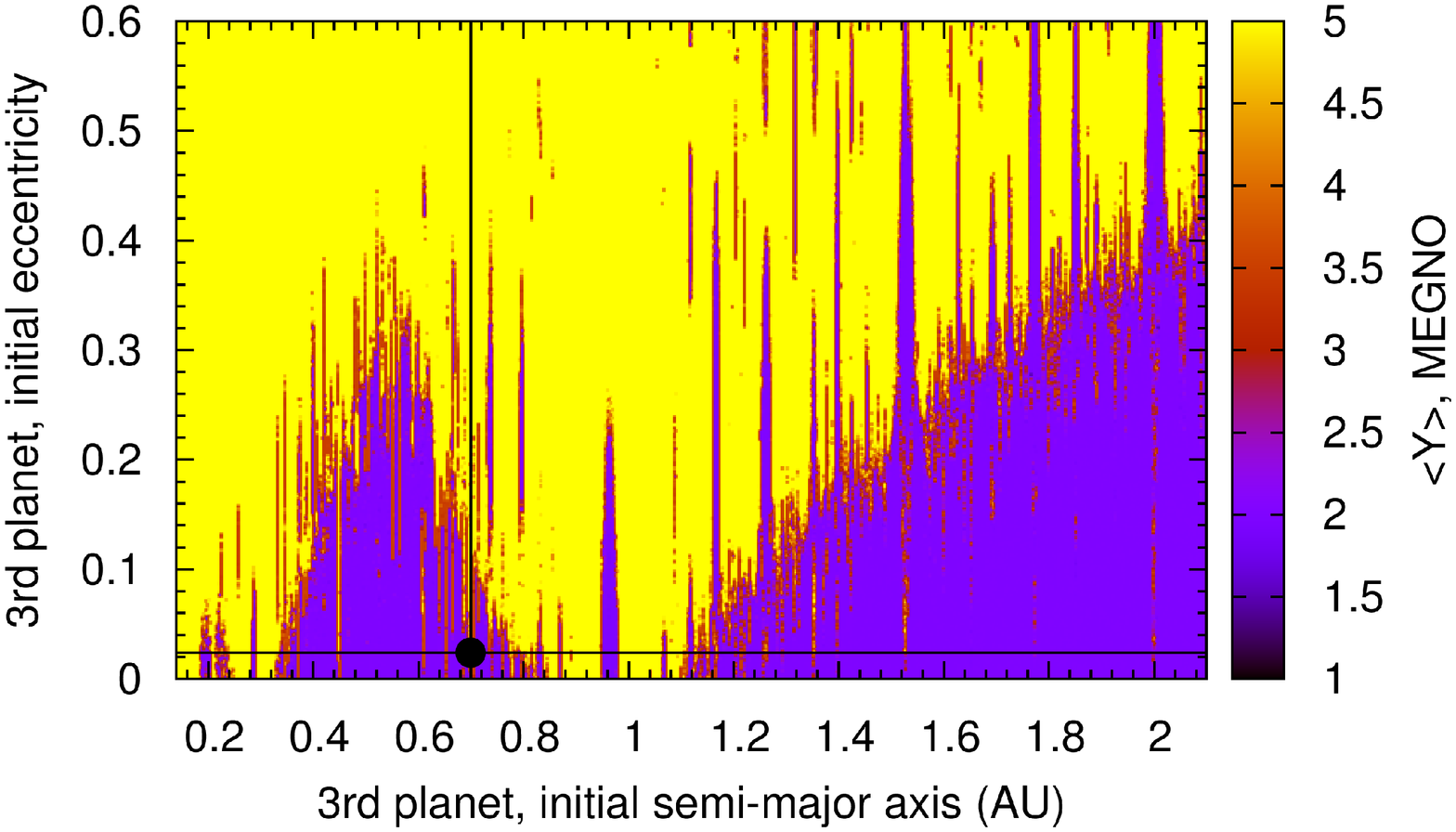}
\caption{A dynamical MEGNO map for the five-body system is shown. 
The map shows chaotic
(yellow color code)/quasiperiodic (blue color code) regions for a
third planet being perturbed by the remaining four bodies
(the two stars and the inner and outer planets,
with nominal parameter values). The
resolution is $N_x = 600, N_y = 500$ pixels, 
rendering a total of
30,000 initial conditions. Each orbit is integrated for 5000 yr.
The location of the middle
planet's best-fit osculating element is shown
by a filled circle placing it in a region of quasiperiodicity packed
between the inner and outer planets.
\label{MEGNO}}
\end{center}
\end{figure}

We have calculated a dynamical 
Mean Exponential Growth factor of Nearby Orbits 
(MEGNO) map \citep{Cincotta_2003,Gozdziewski_2001} of the middle
planet using the best-fit osculating (Jacobian) initial
conditions presented in this work. Initial values for the mean
anomalies of the binary and each planet were set to zero. The MEGNO
factor $\langle Y \rangle$ is a numerical technique to detect and
differentiate between aperiodic (chaotic) and quasiperiodic
(regular/stable) motion. The computation
of the variational vector
necessary to calculate MEGNO was done
by solving the variational
equations of motion in parallel to the five-body equations of
motion. These equations of motion were solved using the accurate
adaptive time-step ODEX 
integrator\footnote{http://www.unige.ch/$\sim$hairer/prog/nonstiff/odex.f}. 
A given 
integration was
terminated once $\langle Y \rangle$ was greater
than five. 
The result of computing MEGNO over a grid in semimajor axis
versus
eccentricity space is shown in Figure \ref{MEGNO}. 
The color coding follows
the numerical value of MEGNO attained at the end of the
integration. Yellow colors indicate chaotic and a blue ($|\langle
Y\rangle - 2.0| < 0.001$) color indicate quasiperiodic dynamics of
the third planet. The binary and remaining two planets were started at
their best-fit osculating Jacobian elements.

Recently, the MEGNO technique was used to study possible positions of the
middle planet in the framework of the five-body problem \citep{Hinse_2015}. 
Compared to that study, the quantitative picture of the
semimajor axis vs.\ eccentricity phase-space structure has changed
significantly in the present analysis. The main difference is that
the
eccentricity of the outer planet  has decreased from 0.41 to
almost zero
in the newly found best-fit model. 
This demonstrates the overall destabilizing effect  
an eccentricity 
of 0.41 for
the outer planet orbit 
has, as it would greatly limit the number of possible stable orbits
of the middle planet. This is a fine example of the possibility of
planet packing as demonstrated in \citet{Kratter_2014}. Finally, 
the location of
mean-motion resonances 
(vertical structures in Figure \ref{MEGNO}) might change
for different initial phases of the planets (when varied within their
parameter uncertainties). However, the overall global topology
structure of chaotic/quasiperiodic orbits of the third planet 
would
not change dramatically.

While there is no dynamical evidence for an additional
planet in the Kepler-47 system, another (low-mass)
planet between the inner and middle
planets could exist.  Also, more planets could have stable orbits 
sufficiently
far outside the
orbit of the outer planet.  Thus, 
it would be worthwhile to obtain
additional high-precision photometry of Kepler-47 whenever 
possible, to look
for additional transits.
From Figure \ref{billy2}, we see that the precession 
timescales for the three known planets
are relatively short, and that the middle and outer planets
can transit the primary star for only a small fraction of
their precession cycles
($\sim23\%$ and $\sim12\%$ of the time
for the middle and outer planets,
respectively).  If there are additional planets
in the Kepler-47 system with similar precession cycles,
they might precess into view at a later time.

\subsection{Comparison with Stellar Evolution Models}\label{evol}

\begin{figure} 
\begin{center}
\includegraphics[scale=0.65]{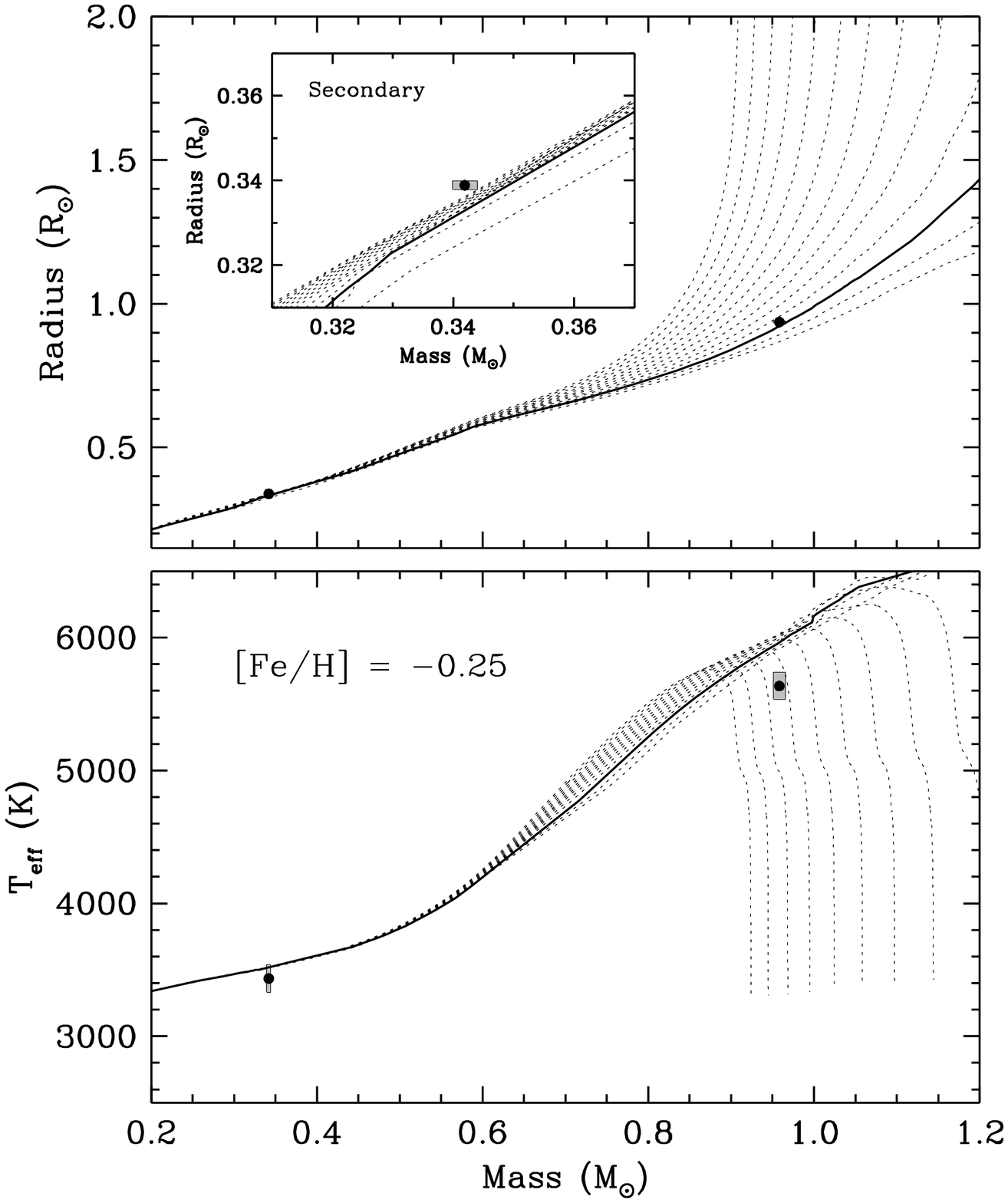}
\caption{Isochrones from the Dartmouth models \citep{Dotter_2008}
corresponding to ages from 1 to 13 Gyr
(the oldest isochrone is the steepest one at the high-mass
end) and metallicity of $[{\rm Fe/H}]=-0.25$
\citep{Orosz_2012b}, compared against the measured
masses, radii, and temperatures of the stars in Kepler-47.
The shaded boxes represent the error bars for the measurements.
The top shows the mass-radius plane, with the inset showing a magnified
view near the secondary.  The bottom shows the mass--temperature diagram.
\label{plotHR01}}
\end{center}
\end{figure} 

\begin{figure} 
\begin{center}
\includegraphics[scale=0.65]{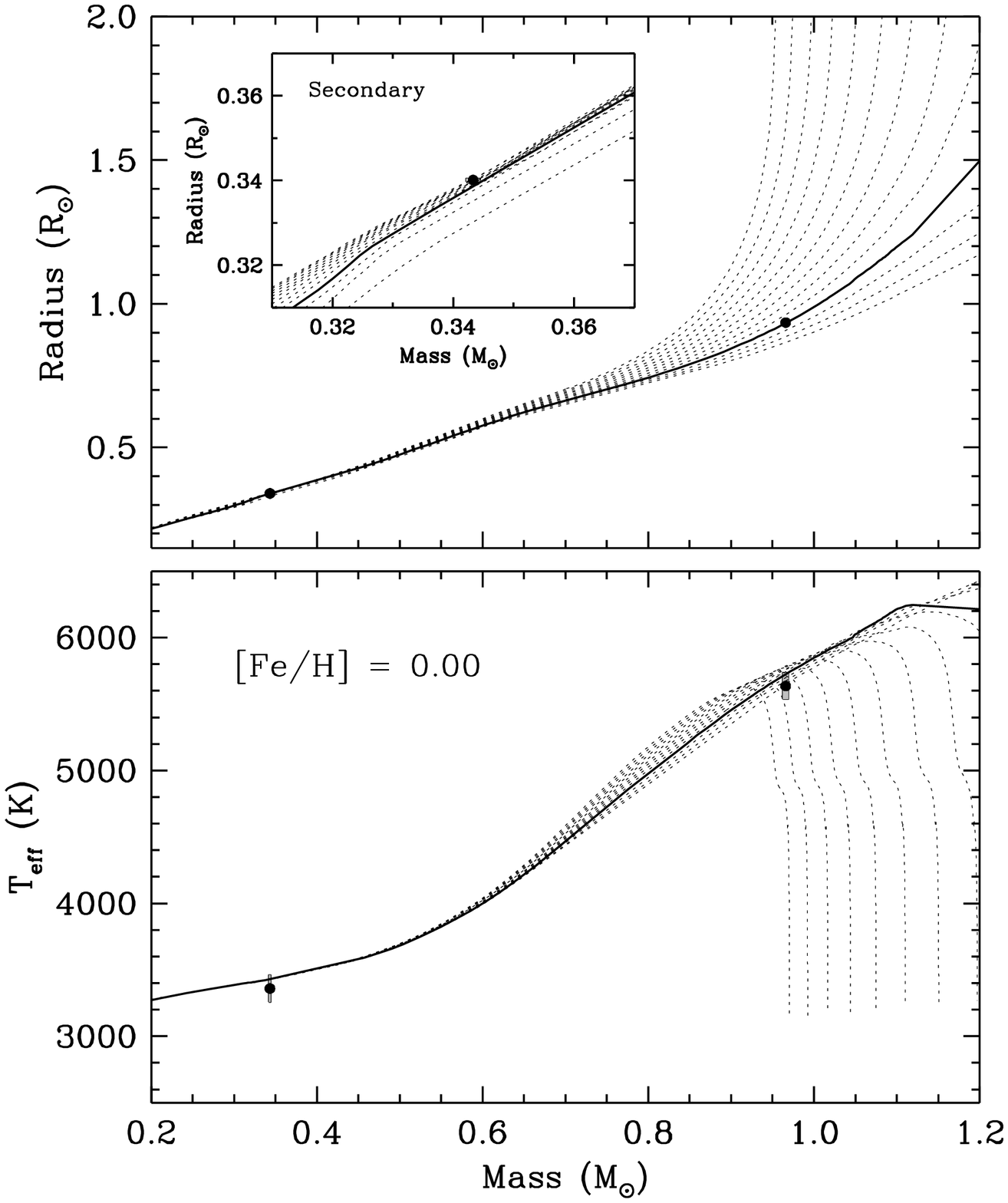}
\caption{Similar to figure \ref{plotHR01}, but for Dartmouth models with solar
metallicity.
\label{plotHR02}}
\end{center}
\end{figure}

Our new values for the
masses of the stars in Kepler-47 are about 6\% 
smaller than in the discovery paper,
and  the new radii are about 3\% 
smaller than previously reported.  We therefore
update our analysis and discussion of the 
evolutionary state of the Kepler-47 binary. 
Figure \ref{plotHR01} compares
the new measurements against Dartmouth models 
\citep{Dotter_2008}
for the nominal spectroscopic metallicity of $[{\rm Fe/H}] = -0.25$ 
\citep{Orosz_2012b},
in the same way as we did previously. 
There is still an age discrepancy for the primary star: the
mass/radius combination yields an age of 3.5 Gyr (solid line in top
panel), whereas the mass/temperature combination
gives 11.5 Gyr. If the radii are
given more weight than the temperatures, then the secondary star is
now slightly larger than the theoretical predictions, as is seen in
many other M dwarfs. As we noted in the discovery paper, the primary
temperature derived from color indices points to a higher value than
the spectroscopy does
(which would actually be consistent with the 3.5 Gyr
age). However, given the
uncertainties in the reddening, the spectroscopy is
perhaps more trustworthy.

By way of comparison, Figure \ref{plotHR02} 
shows the Dartmouth models for solar metallicity. 
The primary age from mass/radius is now 4.5 Gyr (solid line), and
mass/temperature gives an age consistent with this within the error
bar. The secondary radius would be perfectly consistent with the
prediction for this metallicity, as would the secondary
temperature. However, 
solar metallicity is formally ruled out by the
spectroscopic analysis \citep{Orosz_2012b}
at the 3$\sigma$ level, so
a scenario where the stars have solar metallicity is unlikely.

We note also that, in its latest data release (DR2), 
the {\em Gaia} mission
\citep{GaiaCollaboration:2016,
GaiaCollaboration:2018} has provided an accurate parallax for
Kepler-47 (source ID 2080506523540902912) 
of $\pi=0.948 \pm 0.029$ mas, corresponding
to a distance of $1025^{+31.5}_{-29.7}$ pc
\citep{Bailer-Jones_2018},
that enables a check on our
dynamical radius measurement for the primary star. With a reddening
estimate to the system of $E(B-V) = 0.056 \pm 0.020$ from
\cite{Green:2015},  the assumption that $A_V = 3.1 E(B-V)$, a
bolometric correction from \cite{Flower:1996} based on the
temperature, and the visual magnitude of the star \citep[$V = 15.395
  \pm 0.010$,][]{Henden:2015}, we obtain $R = 0.928 \pm
0.072\,R_{\odot}$. This is less precise than our dynamical value of $R =
0.936 \pm 0.005\,R_{\odot}$, but is in excellent agreement.

\subsection{The Habitable Zone}\label{HZdiscussion}

The habitable zone (HZ) is usually defined as the region
around a star such that the stellar incident flux on the planet
(i.e.\ the insolation) would allow an Earth-like planet with a 
CO$_{2}$/H$_{2}$O/N$_{2}$ atmosphere to permanently maintain liquid 
water on its solid surface. The situation is more complicated
for a CBP system, 
in that (i) there are two stars, typically of different temperatures and 
luminosities; and (ii) the orbital motion of the stars causes
the HZ to ``orbit'' with the binary. Even for a planet on a purely
circular orbit, the planet--star 
distances are always rapidly changing.
The HZ is no longer static or spherically symmetric.

Therefore, for a CBP system, the definition of 
the HZ has to be slightly 
generalized.
Here, we define a circumbinary HZ as the region in space where the
time-averaged weighted insolation would allow conditions conducive to 
liquid water on the surface of an Earth-like planet. The weighted 
insolation is not simply the total flux incident
at the top of the
planet's 
atmosphere (though it reduces to this if one of 
the stars' flux contribution is negligible compared to the other,
as is the case for Kepler-47).
Adding the flux from each star is not appropriate because it is not 
simply a matter of the total energy; the spectral energy distribution 
is highly relevant. 
The planet's atmosphere acts to filter the energy received 
from each star before it reaches the planet's surface, and this 
filtering is strongly wavelength-dependent.
The spectral energy distribution of the light emitted by the star
can be reduced to a simple function of the star's effective temperature
($T_{\rm{eff}}$) to a sufficient level of approximation.
Thus, the $T_{\rm{eff}}$-weighted sum of the fluxes from 
each star is used as the total flux received by the planet
\citep{Haghighipour_2013}.
To determine the inner and outer boundaries of the circumbinary HZ, 
we equate the weighted incident flux with the 
insolation corresponding to those boundaries for a single star as 
computed by \citet{Kopparapu_2013,Kopparapu_2014}.

\begin{figure} 
\begin{center}
\includegraphics[scale=0.72]{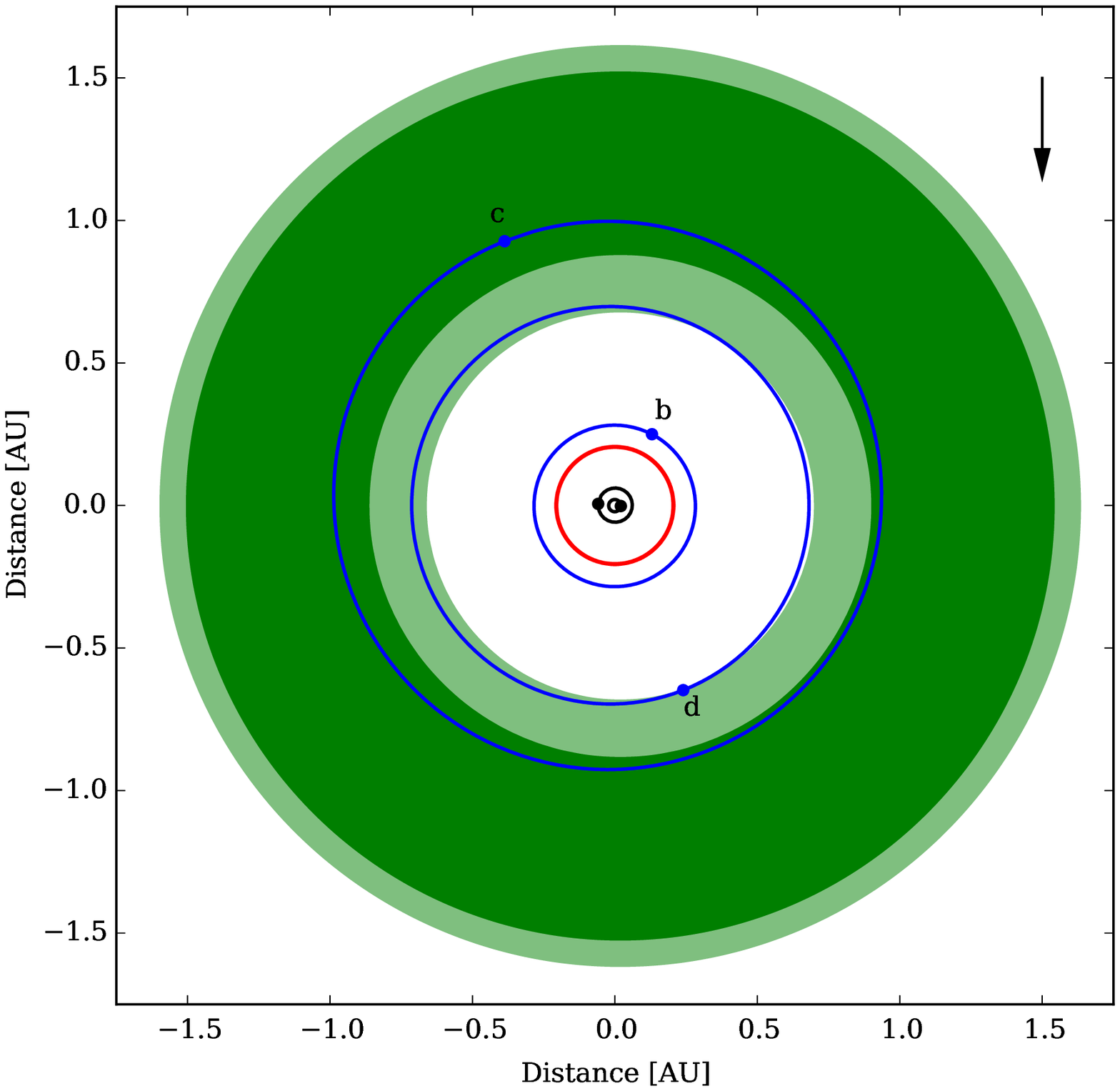}
\caption{The conservative (dark green) and optimistic (light green) habitable 
zone regions are shown for the Kepler-47 system. The red circle shows
the critical stability radius 
\citep{Holman_1999}, 
interior to which 
planetary orbits are most likely unstable.
\label{HZ01}}
\end{center}
\end{figure} 

\begin{figure} 
\begin{center}
\includegraphics[angle = -90,width=1.0\textwidth]{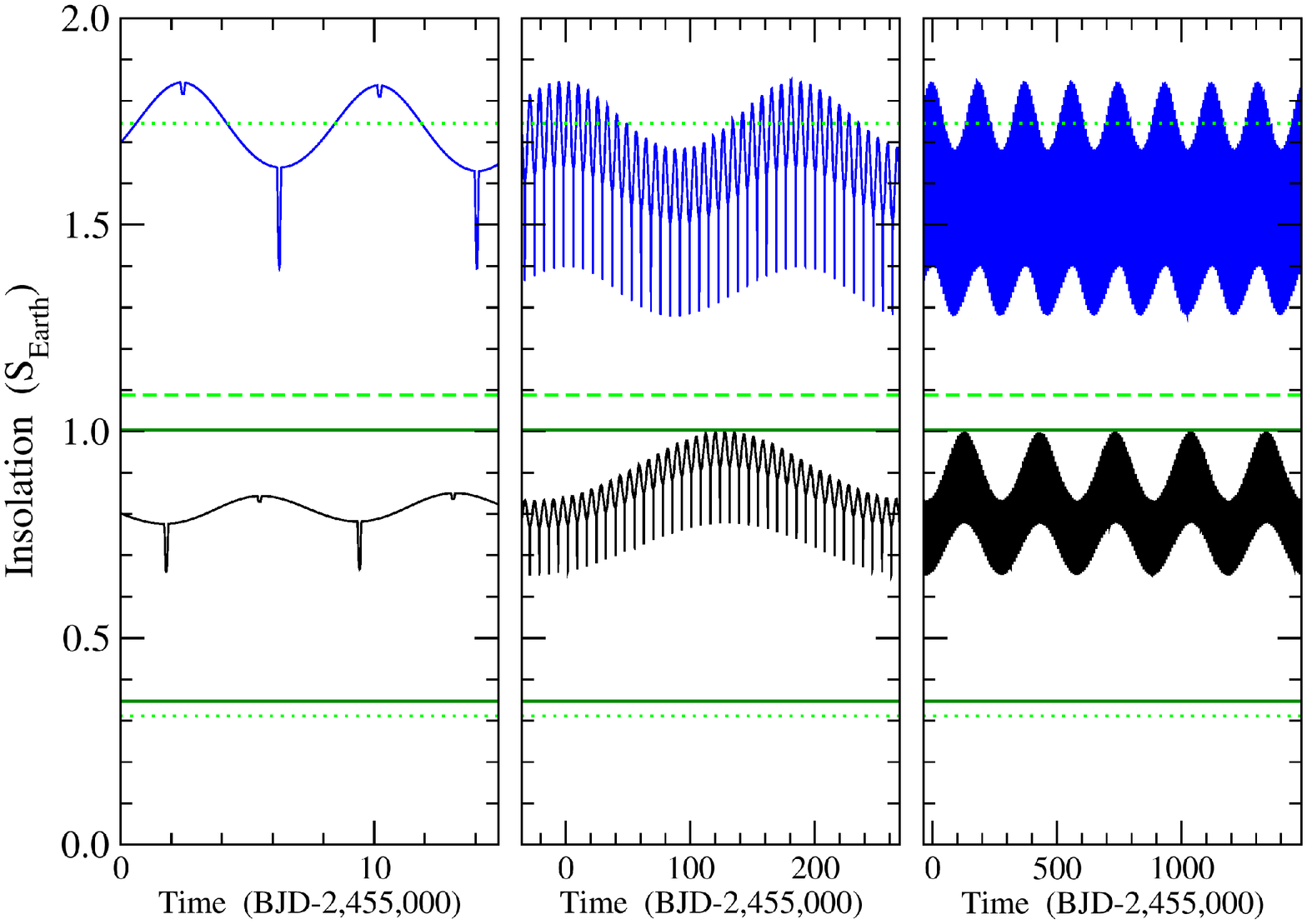}
\caption{The total 
insolation $S$ received by the middle planet (blue) and outer 
planet (black) is shown as a function of time
in days. The solid green lines
mark the boundaries of the conservative HZ, and
the dotted green lines mark 
the optimistic HZ. The dashed green line marks the runaway greenhouse
condition, a less stringent limit than the moist greenhouse condition.
The times spanned in the three panels (left to right) are three binary 
orbits, one outer planet orbit, and five outer planet orbits, 
respectively.
The sharp 
downward spikes show the rapid drop in flux that occurs when the 
stars eclipse one another.
\label{HZ02}}
\end{center}
\end{figure}

Figure \ref{HZ01} shows a face-on view of the Kepler-47 system.
The binary stars are shown in black, the planets in blue,
the HZ in green, and the red circle marks the critical instability 
radius 
\citep{Holman_1999}.
The darker green shaded area corresponds to the conservative HZ as 
defined by the runaway greenhouse and maximum greenhouse conditions.
The lighter green regions show the optimistic HZ boundaries, as
defined by the recent Venus and early Mars conditions \citep{Kopparapu_2013}.
The arrow shows the direction of the line of sight from the observer
to Kepler-47. The sizes of the stars and planets are not to scale, 
and their locations are for the epoch of the orbital elements
presented in Table
\ref{tab:tab2}. The outer planet
is within the limits of the conservative HZ, and Planet D 
(the middle planet) skirts the inner boundary of the optimistic 
HZ\footnote{An animated version of this figure  can be found at
http://astro.twam.info/hz-ptype/}.

In Figure \ref{HZ02}, 
we show the total (unweighted) insolation incident at
the top of the atmosphere of the middle planet (in blue) and the outer
planet (in black)
as a function of time. The dotted lines show the limits of the 
optimistic HZ and the 
solid green lines mark the limits of the HZ defined by the
moist greenhouse condition. The dashed line shows the boundary 
corresponding to the runaway 
greenhouse condition, which is also often used as the boundary
for the inner edge of the conservative HZ.
The three panels show three different time scales.  
The left panel spans two binary star periods, and the variations
in insolation are dominated by the orbital motion of the primary star.
Note the primary and secondary eclipses, as seen from the location of 
the planets. The middle panel shows one orbital period of the outer 
planet, and the insolation variations now show the effect of 
the eccentricities of the planet's
orbit. The right-hand panel shows five
orbits of the outer planet.
The time-averaged total insolation for the outer planet is 
$S=0.865$ in Sun--Earth units, with an rms variation of 7.2\% about 
the mean. For the middle planet (d), the time-averaged insolation is
$S=1.67$, with an rms variation of 5.5\%.
For comparative purposes (only), if one assumes a Bond albedo 
of 0.34 \citep[similar to that of Jupiter or Saturn;
see][]{dePater_2015},
and complete redistribution of the incident energy, 
the equilibrium temperature is $T_{\rm eq} \approx 241$ K for the 
outer planet.
In the course of this discussion, we must keep in mind that these 
are low-density planets, with thick H, He atmospheres,
and therefore not likely to be habitable. 

\subsection{Times, Durations, and Impact Parameters of Future Eclipses
and Planet Transits}

Ground-based observations of eclipse and transit events would be useful
in constraining the dynamical model of the Kepler-47 system.  The
times of primary and secondary eclipses are well-fit
by the linear ephemerides given in Equation (1).  
Because 
the times of the planet transits do not follow a linear ephemeris,
we computed (using ELC) the  times, impact parameters, and depths
of future transits (for the years 2019 through
2026) for each planet using the
best-fitting model given in Table \ref{tab:tab2}.  
The uncertainties
on these quantities were estimated using sets of initial conditions drawn
from the posterior samples of solutions found from the DE-MCMC run, and
are given in Table \ref{tabfuture}.
The formal uncertainties on the predicted times generally
vary between about one to two hours for the inner planet, and 
about 0.75 to two hours for the middle planet and outer planet.
With a depth of about
15\%, the primary eclipses should be observable by modest-sized
ground-based  telescopes.  On the other hand, owing to 
the relative faintness
of the system
\citep[$V = 15.395 \pm 0.010$,][]{Henden:2015}, 
the 0.5\% deep secondary eclipses will be more difficult
to observe.  With depths between about 0.1\% and 0.4\%, the transits
of three planets will be more challenging still.  
According to our best-fitting model, the inner planet  always 
transits during its $\sim10$ yr precession cycle, 
the middle planet transits
about 23\% of the time during its $\sim245$ yr precession cycle, 
and the outer planet transits about 
12\% of the time during its $\sim738$ yr precession cycle.  
We used the REBOUND software \citep{Rein_2012} to integrate a
set of 368 initial conditions drawn from the posterior sample to
estimate the uncertainties on the predicted transit times near the times
when the transits stop.
We find that the uncertainties of the predicted transit times
for the middle planet are about 1.4 days by the time this planet 
stops
transiting in the year $2037.7\pm 3.4$ (because the transits
are grazing before they stop, the predicted stopping date has
a much larger spread). Similarly, the uncertainties
of the predicted transit times for the outer planet are about
2.1 days by the time this planet
stops transiting in the year
 $2049.1\pm 6.8$.

\section{Summary}\label{summary}

We have identified a third planet in the Kepler-47 system.
Although Kepler-47~d is the largest of the three planets, 
it was not previously detected because its transits were much 
weaker in the early {\em Kepler} data, 
as the transit impact parameter was near unity.
The precession of its orbital 
plane is such that the transits are currently
growing deeper with time.
We modeled the {\em Kepler} light curve and radial velocity curve
simultaneously
using a five-body 
photodynamical model with 42 free parameters. 
The revised system parameters are generally in agreement with the
previously published values \citep{Orosz_2012b}.  The tight
constraints placed by the planets on the orbits of the stars enable
highly precise measurements of their masses and radii (see Table
\ref{tab:tab3}).
The stars have masses of $0.957 \pm 0.014$ and $0.342 \pm 0.003$
$M_{\odot}$ and radii of $0.936 \pm 0.005$ and $0.338 \pm 0.002$
$R_{\odot}$, consistent with an age of 3.5-11.5 Gyrs and slightly
favoring an 
age in the  younger portion of its range.  
Uncertainties in the planet masses have improved by over an order of 
magnitude compared with the 2012 discovery paper \citep{Orosz_2012b}.
The $1\sigma$ mass estimates for the planets are
now $< 26 \ M_{\oplus}$,
$\sim 7$-$43 \ M_{\oplus}$, and $\sim 2$-$5 \ M_{\oplus}$
for the inner, middle, and outer planets 
respectively.
The radii are much better determined than the masses, and we find
radii of 
3.05, 7.0, and 4.7 $R_{\oplus}$ for the inner, middle, and
outer planets, respectively.
The middle and outer planets have low bulk densities, 
$<$~0.68 and $<$~0.26 g cm$^{-3}$ at the 
$1\sigma$ level.  With such low densities,
these two planets must have substantial hydrogen
and helium atmospheres \citep{Lopez_2014}.

All of the orbits in the Kepler-47
system have eccentricities less than 0.07 at the $+1\sigma$
level, and all four orbits have mutual inclinations aligned to within 
1.6 degrees of one another at the $+1\sigma$ level.
This nearly circular, co-planar, packed configuration is unlikely to 
have arisen as an outcome of strong gravitational scattering
of the planets into their current orbits.
Rather, the observations suggest that the planetary configuration 
is the result of 
relatively gentle migration in a circumbinary protoplanetary disk.

We found that the configuration is
dynamically stable for at least 100 Myr.  The new planet conforms with
expectations based on an earlier analysis of a two-planet system
\citep{Hinse_2015}.  As clearly demonstrated by the appearance and
growth of the
middle planet's transits, the planet's orbit precesses. The
precession time scales for the inner, middle, and outer planets are
calculated to be $\sim$ 10, 245, and 738 yr.  The middle and outer
planets are separated by $\sim$ 12 mutual Hill radii, and we verify
the earlier estimation \citep{Kratter_2014}
that they are dynamically packed.  This is
the first detection of a dynamically packed region in a circumbinary
system, and it further
confirms suspicions that planet formation and
subsequent migration can proceed much like that around a single star,
at least when far from the binary
\citep{Pierens_2008,Pierens_2013,Kley_2014,Kley_2015}.  We also find
that, although they are close to having integer commensurate periods,
the middle and outer planets are not in a mean-motion resonance--and
yet they
are gravitationally interacting and exchanging angular momentum, as
indicated by their antiphased oscillations in inclination and
eccentricity.

The average insolation (incident radiative
flux) the outer planet receives from its two suns is 86.5\% of the
Sun--Earth value, placing it within the boundaries of the circumbinary
habitable zone \citep{Haghighipour_2013}.  The middle planet's orbit
straddles the ``recent Venus'' hot edge of the system's habitable
zone.  However, given their densities, neither planet is likely to host
life.

\acknowledgments
{\em Kepler} was selected as the 10th mission of the Discovery Program.
Funding for this mission is provided by the NASA
Science Mission Directorate. 
The {\em Kepler} data were obtained from the Mikulski
Archive for Space Telescopes (MAST). 
The Space Telescope Science Institute
(STScI) is operated by the Association of Universities for Research
in Astronomy, Inc., under NASA contract NAS5-26555.
Support for MAST for non-HST data is provided by the NASA Office
of Space Science via grant NXX09AF08G and by other grants and contracts.  
This work is based in part on observations obtained with the 
Hobby--Eberly Telescope, which is a joint project of the University of 
Texas at Austin, the Pennsylvania State University, Stanford University, 
Ludwig-Maximilians-Universit\"at M\"unchen, and Georg-August-Universit\"at 
G\"ottingen.
This research made use of data from the AAVSO Photometric 
All-Sky Survey (APASS), funded by the Robert Martin Ayers Sciences Fund and NSF AST-1412587.
This work presents results from the European Space Agency (ESA)
  space mission {\em Gaia}. 
{\em Gaia} data are being processed by the 
{\em Gaia} Data
  Processing and Analysis Consortium (DPAC). Funding for the DPAC is
  provided by national institutions, particularly the institutions
  participating in the {\em Gaia} 
MultiLateral Agreement (MLA). The {\em Gaia}
  mission website is https://www.cosmos.esa.int/gaia. The {\em Gaia} archive
  website is https://archives.esac.esa.int/gaia.
Some of the simulations in this paper made use of the 
REBOUND code, 
which can be downloaded freely at http://github.com/hannorein/rebound.
J.A.O.\ and W.F.W.\
acknowledge support from the Kepler Participating Scientist 
Program via NASA grant NNX14AB91G and support from NSF grant AST-1617004.
J.A.O., W.F.W., G.W., and B.Q.\ 
also gratefully acknowledge support from NASA via
grant NNX13AI76G.
D.C.F.\ acknowledges support from the Alfred P. Sloan Research Fellowship.  
D.C.F.\ and S.M.M.\ were 
supported by NASA through the {\em Kepler} Participating 
Scientist Program award NNX14AB87G.
N.H.\ 
acknowledges support from the NASA ADAP program under grant 
NNX13AF20G, NASA PAST program grant NNX14AJ38G, and NASA XRP 
program under grant 80NSSC18K0519.
T.C.H.\ 
acknowledges KASI research grant \#2015-1-850-04.
B.Q.\ gratefully acknowledges support by an appointment to the NASA 
Postdoctoral Program at the Ames Research Center, administered by Oak 
Ridge Associated Universities through a contract with NASA.
G.T.\ acknowledges partial support for this work from NSF grant
AST-1509375.
E.B.F.\ 
acknowledges support from NASA Participating Scientists Program
awards NNX12AF73G and NNX14AN76G, NASA ADAP program under grant 
NNX13AF20G, as well as the University of Florida
and the Pennsylvania State University's Center for Exoplanets and
Habitable Worlds.  The Center for Exoplanets and Habitable Worlds is
supported by the Pennsylvania State University, the Eberly College of
Science, and the Pennsylvania Space Grant Consortium.
J.J.L.'s work on this project was supported by 
NASA's Astrophysics Data Analysis Program under grant
no.\ 16-ADAP16-0034. 
Finally, we are deeply grateful to John Hood Jr.\ 
for his generous support of exoplanet research at San Diego
State University.

\clearpage

\begin{deluxetable}{cccccc}
\tablecaption{Radial velocities for Kepler-47\label{tabRV}}
\tablewidth{0pt}
\tablehead{
\colhead{Date}   & 
\colhead{UT Time}    &
\colhead{BJD} & 
\colhead{RV$_{A}$}  &
\colhead{RV$_{A}$}  &
\colhead{Telescope} \\
\colhead{YYYY-MM-DD}   & 
\colhead{}    &
\colhead{(2,455,000+)} & 
\colhead{(km s$^{-1}$)\tablenotemark{a}}  &
\colhead{(km s$^{-1}$)\tablenotemark{b}}  &
\colhead{}
}
\startdata
2012 Apr 10 &  13:25:48.68  & 1028.05942 & $11.442  \pm 0.011$   & \nodata  & Keck \\
2012 Apr 23 &  09:11:27.36  & 1040.90325 & $33.534  \pm 0.091$   & $28.934\pm 0.164$  & HET \\
2012 May 1 &  09:52:55.08  & 1048.93237 & $35.458  \pm 0.171$   & $30.890\pm 0.308$  & HJST \\
2012 May 2 &  07:23:45.95  & 1049.82882 & $24.430  \pm 0.440$   & $19.862\pm 0.792$  & HJST \\
2012 May 4 &  08:34:10.40  & 1051.88474 & $-21.957 \pm 0.159$   & $-26.525\pm 0.286$  & HJST \\
2012 May 5 &  08:08:26.83  & 1052.86692 & $-26.719 \pm 0.178$   & $-31.287\pm 0.320$  & HJST \\
2012 May 6 &  08:08:55.42  & 1053.86729 & $-9.150  \pm 0.122$   & $-13.718\pm 0.220$  & HJST \\
2012 May 18 &  07:35:21.15  & 1065.83749 & $-1.843  \pm 0.060$   & $-6.443\pm 0.108$  & HET \\
\enddata
\tablenotetext{a}{Uncorrected radial velocities.}
\tablenotetext{b}{Corrected radial velocities with scaled error bars.}
\end{deluxetable}

\begin{deluxetable}{lccccccl}
\tabletypesize{\scriptsize}
\tablecaption{Times and Durations of Planetary Transits \label{measuredtransits}}
\tablewidth{0pt}
\tablehead{
\colhead{}   & 
\colhead{Measured}    &
\colhead{Uncertainty} & 
\colhead{Duration}  & 
\colhead{Model}  &
\colhead{Model}  &
\colhead{Impact}  &
\colhead{} \\
\colhead{cycle\#}    &
\colhead{Time\tablenotemark{a}} & 
\colhead{(minute)}  & 
\colhead{(hr)}  &
\colhead{Time$^a$}  &
\colhead{Duration (hr)}  &
\colhead{Parameter} &
\colhead{Note} 
}
\startdata
{\underline{\bf Inner Planet}} &   &   &   &   &   &   &   \\
1.0   &  -30.80969 &  4.46  & $ 4.55 \pm 0.19$ &  -30.80876 & 4.76  & 0.29  & \\
  2.0 &      ...   &  ...   &            ...   &   16.27666 & 4.10  & 0.30  & data gap\\
  3.0 &   65.24832 & 22.29  & $ 5.81 \pm 1.40$ &   65.24587 & 6.41  & 0.29  & \\
  4.0 &  112.54179 & 10.51  & $ 2.94 \pm 0.69$ &  112.54644 & 3.76  & 0.33  & \\
  5.0 &  160.89953 & 32.75  & $10.83 \pm 2.66$ &  160.90621 & 10.45 & 0.28  & \\
  6.0 &  208.84796 &  7.54  & $ 3.47 \pm 0.66$ &  208.83932 & 3.69  & 0.36  & \\
  7.0 &  256.34222 & 11.56  & $ 8.86 \pm 0.69$ &  256.32139 & 7.90  & 0.30  & \\
\enddata
\tablenotetext{a}{BJD-2,455,000}
\end{deluxetable}

\clearpage

\startlongtable
\begin{deluxetable}{lcccccc}
\tablecaption{Times of stellar eclipses \label{eclipsetimes}}
\tablewidth{0pt}
\tablehead{
\colhead{cycle \#}   & 
\colhead{primary}    &
\colhead{corrected} & 
\colhead{uncertainty}  &
\colhead{cycle \#}  &
\colhead{secondary}  &
\colhead{uncertainty} \\
\colhead{}   & 
\colhead{time\tablenotemark{a}}    &
\colhead{time\tablenotemark{a}} & 
\colhead{(min)}  &
\colhead{} &
\colhead{time\tablenotemark{a}}  &
\colhead{(min)}  
}
\startdata
  0.0  &   ... &  ...  &  ...  &    0.487385   &   $ -33.12242$  &  2.55  \\
  1.0  &   $ -29.30551$   &  $ -29.30557$  &  0.51  &    1.487385   &   $ -25.67575$  &  2.65 \\ 
  2.0  &   $ -21.85721$   &  $ -21.85712$  &  0.51  &    2.487385   &   $ -18.22955$  &  2.65 \\ 
  3.0  &   ... &  ...  &  ...  &    3.487385   &   $ -10.77950$  &  2.54  \\
  4.0  &   $  -6.96069$   &  $  -6.96058$  &  0.49  &    4.487385   &   ...  &  ...  \\
  5.0  &   ... &  ...  &  ...  &    5.487385   &   $   4.11925$  &  3.00  \\
  6.0  &   $   7.93602$   &  $   7.93622$  &  0.48  &    6.487385   &   $  11.56766$  &  2.89 \\ 
  7.0  &   ... &  ...  &  ...  &    7.487385   &   $  19.01206$  &  2.77  \\
\enddata
\tablenotetext{a}{BJD-2,455,000}
\end{deluxetable}

\clearpage

\startlongtable
\begin{deluxetable}{llcc}
\tabletypesize{\small}
\tablecaption{Priors for fitting parameters\label{priors}}
\tablewidth{0pt}
\tablehead{
\colhead{Orbit}   & 
\colhead{Parameter}    &
\colhead{Lower}  &
\colhead{Upper}  
 }
\startdata
\underline{\bf Binary and Stellar}  &   Period (days)         & 7.4480  & 7.44899  \\
                        &   $T_{\rm conj}$ (BJD 2,455,000+) & $-29.316$ & $-29.300$ \\ 
                        &   $e\cos\omega$         & $-0.1$  &  $0.0$  \\
                        &   $e\sin\omega$         & $-0.05$ &    $0.01$   \\
                        &   Inclination (deg)     & 89.00      &   90.25   \\
                        &   $M_1$ ($M_{\odot}$)   & 0.80 & 1.0   \\
                        &   $Q$   & 0.340  &   0.373   \\  
                        &   $R_1/a$   & 0.046  &       0.061  \\
\enddata
\tablenotetext{a}{{\em Kepler} Quarters Q1, Q5, ...}
\tablenotetext{b}{{\em Kepler} Quarters Q2, Q6, ...}
\tablenotetext{c}{{\em Kepler} Quarters Q3, Q7, ...}
\tablenotetext{d}{{\em Kepler} Quarters Q4, Q8, ...}
\end{deluxetable}

\clearpage

{\renewcommand{\arraystretch}{1.3}%
\begin{deluxetable}{llllll}
\tabletypesize{\scriptsize}
\rotate
\tablecaption{Fitted Parameters from the Photometric-dynamical 
model\tablenotemark{a} \label{tab:tab2}}
\tablewidth{0pt}
\tablehead{
\colhead{Parameter}   & 
\colhead{Binary Orbit}    &
\colhead{Inner Orbit} & 
\colhead{Middle Orbit}  &
\colhead{Outer Orbit} &
\colhead{} 
}
\startdata
Period (days) & $7.4483648^{+0.0000038}_{-0.0000270}$ & $49.4643^{+0.0081}_{-0.0074}$ 
              & $187.366^{+0.069}_{0.051}$ & $303.227^{+0.062}_{-0.027}$ &  \\
$T_{\rm conj}$ (BJD + 2,455,000) & $-29.305248^{+0.000031}_{-0.000030}$ & $-31.3573^{+0.0043}_{-0.0049}$ 
                             & $ 45.428^{+0.044}_{-0.047}$ & $-55.638^{+0.067}_{-0.046}$ &  \\
$e\cos\omega$ & $-0.019895^{+0.000025}_{-0.000024}$ & $ 0.0139^{+0.0020}_{-0.0025}$ 
                             & $  0.024^{+0.038}_{-0.036}$ & $  0.026^{+0.029}_{-0.034}$  &   \\
$e\sin\omega$ & $-0.0208^{+0.0019}_{-0.0020}$ & $ 0.0157^{+0.0015}_{-0.0011}$ & $ -0.004^{+0.023}_{-0.028}$ 
                           & $ -0.035^{+0.026}_{-0.025}$ &  \\
$i$ (deg) & $89.613^{+0.045}_{-0.040}$ & $89.752^{+0.063}_{-0.045}$ 
                           & $90.395^{+0.009}_{-0.012}$ & $90.1925^{+0.0055}_{-0.0042}$ &   \\
$\Omega$ (deg) & $0.0$ & $ -0.088^{+0.056}_{-0.053}$ & $ -0.87^{+0.11}_{-0.10}$ & $ -1.28^{+0.23}_{-0.25}$ &  \\
$Q_{2,1}\equiv M_2/M_1$  & $0.3569^{+0.0021}_{-0.0020}$ & ... & ...& ...  &  \\
$M_i$ ($M_{\oplus}$)  &  ... &  $2.07^{+23.70}_{-2.07}$ 
                          & $19.02^{+23.84}_{-11.67}$ & $ 3.17^{+2.18}_{-1.25}$  &  \\
\enddata
\tablenotetext{a}{$T_{\rm ref} =$ BJD 2,454,965.000}
\tablenotetext{b}{Fractional stellar radii}
\tablenotetext{c}{Ratio of primary radius to planet radius}
\tablenotetext{d}{Kipping's first quad-law triangular limb darkening coefficient}
\tablenotetext{e}{Kipping's second quad-law triangular limb darkening coefficient}
\end{deluxetable}
}

\clearpage

{\renewcommand{\arraystretch}{1.4}%
\begin{deluxetable}{llllll}
\tabletypesize{\scriptsize}
\rotate
\tablecaption{Derived Parameters from the Photometric-dynamical Model\label{tab:tab3}}
\tablewidth{0pt}
\tablehead{
\colhead{Parameter}   & 
\colhead{Binary Orbit}    &
\colhead{Inner Orbit} & 
\colhead{Middle Orbit}  &
\colhead{Outer Orbit} &
\colhead{} 
}
\startdata
$a$ (au) & $0.08145^{+0.00036}_{-0.00037}$ & $0.2877^{+0.0014}_{-0.0011}$ & $0.6992^{+0.0031}_{-0.0033}$ & $0.9638^{+0.0041}_{-0.0044}$  &   \\
$e$  & $0.0288^{+0.0015}_{-0.0013}$ & $0.0210^{+0.0025}_{-0.0022}$ & $0.024^{+0.025}_{-0.017}$ & $0.044^{+0.029}_{-0.019}$ & \\
$\omega$ (deg)  & $ 226.3^{+2.8}_{-2.6}$ & $ 48.6^{+3.0}_{-2.3}$ & $352^{+96}_{-166}$ & $306^{+40}_{-67}$ &  \\
$I_{1,i}$ (deg)\tablenotemark{a}  & ...   & $0.166^{+0.037}_{-0.041}$ & $1.165^{+0.101}_{-0.093}$ & $1.38^{+0.20}_{-0.24}$ &  \\
$I_{2,i}$ (deg)\tablenotemark{b}  &  ...   & ... & $1.006^{+0.089}_{-0.050}$ & $1.247^{+0.024}_{-0.018}$ &  \\
$I_{3,i}$ (deg)\tablenotemark{c}  &  ...   & ...  & ... & $0.442^{+0.028}_{-0.102}$ &  \\ 
          &          &          &              &               &           \\
Parameter & Primary & Secondary & Inner Planet & Middle Planet & Outer Planet \\
\enddata
\tablenotetext{a}{Mutual inclination between the orbital planes of the binary and the planets, given by 
$\cos I  =  \sin i_p \sin i_b \cos \Delta \Omega + \cos i_p \cos i_b$}
\tablenotetext{b}{Mutual inclination between the orbital planes of the inner planet and the other planets}
\tablenotetext{c}{Mutual inclination between the orbital planes of the middle planet and the outer planet}
\end{deluxetable}}

\clearpage

\begin{deluxetable}{lrrrrr}
\tabletypesize{\scriptsize}
\rotate
\tablecaption{Initial Cartesian Barycentric Conditions$^a$ for Kepler-47 \label{cartcoords}}
\tablewidth{0pt}
\tablehead{
\colhead{Parameter}   & 
\colhead{Primary}    &
\colhead{Secondary}  &
\colhead{Inner Planet} & 
\colhead{Middle Planet}  &
\colhead{Outer Planet} 
 }
\startdata
Mass ($M_{\odot}$) &  9.573912601143495E$-$01  &   3.416626576742950E$-$01  &    6.208164703811071E$-$06  &  5.711706000000001E$-$05  &  9.530601983442541E$-$06 \\ 
  $x$ (au)  &    2.092317942847221E$-$02  & $-5$.866162141774457E$-$02  &    1.302102743834203E$-$01  &    2.398289447308760E$-$01  & $-3$.872087634688049E$-$01  \\
  $y$ (au)  & $-1$.364205200783173E$-$05  &    3.791628623967441E$-$05  &    8.812713793817102E$-$04  &    8.391663271667140E$-$04  &    5.540095504781927E$-$03  \\ 
  $z$ (au)  & $-1$.988193746560007E$-$03  &    5.649120240556342E$-$03  &    2.500211844794519E$-$01  & $-6$.478290001132635E$-$01  &    9.270974333301998E$-$01  \\ 
$v_x$ (au)  &    1.350420386068420E$-$03  & $-3$.786681230811973E$-$03  & $-3$.299902928667536E$-$02  &    2.207346827176963E$-$02  & $-1$.773342551338714E$-$02  \\ 
$v_y$ (au)  &    1.239137665559171E$-$04  & $-3$.471736278879929E$-$04  &    1.256761297656883E$-$04  & $-3$.936013341278077E$-$04  &    4.205823582940745E$-$04  \\ 
$v_z$ (au)  &    1.835286133723581E$-$02  & $-5$.142910738078619E$-$02  &    1.739447523236208E$-$02  &    8.701783937203954E$-$03  & $-7$.191666020340975E$-$03  \\ 
\enddata
\tablenotetext{a}{$T_{\rm ref} =$ BJD 2,454,965.000}
\end{deluxetable}

\clearpage

\begin{deluxetable}{lrrrr}
\tablecaption{Initial Orbital Elements$^a$ for Kepler-47 \label{elements}}
\tablewidth{0pt}
\tablehead{
\colhead{parameter}   & 
\colhead{binary}    &
\colhead{inner}  &
\colhead{middle} & 
\colhead{outer}  \\
\colhead{ } &
\colhead{orbit} &
\colhead{orbit} &
\colhead{orbit} &
\colhead{orbit} 
 }
\startdata
$a$ (AU)       & 8.144243202775878E$-$02  & 2.877434527504005E$-$01  & 6.992154891162726E$-$01  & 9.638218771090775E$-$01  \cr
$e$            & 2.877193854136767E$-$02  & 2.095514505946130E$-$02  & 2.416231575331903E$-$02  & 4.371716413425727E$-$02  \cr
$i$ (deg)      & 8.961321083430335E$+$01  & 8.975235828571519E$+$01  & 9.039477337702283E$+$01  & 9.019249877520613E$+$01  \cr
$\omega$ (deg) & 2.262537367634179E$+$02  & 4.859496576626187E$+$01  & 3.517243342867096E$+$02  & 3.059357996708584E$+$02  \cr
$\Omega$ (deg) & $~0.$000000000000000E$+$00 & $-8.$764673739309498E$-$02 & $-8.$659498733670760E$-$01 & $-1.$280580223423379E$+$00 \cr     
$f$ (deg)      & 3.082645652441405E$+$02  & 1.389173589110049E$+$01  & 2.985868391504580E$+$02  & 1.667337092860433E$+$02  \cr
$M$ (deg)      & 3.108188471954695E$+$02  & 1.332389759998393E$+$01  & 3.009969996738955E$+$02  & 1.655463689900479E$+$02  \cr
$\lambda$ (deg)& 1.745183020075583E$+$02  & 6.239905491996927E$+$01  & 2.894452235638005E$+$02  & 1.113889287334783E$+$02  \cr
\enddata
\tablenotetext{a}{$T_{\rm ref} =$ BJD 2,454,965.000.
The parameters are the semimajor axis, the eccentricity, the inclination,
the argument of periastron, the nodal angle, the true anomaly,
the mean anomaly, and the mean longitude.}
\end{deluxetable}

\clearpage

\begin{deluxetable}{llcc}
\tabletypesize{\small}
\tablecaption{Parameter Ranges Displayed in Figures 
\protect{\ref{quad1}-\ref{quad3}}\label{tab_appendix}}
\tablewidth{0pt}
\tablehead{
\colhead{Orbit}   & 
\colhead{Parameter}    &
\colhead{Lower}  &
\colhead{Upper}  
 }
\startdata
\underline{Binary}  &   Period (days)         & 7.44820  & 7.44838  \\
                        &   $T_{\rm conj}$ (BJD 2,455,000+) & $-29.30541$ & $-29.30515$ \\ 
                        &   $e\cos\omega$         & $-0.02000$  &  $-0.01979$  \\
                        &   $e\sin\omega$         & $-0.028$ &    $-0.012$   \\
                        &   Inclination (deg)     & 89.5      &   89.8   \\
                        &   $M_1$ ($M_{\odot}$)   & 0.8950 & 0.9905   \\
                        &   $M_2$ ($M_{\odot}$)   & 0.326  &   0.352   \\  
\underline{Inner}  &       Period (days)         & 49.440  &  49.495   \\
\enddata
\end{deluxetable}

\clearpage

\startlongtable
\begin{deluxetable}{lcccccc}
\tabletypesize{\small}
\tablecaption{Times, durations, and impact parameters 
of future planet transits\label{tabfuture}}
\tablewidth{0pt}
\tablehead{
\colhead{ BJD--2,455,000}  & 
\colhead{Year}  &
\colhead{Month}  &
\colhead{Day}  &
\colhead{UT}  &
\colhead{Impact}  &
\colhead{Duration} \\
\colhead{}  & 
\colhead{}  &
\colhead{}  &
\colhead{}  &
\colhead{}  &
\colhead{Parameter}  &
\colhead{(hr)}  
 }
\startdata
{\underline{Inner Planet}}      &      &     &    &             &                  &    \\
$3522.04361\pm 0.07701$ & 2019 & Feb& 7  & 13:02:47.6 & $0.258\pm 0.033$ & $ 4.86\pm 0.13$ \\ 
$3569.12508\pm 0.06896$ & 2019 & Mar& 26  & 15:00:06.6 & $0.269\pm 0.034$ & $ 4.12\pm 0.07$ \\ 
$3618.07766\pm 0.10808$ & 2019 & May& 14  & 13:51:49.8 & $0.245\pm 0.027$ & $ 6.66\pm 0.30$ \\ 
$3665.39077\pm 0.06593$ & 2019 & Jun& 30  & 21:22:42.5 & $0.286\pm 0.030$ & $ 3.81\pm 0.04$ \\ 
$3713.68886\pm 0.18417$ & 2019 & Aug& 18  & 04:31:57.6 & $0.233\pm 0.023$ & $10.88\pm 0.33$ \\ 
$3761.67854\pm 0.06705$ & 2019 & Oct& 5  & 04:17:05.5 & $0.303\pm 0.025$ & $ 3.77\pm 0.03$ \\ 
$3809.11491\pm 0.14101$ & 2019 & Nov& 21  & 14:45:28.6 & $0.252\pm 0.023$ & $ 7.74\pm 0.54$ \\
\enddata
\end{deluxetable}

\clearpage

\startlongtable
{\renewcommand{\arraystretch}{1.3}%
\begin{deluxetable}{lllll}
\tablecaption{Planetary mass and radius data\label{massradiusdata}}
\tablewidth{0pt}
\tablehead{
\colhead{Name}   & 
\colhead{Mass ($M_{\oplus}$)}    &
\colhead{Radius ($R_{\oplus}$)} & 
\colhead{Insolation ($S_{\oplus}$)}  &
\colhead{References}}
\startdata
        55Cnc~e       &  $ 7.81^{+0.58}_{-0.53}$   &  $ 2.17^{+0.10}_{-0.10}$   &  $2442.2^{+ 54.2}_{- 54.2}$   &     13, 39           \\ 
       GJ1214~b       &  $ 6.55^{+0.98}_{-0.98}$   &  $ 2.68^{+0.13}_{-0.13}$   &  $  16.3^{+  2.4}_{-  2.4}$   &        7             \\ 
       GJ3470~b       &  $14.00^{+1.80}_{-1.80}$   &  $ 4.20^{+0.60}_{-0.60}$   &  $  29.7^{+  6.8}_{-  6.8}$   &        4             \\ 
       GJ436~b        &  $24.79^{+2.22}_{-2.54}$   &  $ 4.05^{+0.17}_{-0.17}$   &  $  28.4^{+  3.7}_{-  2.0}$   &       36             \\ 
       GJ1132~b       &  $ 1.63^{+0.54}_{-0.45}$   &  $ 1.44^{+0.21}_{-0.21}$   &  $  10.5^{+  2.2}_{-  2.2}$   &        40            \\ 
        GJ9827~K2-133  &  $ 3.74^{+0.50}_{-0.48}$   &  $ 1.62^{+0.17}_{-0.16}$   &  $ 311.6^{+ 50.6}_{- 50.6}$   &    41                \\ 
        GJ9827~K2-13  &  $ 1.47^{+0.59}_{-0.58}$   &  $ 1.27^{+0.13}_{-0.13}$   &  $  71.5^{+ 11.6}_{- 11.6}$   &     41               \\ 
        GJ9827~K2-13  &  $ 2.38^{+0.71}_{-0.69}$   &  $ 2.09^{+0.22}_{-0.21}$   &  $  35.2^{+  5.7}_{-  5.7}$   &       41             \\
\enddata
\tablecomments{References.
(1) \citet{Almenara_2015}; 
(2) \citet{Batalha_2011}; 
(3) \citet{Becker_2016}; 
(4) \citet{Bonfils_2012}; 
(5) \citet{Borucki_2010a}; 
(6) \citet{Carter_2012}; 
(7) \citet{Charbonneau_2009}; 
(8) \citet{Cochran_2011}; 
(9) \citet{Dressing_2015}; 
(10) \citet{Dumusque_2014};
(11) \citet{Fressin_2012}, 
(12) \citet{Gautier_2012}; 
(13) \citet{Gillon_2012}; 
(14) \citet{Hadden_2016}; 
(15) \citet{Hartman_2011}; 
(16) \citet{Haywood_2014}; 
(17) \citet{Howard_2013}; 
(18) \citet{Jontof_2014}; 
(19) \citet{Jontof_2015}; 
(20) \citet{Jontof_2016}; 
(21) \citet{Kipping_2014}; 
(22) \citet{Lissauer_2012}; 
(23) \citet{Lissauer_2013};
(24) \citet{MacDonald_2016},
(25) \citet{Marcy_2014}; 
(26) \citet{Masuda_2014};
(27) \citet{Mills_2016}
(28) \citet{Motalebi_2015}; 
(29) \citet{Ofir_2014}; 
(30) \citet{Pepe_2013}; 
(31) \citet{Sanchis_2012}; 
(32) \citet{Schmitt_2014}; 
(33) \citet{Silva_Aguirre2015}; 
(34) \citet{Vanderburg_2015}; 
(35) \citet{vanGrootel_2014}; 
(36) \citet{vonBraun_2012};
(37) \citet{Weiss_2013};
(38) \citet{Weiss_2016};  
(39) \citet{Winn_2011}; 
(40) \citet{Southworth_2017};
(41) \citet{Prieto_2018};
(42) \citet{Southworth_2011};
(43) \citet{Hadden_2017};
(44) \citet{Hadden_2016};
(45) \citet{Mills_2017};
(46) \citet{Vanderburg_2015};
(47) \citet{Damasso_2018};
(48) \citet{Cloutier_2017};
(49) \citet{Narita_2015};
(50) \citet{Sinukoff_2016};
(51) \citet{Sinukoff_2017};
(52) \citet{Osborn_2017};
(53) \citet{Dai_2017};
(54) \citet{Fridlund_2017};
(55) \citet{Dittmann_2017};
(56) \citet{Grimm_2018};
(57) \citet{Santerne_2018}}
\end{deluxetable}}

\clearpage

\appendix

\section{Planetary mass and radius data}\label{massraddiscussion}

Table \ref{massradiusdata} lists
the planetary masses, radii, and insolation values used in 
Figure \ref{massrad}. 
Note that the \citet{Marcy_2014} sample
includes many nondetections and upper limits 
that we do not plot. 

We estimate the insolation each planet received from its host star
using the cited $T_{\rm eff}$, 
$R_{*}$ and $M_{*}$ of each host such that: 
$$
{S\over S_{\oplus}} = \left({T_{\rm eff}\over T_{\odot}}\right)^4 
\left({1\,{\rm AU}\over a}\right)^2
\left({R_{\odot}\over R_{*}}\right)^2
$$
where the semimajor axis
$a$ is calculated from the measured mass and orbital period
(eccentricities are neglected). 
Uncertainties are added in quadrature.

\listofchanges


\begin{thebibliography}{}

\bibitem[Agol et al.(2005)]{Agol_2005}
Agol, E., Steffen, J., Sari, R., \& Clarkson, W. 2005,
\mnras, 359, 567

\bibitem[Allard et al.(1997)]{Allard_1997}
Allard, F., Hauschildt, P. H., Alexander, D. R., Starrfield, S. 1997, ARA\&A,
35, 137

\bibitem[Almenara et al.(2015)]{Almenara_2015}
Almenara, J. M., Astudillo-Defru, N., Bonfils, X., et al. 2015, \aap, 581, L7

\bibitem[Bailer-Jones et al.(2018)]{Bailer-Jones_2018}
Bailer-Jones, C. A. L., Rybizki, J., Fouesneau, M., Mantelet, G.,
\& Andrae, R. 2018, \aj, 156, 58

\bibitem[Bass et al.(2012)]{Bass_2012}
Bass, G. P., Orosz, J. A., Welsh, W. F., Windmiller, G., Gregg, T. A.,  
Fetherolf, T.,
Wade, W. A., \& Quinn, S. N. 2012, \apj, 761, 157

\bibitem[Batalha et al.(2011)]{Batalha_2011}
Batalha, N. M., Borucki, W. J., Bryson, S. T., et al. 2011, \apj, 729, 27

\bibitem[Becker et al.(2016)]{Becker_2016}
Becker, J. C., Vanderburg, A., Adams, F. C., et al. 2015, \apjl, 812, L18

\bibitem[Bevington(1969)]{Bevington_1969}
Bevington, P. R. 1969, 
Data Reduction and Error Analysis for the Physical
Sciences (New York:  McGraw-Hill)

\bibitem[Bonfils et al.(2012)]{Bonfils_2012}
Bonfils, X., Gillon, M., Udry, S., et al. 2012,  \aap, 546, A27

\bibitem[Borucki et al.(2010a)]{Borucki_2010a}
Borucki, W. J., Koch, D. G., Basri, G., et al. 2010, Sci, 327, 977 

\bibitem[Borucki et al.(2010b)]{Borucki_2010b}
Borucki, W. J., Koch, D. G., Brown, T. M., et al. 2010, \apjl, 713, L126

\bibitem[Carter et al.(2011)]{Carter_2011}
Carter, J. A., Fabrycky, D. C., Ragozzine, D. 2011, Sci, 331, 562 

\bibitem[Carter et al.(2012)]{Carter_2012}
Carter, J. A., Agol, E., Chaplin, W. J., et al. 2012, Sci, 337, 556 

\bibitem[Chambers, Wetherill \& Boss(1996)]{Chambers_1996}
Chambers, J. E., Wetherill, G. W., \& Boss, A. P. 1996, Icar, 119, 261

\bibitem[Chambers et al.(2002)]{Chambers_2002}
Chambers, J. E., Quintana, E. V., Duncan, M. J., \& Lissauer, J. J. 
2002, AJ, 123, 2884

\bibitem[Charbonneau(1995)]{Charbonneau_1995}
Charbonneau, P. 1995, \apjs, 101, 309

\bibitem[Charbonneau et al.(2009)]{Charbonneau_2009}
Charbonneau, D., Berta, Z. K., Irwin, J., et al. 2009, Natur, 462, 891 

\bibitem[Cincotta, Giordano \& Sim{\'o}(2003)]{Cincotta_2003}
Cincotta, P. M., Giordano, C. M., \& Sim{\'o}, C. 2003,  PhyD, 
182, 151

\bibitem[Clemence(1965)]{Clemence_1965}
Clemence, G. M. 1965, ARA\&A, 3, 93

\bibitem[Cloutier et al.(2017)]{Cloutier_2017}
Cloutier, R., Astudillo-Defru, N., Doyon, R., 
Bonfils, X., Almenara, J.-M., et al.\ 2017, \aap, 608, A35

\bibitem[Cochran et al.(2011)]{Cochran_2011}
Cochran, W. D., Fabrycky, D. C., Torres, G., et al. 2011, \apjs, 197, 7

\bibitem[Dai et al.(2017)]{Dai_2017}
Dai, F., Winn, J. N., Gandolfi, D.,  Wang, S. X., 
Teske, J. K., et al.\ 2017, \aj, 154, 226 

\bibitem[Damasso et al.(2018)]{Damasso_2018}
Damasso, M., Bonomo, A. S., Astudillo-Defru, N.,
Bonfils, X., Malavolta, L., et al.\ 2018, \aap, 615, A69

\bibitem[de Pater \& Lissauer(2015)]{dePater_2015}
de Pater, I., \& Lissauer, J. J. 2015, Planetary Sciences,
(Cambridge, UK: Cambridge Univ.\ Press)

\bibitem[Dittmann et al.(2017)]{Dittmann_2017}
Dittmann, J. A., Irwin, J. M., Charbonneau, D.,
Bonfils, X., Astudillo-Defru, N., et al.\ 2017, Natur, 544, 333 

\bibitem[Dotter et al.(2008)]{Dotter_2008} 
Dotter, A., Chaboyer, B., Jevremovi\'c, D., Kostov, V., Baron, E., 
\& Ferguson, J. W. 
2008, \apjs, 178, 89

\bibitem[Doyle et al.(2011)]{Doyle_2011}
Doyle, L. R., Carter, J. A., Fabrycky, D. C., Slawson, R. W., Howell, 
S. B., et al. 
2011, Sci, 333, 1602

\bibitem[Dressing et al.(2015)]{Dressing_2015}
Dressing, C. D., Charbonneau, D., Dumusque, X., et al. 2015, \apj, 800, 135

\bibitem[Dumusque et al.(2014)]{Dumusque_2014}
Dumusque, X., Bonomo, A. S., Haywood, R. D., et al. 2014, \apj, 789, 154

\bibitem[Dunhill et al.(2013)]{Dunhill_2013}
Dunhill, A. C., \& Alexander, R. D. 2013, \mnras, 435, 2328

\bibitem[Dvorak(1982)]{Dvorak_1982}
Dvorak, R. 1982, OAWMN, 191, 423

\bibitem[Eggleton, Kiseleva \& Hut(1998)]{Eggleton_1998}
Eggleton, P. P., Kiseleva, L. G., \& Hut, P. 1998, \apj, 499, 853

\bibitem[Flower(1996)]{Flower:1996} 
Flower, P.\ J.\ 1996, \apj, 469, 355 

\bibitem[Fressin et al.(2012)]{Fressin_2012}
Fressin, F., Torres, G., Rowe, J. F., et al. 2012, Natur, 482, 195

\bibitem[Fridlund et al.(2017)]{Fridlund_2017}
Fridlund, M., Gaidos, E., Barrag{\'a}n, O., Persson, C. M., 
Gandolfi, D., et al.\ 2017, \aap, 604, A16

\bibitem[{\em Gaia} Collaboration et al.(2016)]{GaiaCollaboration:2016} 
Gaia Collaboration, Prusti, T.,  de Bruijne, J.H.J.,
Brown, A., Vallenari, A., Babusiaux, C., et al.\ 2016, \aap, 616, A1 

\bibitem[Gaia Collaboration et al.(2018)]{GaiaCollaboration:2018} 
Gaia Collaboration, Brown, A.\ G.\ A., Vallenari, A., 
Prusti, T., de Bruijne, J. H. J., Babusiaux, C., 
et al.\ 2018, \aap, 616, A1 

\bibitem[Gautier III(2012)]{Gautier_2012}
Gautier III, T. N., Charbonneau, D., Rowe, J. F., et al. 2012, \apj, 749, 15

\bibitem[Gillon et al.(2012)]{Gillon_2012}
Gillon, M., Demory, B. O., Benneke, B., et al. 2012,  \aap, 539, A28

\bibitem[Gim\'enez(2006a)]{Gimenez_2006a}
Gim\'enez, A. 2006a, \aap, 450, 1231

\bibitem[Gim\'enez(2006b)]{Gimenez_2006b}
Gim\'enez, A., 2006b, \apj, 650, 408

\bibitem[Gladman(1993)]{Gladman_1993}
Gladman, B. 1993, Icar, 106, 247 

\bibitem[Go{\'z}dziewski et al.(2001)]{Gozdziewski_2001}
Go{\'z}dziewski, K., Bois, E., Maciejewski, A. J., \& Kiseleva-Eggleton, 
L. 2001,
\aap, 378, 569

\bibitem[Green et al.(2015)]{Green:2015} 
Green, G.~M., Schlafly, E.~F., Finkbeiner, D.~P., et al.\ 2015, \apj, 810, 25

\bibitem[Grimm et al.(2018)]{Grimm_2018}
Grimm, S. L., Demory, B.-O., Gillon, M., Dorn, C., 
Agol, E., et al.\ 2018, \aap, 613, A68

\bibitem[Hadden \& Lithwick(2016)]{Hadden_2016}
Hadden, S., \& Lithwick, Y. 2016, \apj, 828, 44

\bibitem[Hadden \& Lithwick(2017)]{Hadden_2017}
Hadden, S., \& Lithwick, Y. 2017, AJ, 154, 5

\bibitem[Haghighipour \& Kaltenegger(2013)]{Haghighipour_2013}
Haghighipour, N., \& Kaltenegger, L. 2013, \apj, 777, 166

\bibitem[Hairer, Lubich \& Wanner(2006)]{Hairer_2006}
Hairer, E., Lubich, C., \& Wanner, G. 2006, Geometric Numerical Integration. 
Structure-Preserving Algorithms for Ordinary Differential Equations, 2nd Ed.
Springer Series in Computational Mathematics, 31 (Berlin: Springer-Verlag)

\bibitem[Hartman et al.(2011)]{Hartman_2011}
Hartman, J. D., Bakos,  G. \'{A}, Kipping, D. M., et al. 2011, \apj, 728, 138

\bibitem[Hauschildt, Allard, \& Baron(1999)]{Hauschildt_1999}
Hauschildt, P. H., Allard, F., \& Baron, E. 1999,
\apj, 512, 377

\bibitem[Henden et al.(2015)]{Henden:2015} 
Henden, A.\ A., Levine, S., Terrell, D., \& Welch, D.\ L.\ 2015, 
American Astronomical Society Meeting Abstracts \#225, 225, 336.16 

\bibitem[Heywood et al.(2014)]{Haywood_2014}
Haywood, R. D., Cameron, A. C., Queloz, D.,  Barros, S. C. C., Deleuil, M.
2014, \mnras, 443, 2517

\bibitem[Hinse et al.(2015)]{Hinse_2015}
Hinse, T. C., Haghighipour, N., Kostov, V. B., \& Go\'{z}dziewski, K. 2015,
\apj, 799, 88

\bibitem[Holczer et al.(2015)]{Holczer_2015}
Holczer, T., Shporer, A., Mazeh, T., et al. 2015, \apj, 807, 170

\bibitem[Holman \& Wiegert(1999)]{Holman_1999}
Holman, M. J., \&  Wiegert, P. A. 1999, AJ, 117, 621

\bibitem[Howard et al.(2013)]{Howard_2013}
Howard, A. W., Sanchis-Ojeda, R., Marcy, G. W., et al. 2013, Natur, 
503, 381

\bibitem[Jontof-Hutter et al.(2014)]{Jontof_2014}
Jontof-Hutter, D., Lissauer, J. J., Rowe, J. F., \& Fabrycky, D. C. 
2014, \apj, 785, 15

\bibitem[Jontof-Hutter et al.(2015)]{Jontof_2015}
Jontof-Hutter, D., Rowe, J. F., Lissauer, J. J., Fabrycky, D. C., 
Ford, E. B. 2015, 
Natur, 522, 321

\bibitem[Jontof-Hutter et al.(2016)]{Jontof_2016}
Jontof-Hutter, D., Ford, E. B., Rowe, J. F., 
Lissauer, J. J.,  Fabrycky, D. C., 
Van Laerhoven, C.,  
Agol, E.,  Deck, K. M., Holczer, T., \&  Mazeh, T. 2016,
\apj, 820, 39

\bibitem[Kipping(2013)]{Kipping_2013}
Kipping, D. M. 2013, \mnras, 435, 2152

\bibitem[Kipping et al.(2014)]{Kipping_2014}
Kipping, D. M., Nesvorn{\'y}, D., Buchhave, L. A., et al. 2014, \apj, 784, 28

\bibitem[Kirk et al.(2016)]{Kirk_2016}
Kirk, B., Conroy, L., Pr{\v{s}}a, A., Abdul-Mashih, M.,
Kockoska, A., et al.\ 2016, AJ, 151, 68

\bibitem[Kley \& Haghighipour(2014)]{Kley_2014}
Kley, W. \& Haghighipour, N. 2014, \aap, 564, A72 

\bibitem[Kley \& Haghighipour(2015)]{Kley_2015}
Kley, W., \& Haghighipour, N. 2015, \aap, 581, A20

\bibitem[Koch et al.(2010)]{Koch_2010}
Koch, D. G., Borucki, W. J., Basri, G., et al.
2010, \apjl, 713, L79

\bibitem[Kopparapu et al.(2013)]{Kopparapu_2013}
Kopparapu, R. K., Ramirez, R., Kasting, J. F., et al. 2013, \apj, 765, 131


\bibitem[Kopparapu et al.(2014)]{Kopparapu_2014}
Kopparapu, R. K., Ramirez, R. M., SchottelKotte, J. 2014,  \apjl, 787, L29

\bibitem[Kostov et al.(2013)]{Kostov_2013}
Kostov, V. B., McCullough, P. R., Hinse, T. C., Tsvetanov, Z, I., 
H\'ebrard, G., 
D\'iaz, R. F., Deleuil, D., Valenti, J. A., 2013, \apj, 770, 52 

\bibitem[Kostov et al.(2014)]{Kostov_2014}
Kostov, V. B., McCollough, P. R., Carter, J. A., Deleuil, 
M., D\'iaz, R. F.,  et al.
2014, \apj, 784, 14 

\bibitem[Kostov et al.(2016)]{Kostov_2015}
Kostov, V. B., Orosz, J. A., Welsh, W. F., et al. 2016, 
\apj, 827, 86

\bibitem[Kratter \& Shannon(2014)]{Kratter_2014}
Kratter, K. M., \& Shannon, A. 2014, \mnras, 437, 3727

\bibitem[Lee et al.(2014)]{Lee_2014}
Lee, J. W., Hinse, T. C., Youn, J.-H.,
\& Han, W. 2014, \mnras, 445, 2331

\bibitem[Lines et al.(2015a)]{Lines_2014}
Lines, S., Leinhardt, Z. M., Paardekooper, S-J., Baruteau, C., \& 
Thebault, P. 2015,
\apjl, 782, L11 

\bibitem[Lines et al.(2015b)]{Lines_2015}
Lines, S., Leinhardt, Z. M., Baruteau, C., Paardekooper, S.-J.,
\& Carter, P. J. 2015, \aap, 582, A5

\bibitem[Lines et al.(2016)]{Lines_2016}
Lines, S., Leinhardt, Z. M., Baruteau, C., 
Paardekooper, S.-J., \& Carter, P. J. 2016, \aap, 590, A62

\bibitem[Lissauer et al.(2012)]{Lissauer_2012}
Lissauer, J. J., Marcy, G. W., Rowe, J. F., et al. 2012, \apj, 750, 112

\bibitem[Lissauer et al.(2013)]{Lissauer_2013}
Lissauer, J. J., Jontof-Hutter, D., Rowe, J. F., et al. 2013, \apj, 770, 131

\bibitem[Lopez \& Fortney(2014)]{Lopez_2014}
Lopez, E. D., \& Fortney, J. J. 2014, \apj, 792, 1

\bibitem[MacDonald et al.(2016)]{MacDonald_2016}
MacDonald, M. G., Ragozzine, D., Fabrycky, D. C., 
Ford, E. B, Holman, M. J., Isaacson, H. T., 
Lissauer, J. J., Lopez, E. D., Mazeh, T., Rogers, L.,  
Rowe, J. F., Steffen, J. H., \& Torres, G. 2016, AJ,
152,105

\bibitem[Mandel \& Agol(2002)]{Mandel_2002}
Mandel, K., \& Agol, E. (2002) \apjl, 580, L171

\bibitem[Marcy et al.(2014)]{Marcy_2014}
Marcy, G. W., Isaacson, H., Howard, A. W., et al. 2014, \apjs, 210, 20

\bibitem[Mardling \& Lin(2002)]{Mardling_2002}
Mardling, R. A., \& Lin, D. N. C. 2002, \apj, 573, 829

\bibitem[Marsh(2018)]{Marsh_2017}
Marsh, T. R. 2018, in Handbook of Exoplanets, eds.\
Hans J. Deeg \& Juan Antonio Belmonte (Springer: Cham)

\bibitem[Martin, Armitage \& Alexander(2013)]{Martin_2013}
Martin, R. G., Armitage, P. J., \& Alexander, R. D. 2013, \apj, 773, 74 

\bibitem[Marzari et al.(2013)]{Marzari_2013}
Marzari, F., Thebault, P., Scholl, H., Picogna, G.,
\& Baruteau, C. 2013,  \aap, 553, A71


\bibitem[Masuda(2014)]{Masuda_2014}
Masuda, K. 2014, \apj, 783, 53 


\bibitem[Mazeh, Holczer \& Shporer(2015)]{Mazeh_2015}
Mazeh, T., Holczer, T., \& Shporer, A. 2015, \apj, 800, 42

\bibitem[Meschiari(2012)]{Meschiari_2012}
Meschiari, S. 2012, \apj, 752, 71 

\bibitem[Mills \& Fabrycky(2017)]{Mills_2017}
Mills, S. M., \& Fabrycky, D. C. 2017, \apjl, 838, L11

\bibitem[Mills et al.(2016)]{Mills_2016}
Mills, S. M., Fabrycky, D. C., Migaszewski, C.,  
Ford, E. B., Petigura, E., \&  Isaacson, H. 2016,
Natur, 533, 509

\bibitem[Motalebi et al.(2015)]{Motalebi_2015}
Motalebi, F., Udry, S., Gillon, M., et al. 2015, \aap, 584, A72

\bibitem[Murray \& Dermott(1999)]{Murray_1999}
Murray, C. D., \& Dermott, S. F. 1999, Solar System Dynamics
(Cambridge, UK: Cambridge Univ.\ Press)

\bibitem[Narita et al.(2015)]{Narita_2015}
Narita , N.,  Hirano, T., Fukui, A., Hori, Y., 
Sanchis-Ojeda, R., et al.\ 2015, \apj, 815, 47

\bibitem[Nelson et al.(2014)]{Nelson_2014}
Nelson, B., Ford, E. B., \& Payne, M. J. 2014, \apjs, 210, 11

\bibitem[Ofir et al.(2014)]{Ofir_2014}
Ofir, A., Dreizler, S., Zechmeister, M., \& Husser, T. O. 2014, 
\aap, 561, A103

\bibitem[Orosz \& Hauschildt(2000)]{Orosz_2000}
Orosz, J. A., \& Hauschildt, P. H. 2000, \aap, 364, 265

\bibitem[Orosz et al.(2012a)]{Orosz_2012a}
Orosz, J. A., Welsh, W. F., Carter, J. A., Brugamyer, E., 
Buchhave, L. A., et al.\
2012a,
\apj, 758, 87

\bibitem[Orosz et al.(2012b)]{Orosz_2012b}
Orosz, J. A., Welsh, W. F., Carter, J. A., Fabrycky, D. C., 
Cochran, W.,  et al.\ 
2012b,
Sci, 337, 1511 

\bibitem[Osborn et al.(2017)]{Osborn_2017}
Osborn, H. P., Santerne, A.,  Barros, S. C. C., 
Santos, N. C., Dumusque, X., et al.\ 2017, \aap, 604 A19

\bibitem[Paardekooper et al.(2012)]{Paardekooper_2012}
Paardekooper, S.-J., Leinhardt, Z. M., Th\'ebault, P., \&  Baruteau, C. 2012,  
\apjl, 754, L16

\bibitem[Pelupessy \& Portegies Zwart(2013)]{Pelupessy_2013}
Pelupessy, F. I., \& Portegies Zwart, S. 2013, \mnras, 429, 895 

\bibitem[Pepe et al.(2013)]{Pepe_2013}
Pepe, F., Cameron, A. C., Latham, D. W., et al. 2013, Natur, 503, 377

\bibitem[Pierens \& Nelson(2008)]{Pierens_2008}
Pierens, A., \& Nelson, R. P. 2008, \aap, 483, 633

\bibitem[Pierens \& Nelson(2013)]{Pierens_2013}
Pierens, A., \& Nelson, R. P. 2013, a\&A, 556, A134 


\bibitem[Prieto-Arranz et al.(2018)]{Prieto_2018}
Prieto-Arranz, J., Palle, E., Gandolfi, D., Barrag{\'a}n, O., 
Guenther, E.~W., et al.\ 2018, \aap, 618, A116

\bibitem[Pr{\v s}a et al.(2011)]{Prsa_2011}
Pr{\v s}a, A., Batalha, N. M., Slawson, R. W., et al. 2011, AJ, 141, 83

\bibitem[Qian et al.(2012a)]{Qian_2012a}
Qian, S.-B., Liu, L., Zhu, L.-Y., Dai, Z.-B.,
Fernandez Lajus, E., \& Baume, G.
L. 2012a, \mnras, 422, L24

\bibitem[Qian et al.(2012b)]{Qian_2012b}
Qian, S.-B., Zhu, L.-Y., Dai, Z.-B., Fernandez Lajus, E.,
Xian, F.-Y., \& He, J.-J. 2012b,
\apjl, 745, L23

\bibitem[Qian et al.(2013)]{Qian_2013}
Qian, S.-B., Shi, G., Zola, S., Koziel-Wierzbowska,
D., Winiarski, M., et al.\ 2013, \mnras, 436, 1408

\bibitem[Quarles et al.(2018)]{Quarles_2018}
Quarles, B., Satyal, S.,  Kostov, V.,
Kaib, N., \&  Haghighipour, N. 2018, 
\apj, 856, 150

\bibitem[Quarles \& Lissauer(2018)]{Quarles_2018a}
Quarles, B., \& Lissauer, J. J. 2018, \aj, 155, 130

\bibitem[Rafikov(2013)]{Rafikov_2013}
Rafikov, R. R. 2013, \apjl, 764, L16

\bibitem[Raghavan et al.(2010)]{Raghavan2010}
Raghavan, D., McAlister, H. A., 
Henry, T. J., Latham, D. W., Marcy, G. W., Mason, B. D., 
Gies, D. R., White, R. J., ten Brummelaar, T. A. 2010, \apjs, 190, 1 

\bibitem[Rein \& Liu(2012)]{Rein_2012}
Rein, H., \& Liu, S. 2012, \aap, 537, A128

\bibitem[Sanchis-Ojeda et al.(2012)]{Sanchis_2012}
Sanchis-Ojeda, R., Fabrycky, D. C., Winn, J. N., et al.\ 
2012,  Natur, 487, 449

\bibitem[Santerne et al.(2018)]{Santerne_2018}
Santerne, A., Brugger, B., Armstrong, D. J., Adibekyan, V., Lillo-Box, J.,
et al. 2018, NatAs, 2, 393 

\bibitem[Schmitt et al.(2014)]{Schmitt_2014}
Schmitt, J. R., Agol, E., Deck, K. M., et al. 2014, \apj, 795, 167

\bibitem[Schneider(1994)]{Schneider_1994}
Schneider, J. 1994, P\&SS, 42, 539

\bibitem[Schwarz et al.(2011)]{Schwarz_2011}
Schwarz, R.,  Haghighipour, N.,
Eggl, S.,  Pilat-Lohinger, E. \& Funk, B.,
\mnras, 414, 2763

\bibitem[Schwamb et al.(2013)]{Schwamb_2013}
Schwamb, M. E., Orosz, J. A., 
Carter, J. A., Welsh, W., F., Fischer, D. A., et al.
2013, \apj, 768, 127

\bibitem[Silva Aguirre et al.(2015)]{Silva_Aguirre2015}
Silva Aguirre, V., Davies, G. R., Basu, S., et al. 2015, \mnras, 452, 2127

\bibitem[Sinukoff et al.(2016)]{Sinukoff_2016}
Sinukoff, E., Howard, A. W., Petigura, E. A., 
Schlieder, J. E.,  Crossfield, I. J. M., et al.\ 2016, \apj, 827, 785

\bibitem[Sinukoff et al.(2017)]{Sinukoff_2017}
Sinukoff, E., Howard, A. W., Petigura, E. A., 
Fulton, B. J., Crossfield, I. J. M., et al.\ 2017, \aj, 153, 271 

\bibitem[Slawson et al.(2011)]{Slawson_2011}
Slawson, R. W., Pr{\v s}a, A., Welsh, W., F. 2011, \aj, 142, 160

\bibitem[Smith \& Lissauer(2009)]{Smith_2009}
Smith, A. W., \& Lissauer, J. J. 2009, Icar, 201, 381

\bibitem[Southworth(2011)]{Southworth_2011}
Southworth, J. 2011, \mnras, 417, 2166

\bibitem[Southworth et al.(2017)]{Southworth_2017}
Southworth, J., Mancini, L., Madhusudhan, N.,
Molli{\`e}re, P., Ciceri, S., et al.\ 2017, \aj, 153, 191

\bibitem[Tegmark et al.(2004)]{Tegmark_2004}
Tegmark, M., Strauss, M. A., Blanton, M. R., 
et al. 2004, PhRvD, 69, 103501

\bibitem[Ter Braak(2006)]{terBraak_2006} 
Ter Braak, C. J. F.  2006, Statistic.\ Comput., 16, 239

\bibitem[van Grootel et al.(2014)]{vanGrootel_2014}
van Grootel, V., Gillon, M., Valencia, D., et al. 2014, \apj, 786, 2 

\bibitem[Vanderburg et al.(2015)]{Vanderburg_2015}
Vanderburg, A., Montet, B. T., Johnson, J. A., et al. 2015, \apj, 800, 59

\bibitem[von Braun et al.(2012)]{vonBraun_2012}
von Braun, K., Boyajian, T. S., Kane, S. R., et al. 2012, \apj, 753, 171

\bibitem[Weiss et al.(2013)]{Weiss_2013}
Weiss, L. M., Marcy, G. W., Rowe, J. F., et al. 2013, \apj, 768, 14

\bibitem[Weiss et al.(2016)]{Weiss_2016}
Weiss, L. M., Rogers, L. A., Isaacson, H. T., 
Agol, E., Marcy, G. W., Rowe, J. F., Kipping, D., 
Fulton, B. J., Lissauer, J. J., Howard, A. W. \& Fabrycky, D. 2016,
\apj, 819, 83

\bibitem[Welsh et al.(2012)]{Welsh_2012}
Welsh, W. F., Orosz, J. A., Carter, J. A., 
Fabrycky, D. C., Ford, E. B., et al. 
2012, Natur, 481, 475

\bibitem[Welsh et al.(2015)]{Welsh_2015}
Welsh, W. F., Orosz, J. A., Short, D. R., et al. 2015, \apj, 809, 26

\bibitem[Wisdom \& Holman(1991)]{Wisdom_1991}
Wisdom, J. \& Holman, M. J. 1991, \aj, 102, 1528

\bibitem[Winn et al.(2011)]{Winn_2011}
Winn, J. N., Matthews, J. M., Dawson, R. I., et al. 2011, \apjl, 737, L18


\end{thebibliography}
\end{document}